# Task Force On Cosmic Microwave Background Research

**Final Report**

July 11, 2005

# MEMBERS OF THE CMB TASK FORCE

| | |
|---|---|
| James Bock | Caltech/JPL |
| Sarah Church | Stanford University |
| Mark Devlin | University of Pennsylvania |
| Gary Hinshaw | NASA/GSFC |
| Andrew Lange | Caltech |
| Adrian Lee | University of California at Berkeley/LBNL |
| Lyman Page | Princeton University |
| Bruce Partridge | Haverford College |
| John Ruhl | Case Western Reserve University |
| Max Tegmark | Massachusetts Institute of Technology |
| Peter Timbie | University of Wisconsin |
| Rainer Weiss (chair) | Massachusetts Institute of Technology |
| Bruce Winstein | University of Chicago |
| Matias Zaldarriaga | Harvard University |

# AGENCY OBSERVERS

| | |
|---|---|
| Beverly Berger | National Science Foundation |
| Vladimir Papitashvili | National Science Foundation |
| Michael Salamon | NASA/HDQTS |
| Nigel Sharp | National Science Foundation |
| Kathy Turner | US Department of Energy |

**Table of Contents**

# EXECUTIVE SUMMARY

One of the most spectacular scientific breakthroughs in past decades was using measurements of the fluctuations in the cosmic microwave background (CMB) to test precisely our understanding of the history and composition of the Universe. This report presents a roadmap for leading CMB research to its logical next step, using precision polarization measurements to learn about ultra-high-energy physics and the Big Bang itself.

*How did the Universe begin?* This question has exercised the human imagination for millennia. In the last 100 years we have been able to address it scientifically. Now, for the first time in history, the possibility exists to explore what transpired in the Universe in the first fraction of a second of its existence.

The Big Bang theory is now well established. When our Universe began about 13.7 billion years ago, it was extremely hot, filled with a myriad of exotic particles and expanding very rapidly. In its first short moments, it produced an excess of matter (from which everything we observe today is composed) over antimatter and synthesized light chemical elements such as helium, deuterium and lithium. Our Universe has expanded and cooled ever since, leaving behind a remnant of its hot past called the cosmic microwave background radiation (CMB). This radiation, discovered in 1965, has a temperature today of only 2.725 Kelvins, just barely above "absolute zero". *The CMB holds a remarkable wealth of information about the early Universe.* Observations of the CMB have recently transformed cosmology into a precision science.

While the basic notion of an expanding Universe is well established, fundamental questions remain, especially about the earliest moments of cosmic history. The prevailing idea is that the "bang" of the Big Bang theory was caused by a burst of nearly exponential expansion early on called *inflation*, after which our Universe coasted into a more leisurely expansion. Inflation elegantly explains why the geometry of space is Euclidean, why a faint pattern of fluctuations with amplitude almost independent of physical scale is seen in the CMB temperature, and why we do not observe relic particles such as magnetic monopoles.

But did inflation really happen and if so, why? How compressed and hot was our Universe when it happened? The simplest versions of inflation predict that it took place when the energy of particles and fields in our Universe was about $10^{16}$ GeV. This energy, 12 orders of magnitude higher than will be attained by the Large Hadron Collider at CERN, is also the energy scale at which the electromagnetic, weak, and strong forces are believed to unify in a so-called Grand Unified Theory (GUT).

Inflation was introduced to explain a variety of puzzling observations and did so with great success and economy. It must now be directly tested. A key prediction of inflation is that its exponential expansion left behind space-time ripples – gravitational waves – with an amplitude that depends on the energy scale of inflation. The direct detection of this gravitational radiation may one day become possible but will be difficult. Fortunately, there is an accessible alternative that is technologically feasible in the near term: gravitational waves from inflation imprint a unique pattern in the CMB polarization. *The accurate measurement of CMB polarization is the next critical step in extending our knowledge of both the early Universe and fundamental physics at the highest energies.*

The importance of the research has been recognized by a number of recent national and interagency reports. The 1999 National Academy of Sciences Board on Physics and Astronomy report, *Gravitational Physics: Exploring the Structure of Space and Time* recommended measurements of "the temperature and polarization fluctuations of the cosmic background radiation from arc minute scales to scales of tens of degrees," and said "Observations of these polarization fluctuations could lead to the detection of a stochastic background of gravitational waves from the early Universe." The 2001 report on *Astronomy and Astrophysics in the New Millennium* said "Gravitational waves excited during the first instants after the Big Bang should have produced effects that polarized the background radiation. More precise measurements of the properties of this polarization—to be made by the generation of CMB missions beyond Planck—will enable a direct test of the current paradigm of inflationary cosmology, and at the same time they will shed light on the physics of processes that occurred in the early Universe at energies far above those accessible to Earth-bound accelerators." Most recently, the 2003 National Research Council report, *Connecting Quarks with the Cosmos*, recommended that NASA, NSF, and DoE "Measure the polarization of the cosmic microwave background with the goal of detecting the signature of inflation" and "undertake research and development to bring the needed experiments to fruition."

In this report, we reaffirm the importance of these recommendations and lay out a roadmap leading to a precision study of the CMB polarization, thereby providing a way to test inflation and the theories on which it is based. Initially carried out with ground-based and balloon-borne experiments, the program culminates in a new space mission toward the end of the next decade. This mission will shed light on the earliest moments of cosmic history and the most fundamental building blocks of matter on a microscopic scale.

The roadmap includes complementary ground-based and balloon-borne observations of small-scale temperature and polarization fluctuations in the CMB. These measurements will refine our understanding of the history and properties of the Universe and of its contents. They will allow us to map the distribution of matter in the Universe by observing gravitational lensing of the CMB caused by large mass concentrations, and to extend measurements of the Sunyaev-Zel'dovich effect to many more galaxy clusters to investigate the evolution of the dark energy in the Universe. These measurements will also help to characterize foreground emissions that interfere with the search for CMB polarization signals. Finally, the ground-based and balloon-borne program will help refine the design and the instrumentation for a space mission.

It is likely that our European colleagues will be thinking along similar lines, and the challenging nature of the problem invites complementary approaches. Cooperation between communities should enable a coordinated attack on this very exciting problem. We hope and expect, however, that the US will maintain its established four decade old leadership in CMB studies.

We present here three recommendations to address the most compelling science we expect to come from observations of the CMB. These are followed by four recommendations on the technical developments that need to be supported to reach the scientific goals.

While the technological demands of this program are considerable, they can be met in the time frame we propose. The technical and scientific skills required to meet this challenge already reside among the scientists and engineers supported by the three agencies that sponsored this report. The search for CMB polarization offers an ideal arena for DoE, NASA, NIST and NSF interagency co-operation. Indeed, given the need for receiver development, ground-based observations, foreground characterization, and a space mission, the roadmap requires such cooperation.

*Science Findings and Recommendations*

**S1) <u>Finding</u>: A unique CMB polarization signal on large angular scales directly tests inflation and probes its energy scale.**

**<u>Recommendation</u>: As our highest priority, we recommend a phased program to measure the large-scale CMB polarization signal expected from inflation. The primary emphasis is to test whether GUT-scale inflation occurred by measuring the signal imprinted by gravitational waves to a sensitivity limited only by our ability to remove the astrophysical foregrounds.**

The phased program, described in §10, begins with a strong ground- and balloon-based program that will make polarization measurements on small and medium angular scales and culminates in a space mission for larger angular scales ($\theta > 1°$) specifically optimized, for the first time, to measure CMB polarization to a sensitivity limited only by our ability to remove the astrophysical foreground emission. We estimate that limits at or below $r = 0.01$ can be set on the amplitude of primordial gravitational waves: to reach this level, a sensitivity at least 10 times that of Planck will be required. The new mission is known as "CMBPOL" and is a candidate *Beyond Einstein* Inflation Probe.

**S2) Finding: The CMB temperature anisotropy on small angular scales contains a wealth of additional information about inflation and the evolution of cosmic structure.**

**Recommendation: We also recommend a program to measure the temperature and polarization anisotropy on small angular scales, including the signals induced by gravitational lensing and by the Sunyaev-Zel'dovich effect.**

The program described in §11 lays out a coherent set of ground-based experiments to measure the small-scale fluctuations in the CMB. The data so obtained will provide a valuable lever arm to help constrain the power spectrum of primordial fluctuations, which will, in turn, provide additional clues to the nature of inflation. Precise small-scale measurements will also help sharpen the constraints on a number of important cosmological parameters. Some of these are of direct interest to high-energy physicists, quantifying properties of dark matter, dark energy and neutrinos.

**S3) Finding: Foreground signals, particularly emission from our Galaxy will limit measurements of polarized fluctuations in the CMB.**

**Recommendation: We recommend a systematic program to study polarized astrophysical foreground signals, especially from our Galaxy.**

To achieve the primary science goals set forth in S1 and S2, a dedicated study of astrophysical foregrounds, described in §4, will be required. Indeed, while foreground signals have perturbed recent CMB *temperature* measurements, they have not dominated them, as they likely will for CMB polarization measurements. Specifically, polarized emission from synchrotron sources and dust clouds in our Galaxy will limit the measurement of the crucial, large scale, CMB polarization.

Since we know much less about the amplitude, scale and frequency dependence of polarized emission from foregrounds than we do about their unpolarized emission, improved knowledge of foreground emission is critical for optimizing the space mission design.

The report outlines the scientific and technical steps needed to carry out these challenging observations and to optimize the chances for success. The roadmap developed here shows how the information collected from current and new near-term experiments will feed into the design and technical base for the space mission. The timeline leads to a CMB polarization mission ready for launch in 2018.

*Technology Recommendations*

To reap these scientific rewards, we will need to improve technology in several areas. The first requirement is the development of arrays of polarization sensitive receivers.

**T1) We recommend technology development leading to receivers that contain a thousand or more polarization sensitive detectors, and adequate support for the facilities that produce these detectors.**

To meet the timeline outlined in this report there is a need to fund the development of polarization sensitive detectors at a level of $7M per year for the next 5 to 6 years. This would roughly restore the pre-2003 level of funding for the field, which has been especially hard-hit by the shift in NASA's priorities toward exploration. It is important to keep open a variety of approaches until a clear technological winner has emerged. *Nevertheless, highest priority needs to be given to the development of bolometer-based polarization sensitive receivers.*

Recommendation T1 requires maintaining core capabilities at NASA-supported centers for detector development and substantial support for detector development at DoE, NIST and University groups as well. Detector development is a particularly appropriate area for increased involvement by DoE in CMB research.

In addition to the development of receivers, we recommend continued NASA support for the development of space-worthy sub-Kelvin coolers to provide cryogenics for the focal planes containing such detector arrays, and interagency support for the development of polarization modulators to reduce systematic errors in the measurements.

As new technologies are developed, they need to be tested, both incrementally, and on a systems level. *The need to field-test emerging technologies is an additional reason for our science recommendations S1 and S2 for phased ground- and balloon-based programs measuring the CMB on both large and small angular scales.* Ground-based and balloon-borne projects will not only permit evaluation of rival technical approaches, but will provide crucial measurements over small areas of the sky or in restricted frequency ranges which can help refine the design of later, more complete, experiments with all-sky coverage.

In addition to the funds for detector development, we estimate the funding needed for the recommended ground- and balloon-based programs (in S1 and S2), including theory, and modeling, to be between $12M and $15M per year for the next 5 to 6 years.

**T2) We recommend a strategy that supports alternative technical approaches to detectors and instruments.**

Advances in CMB science have been based on a variety of technologies. Though we expect that bolometers will be the clear choice for CMBPOL, it is premature to shut down the development of alternatives. We recommend the continued development of HEMT-based detectors as they might lead to an alternative space mission and will certainly be used in ground-based measurements. Other risk management strategies include the development of absorber-coupled bolometers and the application of interferometers, as they offer an alternative approach to controlling systematic errors. These relatively inexpensive enhancements would lower risk by keeping a wider set of technology channels open until an accepted best method has emerged.

**T3) We recommend funding for development of technology and for planning for a satellite mission to be launched in 2018.**

We recommend funding for both development of technology and planning for a satellite mission to be launched in 2018. Background (CMB) noise limited receivers with thousands of elements and the sub-Kelvin cryogenics, required for these detectors, are part of the technical development required for the satellite mission. Another need is for modeling the mission based on improved knowledge of foreground emissions, to decide on the optimal spatial scale and frequency bands to separate the $B$-mode signals from the polarized foreground emission and to control systematic effects. As detailed in §10, preparation for a 2011 AO and a 2018 launch requires adequate funding, starting at $1M in 2007 and rising to $5M per year in 2011, for systems planning and technology development and assessment leading to CMBPOL.

**T4) We recommend strong support for CMB modeling, data analysis and theory.**

Large arrays of receivers will produce large data flows, so efficient data analysis algorithms will be needed. So too will access to high capability computing facilities. Equally important is additional research on efficient data analysis and modeling algorithms and on their implementation at appropriate facilities.

Paralleling the need for technology development is the need for support of theoretical research on the CMB and foregrounds. While the basic paradigm of inflation is clear, there are many details that need attention, spanning the range from fundamental physics research to CMB phenomenology, foreground modeling, and data analysis algorithms.

# 1 Outline of Report

We begin in §2 by presenting the science of inflation, the nature of the gravitational waves originating from the inflation and how they become visualized through the scattering of the CMB. Section 3 provides a pedagogical guide to understanding the CMB polarization *E* and *B* maps. The unique ability of the polarization to determine properties of the inflationary epoch is highlighted here. Next, in §4, astrophysical phenomena that disturb the measurements of the polarization are discussed. The mixing of polarization patterns by the gravitational lensing of intervening matter along the propagation path is discussed. In this section also the estimated amplitude of the CMB polarization is compared with the polarized astrophysical foreground emission. A program to measure these polarized foregrounds is outlined in this section.

The current and near term programs to measure the CMB polarization from the space missions WMAP and Planck, as well as the results from ground-based observations, are discussed in §5. The prospects for new ground-based and balloon-borne efforts to measure the polarization are also brought forward in this section.

Section 6 lists the requirements, as we now understand them, and discusses the observing strategies to make possible the difficult task of mapping the *B* modes. The control of systematic errors is discussed here as well; this control is at the heart of making believable CMB maps. The information in this section is the basis for designing the phased observing program we propose. This program, combined with modeling, will provide the necessary input for the design and planning of a dedicated CMB polarization space mission.

Sections 7 through 9 present examples of the current state of the receiver and telescope technology and the concepts and prospects for achieving the requirements laid out in the earlier sections. A central feature of the technology development is the need for receiver *arrays*. This includes manufacture of detector elements, the schemes to couple the incoming radiation to the detectors, the techniques to multiplex the low noise electronics, and methods to cool the detectors. All of this must be brought to a level where the receivers are limited only by the statistical fluctuations of the CMB radiation field itself.

The timeline to develop the knowledge, techniques and technology required for the task of measuring the *B* modes of the CMB is presented in §10. This section offers a good summary of the logic and the dynamics of the program.

A final section of the report (§11) deals with the plans for, and the expected results from, small-scale CMB (unpolarized) temperature anisotropy measurements. These will enrich our understanding of the inflationary period and provide another means of investigating the equation of state of the mysterious dark energy.

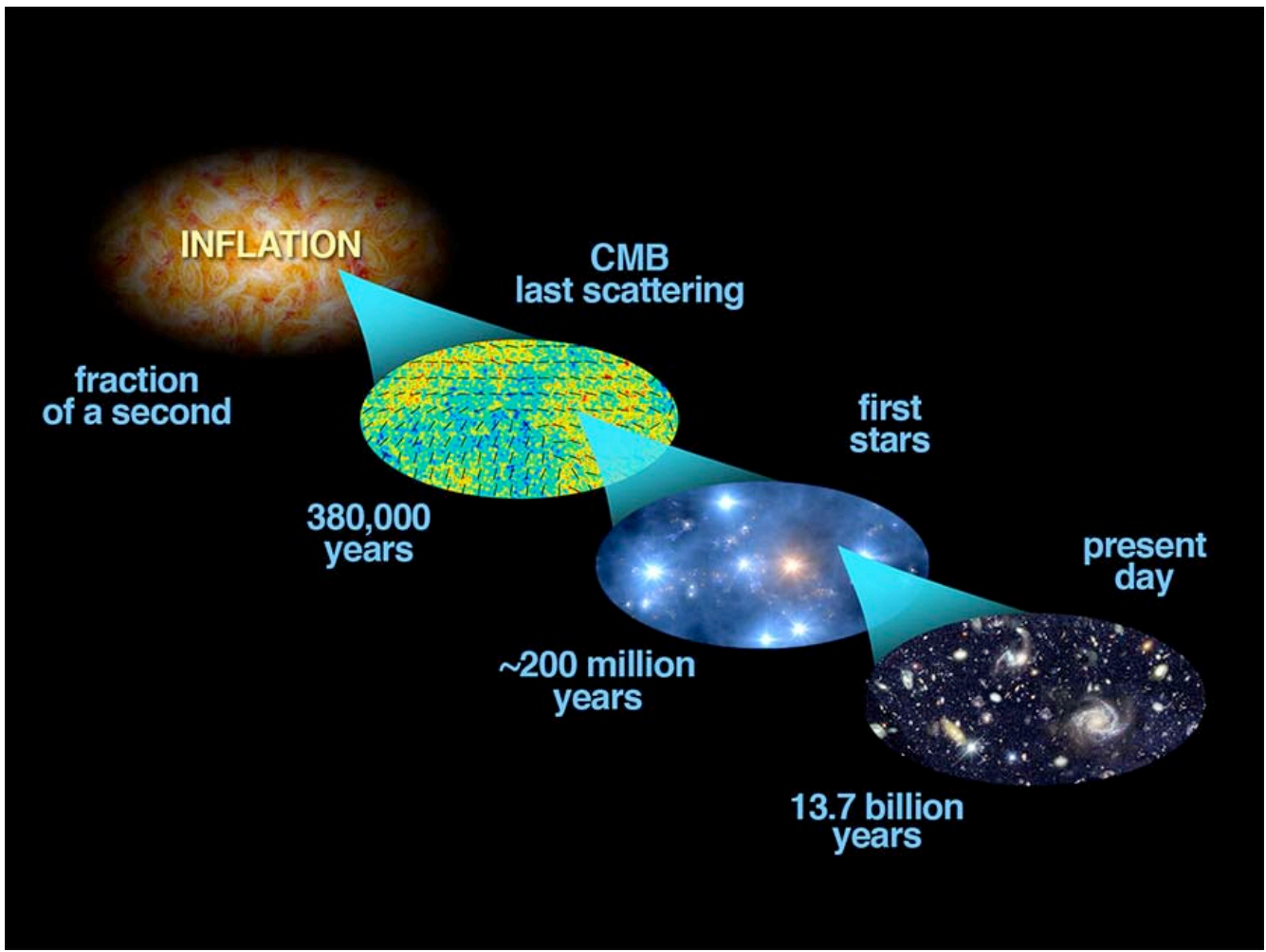

*Figure 1.1: Stages in the evolution of the Universe. According to the cosmological standard model, inflation stretches microscopic quantum fluctuations into astronomical density fluctuations that leave an imprint on the cosmic microwave background (CMB), and then grow into the present day galaxy distribution. This report presents a roadmap for measuring CMB polarization (illustrated by black rods), which encodes a signature of inflation.*

## Some History and Perspective

The discovery of the CMB moved the study of the universe to a central field of fundamental science. The new context for cosmology includes the dynamics and constituents of an evolving universe based on quantum and fundamental particle physics as well as the understanding of gravitation at the deepest level. This is all in addition to the traditional coupling to astronomy and astrophysics. Cosmology has become a branch of science that no longer fits neatly into one academic discipline, nor does it easily conform to the boundaries established in a prior epoch for the responsibilities assigned to the various government agencies. Cosmological research, in particular studies of the CMB, cuts across the missions of DoE, NASA and NSF. There is ample precedent for a joint agency cooperative program in CMB measurements. Measurements of the CMB have improved due to a variety of platforms and different technologies. The CMB is a special field in which much progress can be made from the ground and suborbital platforms. In this regard, it is different from X-ray or many parts of IR astronomy where the atmosphere seriously intervenes and only space missions provide the relevant experience and produce the science. In CMB research space missions have provided the final step and so have been the means to get the definitive results. But the space missions have succeeded only after a significant effort has been mounted from the ground and from balloons to develop the technology, to learn about the systematics and foregrounds, to train observers, and, of course, to produce key science results.

Knowledge gained from ground and balloon based observations is applied to the design and later execution of space missions. Spectrum measurements from the ground and balloons laid the foundations for the COBE FIRAS investigations of the CMB spectrum. The instrument that was so successful on COBE in providing the Planck curve at 2.7K had been flown and tested on a balloon, and was perfected in a ground based program. The space mission then provided an unprecedented platform free of most of the systematic problems encountered on the ground and from balloons. It allowed full spectral coverage to deal with foregrounds known to be a problem from those prior observations. These same observations were used to choose the appropriate observation bands for FIRAS. The full sky and spectral coverage provided by the space mission then clinched the interpretation of the spectrum and set stringent limits on spectral distortions which still stand today.

Much the same sequence applies to measurements of the anisotropy of the CMB. These measurements were first attempted from the ground. As more and more became known about the expected level of anisotropy, however, it became clear that balloon and aircraft measurements would be needed to extend the frequency coverage in order to understand the foregrounds, as well as to reduce the atmospheric fluctuations. The techniques were developed to take differences and to control the systematic problems induced by sidelobes. A critical problem became calibration and, especially, the difficulty of connecting the data from different observing runs to attempt to make wider sky maps. The dipole anisotropy was finally observed from balloons and airplanes but none of the higher moments was reliably measured. The critical intrinsic anisotropies resulting from primeval quantum fluctuations were not observed until full sky coverage and long integration times were provided by the COBE space mission.

The story is much the same with the higher angular resolution maps being provided by the WMAP mission. Ground and balloon based observations opened the way, but it was again the control of the environment and the long integration times with quiet operation and the ability to observe at many wavelengths simultaneously that led to the spectacular results developed by the WMAP space mission.

Given the history of measurements in this field it is particularly appropriate that there be a coordinated effort between DoE, NASA and NSF. In an earlier day NASA supported ground based work in support of planned space missions. While NSF has not been involved with funding a space mission directly, much of the ground- and balloon-based science and technology development it supports has fed into the NASA program in CMB research. A good example is the decades of work by the NSF-supported Princeton group in CMB measurements, which eventually found its way into the NASA programs involving COBE and WMAP. DoE has played a significant role over the years in supporting CMB anisotropy and spectrum observations through the program at Berkeley. The U2 program which firmly established the CMB dipole was a joint DoE and NASA program. These two agencies continue to collaborate in supporting detector development, theory and data analysis.

The program to measure the polarization of the CMB is challenging, but opens up the strong possibility of providing truly remarkable new insights into cosmology and fundamental physics. The program requires significant improvements in receiver technology, understanding of foregrounds, and observing techniques to control systematics. The necessary receiver technology takes the CMB field into an area where DoE, NASA and NIST have prior experience in developing large format arrays. The large volume of data that will come both from the polarization maps and from the experimental modeling to understand systematics, will involve the large computational facilities now being supported by DoE, NASA and NSF. The need to understand the foregrounds and to gain experience with the initial experiments provides the rationale for an active ground and balloon based program supported by NASA, NSF and DoE. It makes excellent sense for the three agencies to coordinate their activities in CMB research. Indeed, the roadmap we lay out requires such collaboration.

## 2 Cosmology and Inflation

Spectacular recent measurements enabled by detector, computer and space technology have given us a consistent, quantitative picture of how our Universe expanded and evolved from a hot, fiery beginning known as the Big Bang. Figure 2.1 quantifies this history by showing how the average energy density of the Universe ρ has decreased over time, continuously diluted by the expansion of space by a factor *a* that grows over time. The figure illustrates that while space has stretched by more than 50 orders of magnitude, the density has dropped by a much larger factor from values vastly above the present. As a consequence, the early Universe is an unmanned physics laboratory probing fundamental physics at density, temperature and energy scales vastly exceeding those accessible in laboratories.

The core subject of this report is how the properties of this cosmic cauldron can be measured from the fluctuations in the cosmic microwave background and their polarization (figure 2.2), and then used to deepen our understanding of fundamental physics.

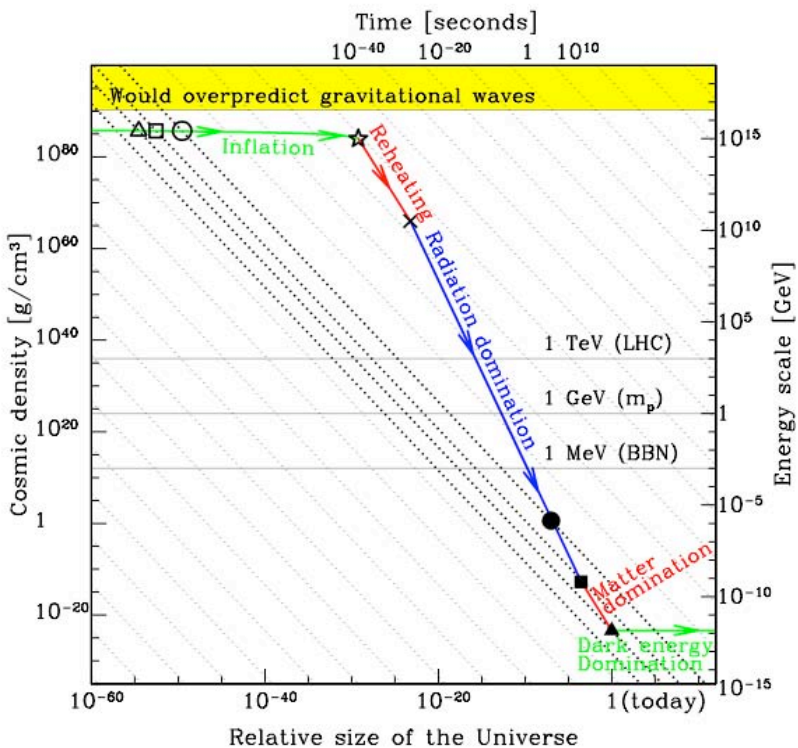

*Figure 2.1: The cosmic mean density (solid curve) is diluted as the Universe expands. Inflation is a period when there is almost no dilution of the cosmic density. Inflation causes the expansion to accelerate by a factor of $>10^{26}$. Acceleration corresponds to the curve decreasing more slowly than the dotted diagonal lines of slope –2. The two triangles lie on the same diagonal, which means that quantum fluctuations generated during inflation at the open triangle have been stretched into the horizon-scale fluctuations that we observe today, shown by the filled triangle. Detecting inflationary gravitational waves with CMB polarization would directly measure the shape of the cosmic density curve in the upper left corner of the plot, just as the proposed Joint Dark Energy Mission would directly measure the same curve in the lower right corner.*

### 2.1 A brief history of the Universe

Our Universe has expanded ever since the Big Bang, and this continuous stretching of space has both diluted and cooled the particles permeating it. As it cooled, particles combined into progressively more complex structures. Around 1 GeV, quarks combined to form protons and neutrons. Around 1 MeV, when the cosmic temperature was comparable to the core of a star, fusion reactions combined neutrons and some of the into light elements like helium, deuterium and lithium in a process known as *Big Bang Nucleosynthesis*. About 400,000 years after the Big Bang, the leftover protons combined with electrons to form electrically neutral hydrogen atoms. This epoch is known as *Last Scattering* (figure 1.1), because neutral hydrogen gas is sufficiently transparent to light that typical photons (particles of light) suddenly ceased to scatter. As a consequence, this is the surface we 'see' when we observe the CMB. Finally, atoms gradually combined into still more complex structures like molecules, stars and galaxies. It is in this epoch when stars began to form that a new but weak plasma is created in the Universe which once again scatters some of the photons, ones that play a significant role in measurements described in this report.

Since light travels with a finite speed, telescopes allow us to see the past. Just as sunlight shows us what the Sun looked like 8 minutes ago, light emitted from the last scattering surface shows us what the Universe looked like some 13 billion years ago. This light is called the Cosmic Microwave Background Radiation (CMB), and provides the centerpiece of this report. Predicted around 1950 and detected in 1965, the CMB reaches us as heat radiation with a temperature around 2.7 Kelvin, just barely above absolute zero.

## 2.2 Cosmic clustering

As our Universe expanded, it grew not only cooler but also clumpier. In the standard model of cosmology, the hot and dense early Universe was almost perfectly uniform, with density variations from place to place of order only $10^{-5}$. These small density variations produce the CMB temperature fluctuations we observe (figure 1.1). These tiny primordial density fluctuations were amplified by gravitational instability and modulated by plasma pressure and a variety of other well-understood effects, growing into the galaxies and the large-scale structure that we observe around us today.

The basic reason for this is that gravity is an attractive force: if a region contains slightly more matter than average, its gravitational pull will make it grow by attracting matter from its surroundings.

## 2.3 Concordance cosmology: successes and puzzles

The standard model of cosmology we have described is remarkably successful. Using standard gravitational, plasma and nuclear physics, it fits an impressive range of cosmological observations, including measurements of the cosmic expansion history $\rho(a)$ shown in figure 2.1 (based on local Hubble constant measurements and type Ia supernovae), the growth and scale dependence of fluctuations (including the CMB, galaxy clustering, galaxy cluster surveys, gravitational lensing, the so-called Lyman $\alpha$ forest and other techniques), and the abundance of light elements from Big Bang Nucleosynthesis. However, this standard model involves a number of free parameters of a rather empirical nature and these raise crucial questions about the underlying physics. One group of these cosmological parameters specifies the cosmic matter budget, consisting of about 5% ordinary baryonic matter (atoms), 25% dark matter and 70% dark energy. The standard model, however, leaves unanswered the pressing questions of the nature of dark matter and dark energy. Other cosmological parameters in the standard model characterize the small primordial seed fluctuations, begging the question of what generated them. The logical next step for cosmology research is therefore to use new measurements to address these open questions, thereby strengthening the bridge between astrophysical observation and fundamental theoretical physics.

## 2.4 Inflation

The leading paradigm for what produced these seed fluctuations is *inflation*. Inflation is defined as an epoch when the cosmic expansion accelerated ($\ddot{a} > 0$) with nearly constant energy density. This corresponds to the density curve $\rho(a)$ in figure 2.1 dropping only slowly for $a < 10^{-30}$, where a is the relative size of the Universe. This has emerged as the leading scenario for what happened early on because (with caveats to which we return below) its predictions fit observations well and solve important cosmological problems:

• Inflation solves the "flatness problem". Space grows more flat during inflation and less flat afterwards, so that without inflation, generic initial conditions would predict curvature growing over time and the density rapidly approaching either zero or infinity.

• Inflation solves the "horizon problem". Figure 2.1 shows that unless the density history $\rho(a)$ crosses the dotted diagonal through the present epoch (filled triangle), regions we see in two opposing directions in a CMB map would never have been in causal contact and we would have no explanation for their nearly identical temperatures.

- Inflation solves the "monopole problem" by diluting away unobserved relics from phase transitions in the early Universe.

Inflation predicts the existence of primordial density fluctuations that have specific properties known as scalar, scale-invariant, adiabatic and Gaussian. This last prediction is crucial, being the key to precision tests of inflation and to deepening our understanding of the underlying physics, as we will now describe.

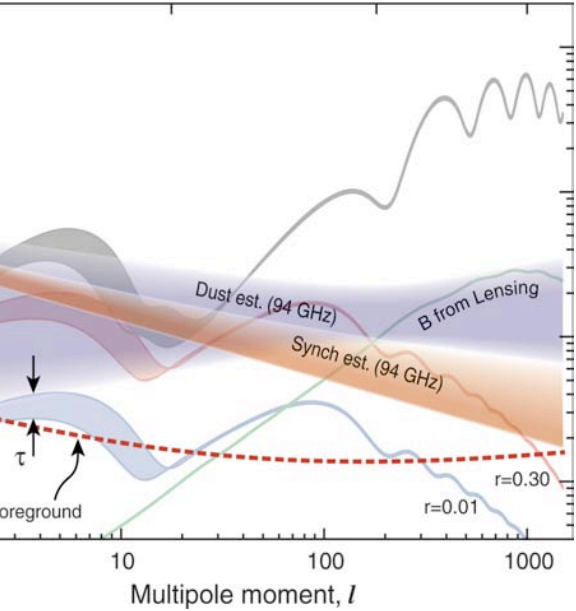
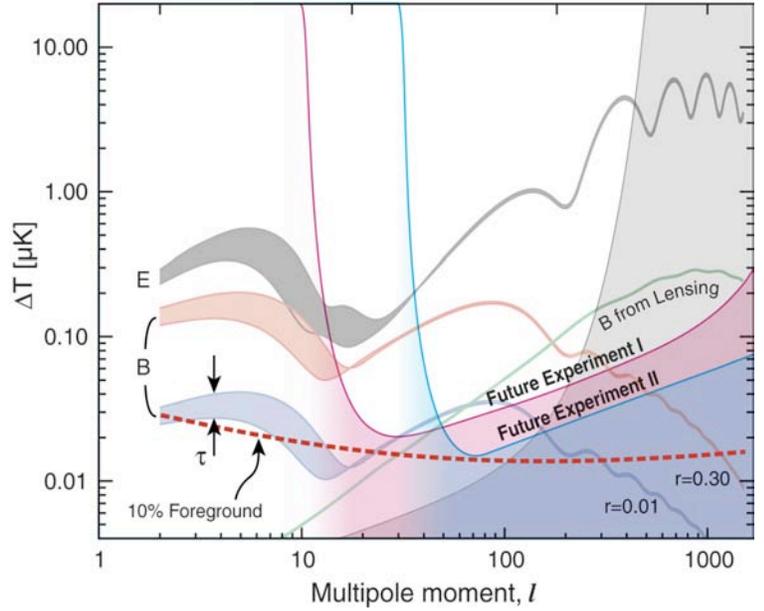

*Figure 2.2: CMB Polarization Power Spectra, Backgrounds and Estimated Sensitivity of Future Experiments. Thin, wiggly curves in the two panels show the predictions for the angular power spectrum of the CMB polarization signal (E modes and B modes) in the standard cosmological model (as of 2005). The E signal is reasonably well predicted, but the B signal depends linearly on the gravitational wave amplitude, as measured by the tensor-to-scalar ratio, r. The B curves shown are for r=0.3 (red) and r=0.01 (blue). For l < 20, the thickness of the theory curves reflects the current degree of uncertainty about the epoch of reionization. The predicted B-mode signal due to the distortion of E modes by weak gravitational lensing is shown in green (see §3).*

*Left: Current estimates of the polarized Galactic foreground signals and their uncertainty, due to synchrotron emission from cosmic ray electrons and thermal emission from interstellar dust grains, as described in §4. The red dashed curve is an estimate for the residual foreground contamination after modeling using multi frequency observations described in §4.*

*Right: Estimated instrumental sensitivities for a space mission of the type called for in our roadmap (grey shading) and two sample ground-based experiments (solid lines), based on the assumptions listed in §10.*

[*]The two ground-based experiments assume 1000 element receivers operating for one year. Experiment I observes 4% of the sky with 6 arc minute resolution while experiment II observes 0.4% of the sky with one arc minute resolution. The experiments lack sensitivity at low $l$ due to their limited sky coverage. The noise estimates are statistical and do not include the fluctuations from the uncertainties of the foreground modeling.

## 2.5 Gravitational waves: the ultimate probe of early Universe physics

Gravitational waves are ripples in spacetime that propagate with the speed of light. They are associated with nonuniformities in spacetime. According to the Heisenberg uncertainty principle, spacetime cannot be completely uniform – instead, quantum fluctuations in the fabric of spacetime will even produce gravitational waves of wavelength comparable to the size of the observable Universe. As we show below, CMB polarization measurements can detect them if their amplitude is sufficiently large.

*Detecting primordial gravitational waves would be one of the most significant scientific discoveries of all time.* The reason for this is that the square of the dimensionless amplitude of these gravitational waves is simply $\rho(a)$, the density during inflation expressed in so-called Planck units. Measuring the gravitational wave power spectrum would provide a direct measurement of the cosmic density history while it remains relatively constant during inflation (as shown in figure 2.1). This would not only demonstrate that something akin to inflation actually happened, but also give tantalizing information about physics on energy scales vastly exceeding those accessible in laboratories.

Figure 2.1 shows that inflation corresponds to a period when the density of the Universe did not get significantly diluted as space expanded. Numerous models have been proposed for substances that are hard to dilute, notably ones involving so-called scalar fields. The energy density in these fields is proportional to the energy density in gravitational radiation. *Many of the most compelling models predict a gravitational wave amplitude r large enough to be detectable with CMB polarization.*

Figure 2.2 illustrates the power of CMB polarization (discussed in more detail in §3) for measuring these inflationary gravitational waves. The so-called power spectrum of primordial gravitational waves is customarily fit by a power law, that is, a line in a log-log plot of power versus spatial frequency, of slope $n_s$ and normalization $r$ relative to the power spectrum of scalar (density) fluctuations. In turn, $r^{1/4}$ is the inflationary energy scale expressed as a fraction of $2 \times 10^{16}$ GeV. Current limits are $r \lesssim 0.5$ (figure 2.3); classic inflation models typically predict $r \sim 0.1$ or larger, detectable by its CMB polarization signature.

## 2.6 Other inflationary observables

Independently of whether primordial gravitational waves are detected, a sensitive, next-generation CMB satellite could dramatically enhance our understanding of the early Universe by measuring three other observables connected with inflation.

1. Departures from scale invariance.
2. Non-Gaussianity.
3. Isocurvature modes.

None of these has been unambiguously detected so far, since, with a few minor caveats of unclear significance, all measurements to date are consistent with "vanilla" primordial fluctuations. By this we mean the very simplest case known as scalar, scale-invariant, Gaussian, adiabatic fluctuations. These are defined by only one free parameter, their amplitude $\sim 10^{-5}$. Although theorists have generalized the classic inflation models (involving a single slow-rolling scalar field) in many different ways over the past two decades, essentially all of these models naturally predict some form of "non-vanilla" behavior that is potentially observable in the power spectrum of primordial density fluctuations, which is customarily fit by a quadratic polynomial in log-log space, specified by its amplitude $\sim 10^{-5}$, slope $(n_s-1)$ and curvature $\alpha$ at wave number $k = 0.05$/Mpc.

**Departures from scale invariance** imply $n_s \neq 1$ or $\alpha \neq 0$. Many classic inflation models predict $n_s \approx 0.96$ (*i.e.*, slightly more large-scale fluctuations than small-scale fluctuations) and $\alpha \sim (1 - n_s^2) \sim 10^{-3}$. Such departures from scale invariance are likely to be detected in the near future (see §11), but the precision measurements permitted by a next-generation CMB satellite would provide crucial clues about inflation.

**Non-Gaussianity** means that the joint probability distribution for the density at different points is not a multivariate Gaussian distribution. Classic inflation models predict almost perfectly Gaussian fluctuations that have their origin in the Gaussian ground state of a quantum harmonic oscillator. Some inflation models, however, predict departures from Gaussianity that are small but detectable with precision CMB measurements.

**Isocurvature modes** are fluctuations in the relative abundances of different substances (photons, dark matter, *etc*.), rather than overall fluctuations of all components together. Some inflation models predict such fluctuations at a level that is small but again detectable with precision CMB measurements.

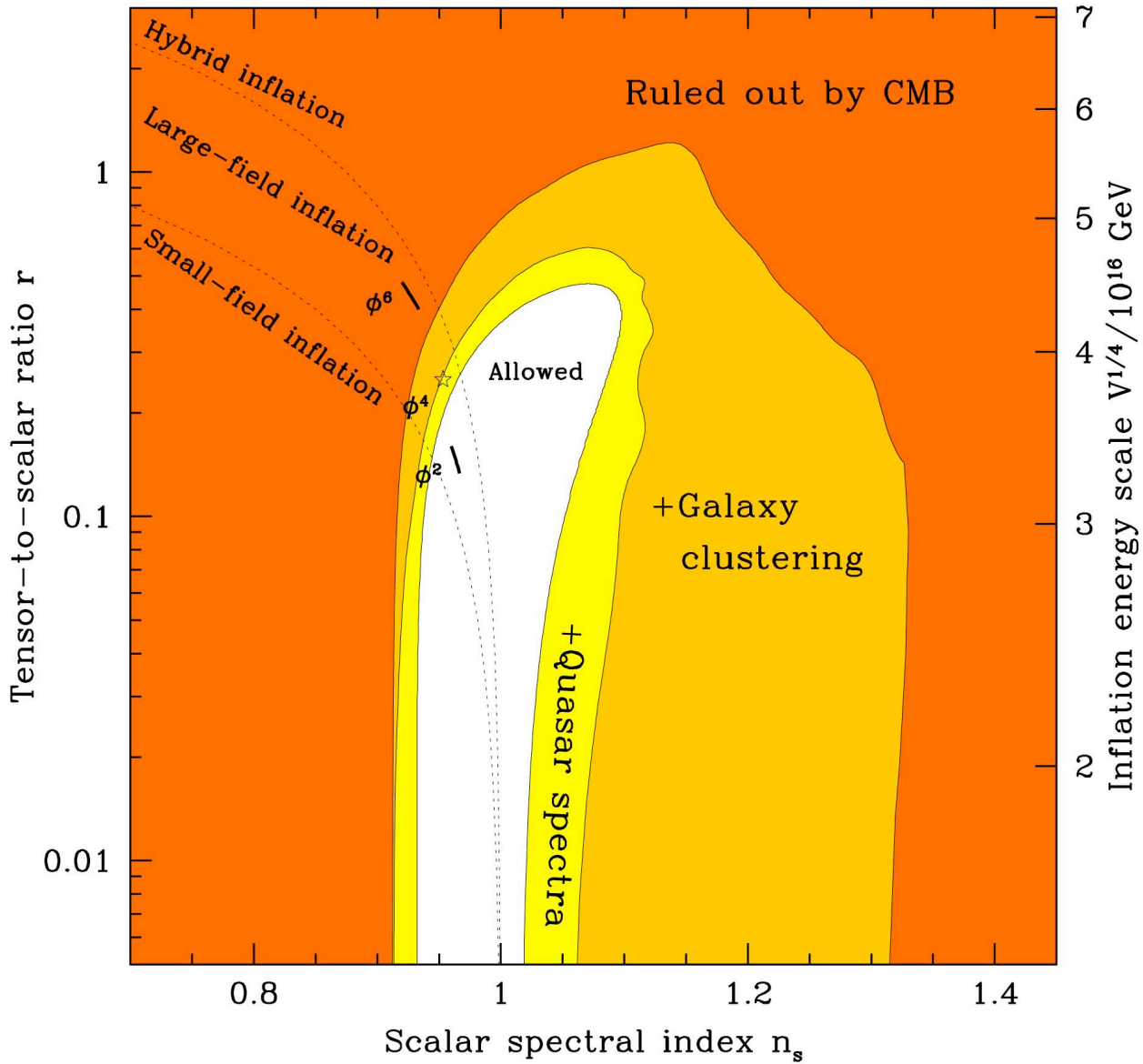

*Figure 2.3: Current constraints and predictions in the ($n_s$, r) plane. The outer regions are ruled out at 95% confidence from CMB measurements alone. Further limits come from adding galaxy clustering information, and finally from adding information from clustering in quasar absorption-line spectra as well. Dotted lines delimit classes of single-field inflation models and solid line segments show predictions from specific models, some already excluded. The CMBPOL satellite mission, recommended by this Task Force, would be designed to reach **r < 0.01**.*

## 2.7 Inflationary taxonomy

To give a flavor of inflationary theory, let us briefly survey different classes of inflation models, and explain how the four above-mentioned observables can test and distinguish among them.

Perhaps the most interesting aspect of inflation as a paradigm for what happened in the first moments of the evolution of the Universe is the fact that it gives a mechanism for generating small density perturbations which, after being amplified by gravity, lead to the structure we observe around us today. During the period of inflation the quantum fluctuations in any massless field freeze in amplitude when their frequency approaches the expansion rate of the Universe. Even though the amplitude becomes constant, their wavelength, or physical size, expands to extremely large scales. *This expansion during inflation and during the subsequent evolution of the Universe takes small quantum fluctuations and stretches them to scales interesting for astronomy.*

Gravitational waves are massless so they are always created during inflation. Another light degree of freedom is needed to generate the density perturbations. The fact that inflation has to end usually implies that all models already have a degree of freedom (the "clock field") that is evolving as inflation proceeds. This brings us to the first fork in classifying inflation models: whether it is the fluctuations in this "clock field" that are responsible for density perturbations, or whether fluctuations in other, additional fields play a significant role.

### 2.7.1 Classic models

Classic inflationary models, the first to be introduced, have the minimal number of ingredients. In these models the fluctuations in the clock field are the ones responsible for the density perturbations. A scalar field rolls down a potential very slowly, braked by the "friction" produced by the rapid expansion of the Universe. This potential is rather shallow, to ensure that the field rolls only slowly. A field with very small kinetic energy acts as a cosmological constant and thus drives the Universe to an accelerated expansion.

All the ingredients for the generation of perturbations are already there. In these classic models, no extra physics is introduced to end inflation either. As the field rolls slowly down the potential, the energy density decreases, slowing the rate of expansion of the Universe and thus lowering the amount of friction. There finally comes a point when the friction is not sufficient to brake the motion of the field, and it starts moving fast enough to end the epoch of inflation.

Since the classic models use the same degree of freedom as the clock, the source of density perturbations and the inflation terminator, they are the ones with the smallest number of free parameters. They generically predict observable departures from scale invariance and an observably large background of gravitational waves. In these models the observed amplitude of the density perturbations fixes the energy scale of inflation to be around $10^{16}$ GeV. This energy scale coincides with the so-called GUT scale where the strengths of the weak, strong and electromagnetic forces appear to unify. This suggests that inflation might be related to GUT-scale physics.

All models where the clock field is responsible for the density perturbations predict purely adiabatic fluctuations. Finally, the shape of the potential needed to guarantee a long period of slow roll also guarantees that the resulting fluctuations are very closely Gaussian.

### 2.7.2 Hybrid models

The first additional complication that can be added is to modify the physics that ends inflation. Well-developed models of this kind are the so-called hybrid models. In this scenario, when the clock field reaches a certain value, it triggers a phase transition in another field that in turn ends inflation. By adding physics at the end of inflation, these models can have more than one scale. The end of inflation can happen while the friction still dominates the evolution of the clock field. As a result, the gravitational wave background is no longer guaranteed to be observably large. The clock field is still rolling down a potential leading generically to departures from scale invariance. Fluctuations are still expected to be Gaussian and adiabatic.

**2.7.3 Other dynamics**

Another modification to the classic models is to change the dynamics of the scalar field, modifying its kinetic energy term. These types of models could be generically called "K-inflation." For example, the fluctuations in the clock field can propagate at a speed different than the speed of light. The fluctuations thus freeze in amplitude earlier in their evolution. As a result, their final amplitude is larger, decreasing the relative importance of gravitational waves.

There are many particle physics realizations of these ideas, sometimes differing dramatically in their details. In many cases, higher order terms in the Lagrangian originating from the same physics that modified the kinetic term lead to observable levels of non-Gaussianity, but fluctuations are still predicted to be adiabatic.

**2.7.4 Other fields**

Different behavior arises when more than one light field is present during inflation. There are several possible scenarios depending on the mechanism by which these additional fields produce the density perturbations. For example, these other fields could act like the Higgs field in the standard model, setting masses and couplings of particles in the early epoch of the hot Big Bang. This would cause these fields to change the efficiency with which reheating occurs in different places, leading to density perturbations. In other scenarios, one of the additional fields could store some energy that could lead to a second reheating later in the evolution of the Universe. Generically the availability of more fields (and therefore more than one clock) implies that not all regions of the Universe necessarily had the same history, which can potentially result in the generation of isocurvature modes. The fluctuations in these other fields need not be Gaussian and departures from scale invariance may also occur.

**2.7.5 Contraction**

There is a class of models that offers an alternative to standard inflation. In them, the perturbations are generated in a phase of the Universe prior to the Big Bang. Various scenarios along this lines have been proposed in the past decade, from the so-called "pre big bang" model, to the more recent "ekpyrotic" and "cyclic" scenarios.

In inflation the perturbations that gave rise to cosmic structure are produced in the earliest epochs of the Universe. The exponential expansion pushes the perturbations outside the horizon during a period when the horizon is changing very slowly, as shown in figure 2.1. As the Universe expands, the perturbations reenter our horizon. In the alternative models, the perturbations form in a contracting phase of the Universe when the horizon shrinks rapidly relative to the size of perturbations. At some point, the contraction is reversed and the standard Big Bang phase begins. Again, in the ensuing expansion, the preexisting perturbations enter our horizon. Whereas in inflation the perturbations are generated at high energies just after the Big Bang, in the alternative models they are generated before the Big Bang at significantly lower energies. As a result gravitational waves of cosmological wavelengths are produced in standard inflation but not in these alternative models. Thus, observations of the $B$ modes would rule out any such alternative. Short wavelength gravitational waves in the absence of cosmological ones would be a signature of these scenarios.

**Further reading:**

*The Cosmic Symphony,* Wayne Hu & Martin White, Scientific American, February 2004

Online cosmology primers:
*http://map.gsfc.nasa.gov/m_uni.html*

*http://astro.ucla.edu/~wright/cosmolog.html*

**Direct measurement of primeval gravitational waves**

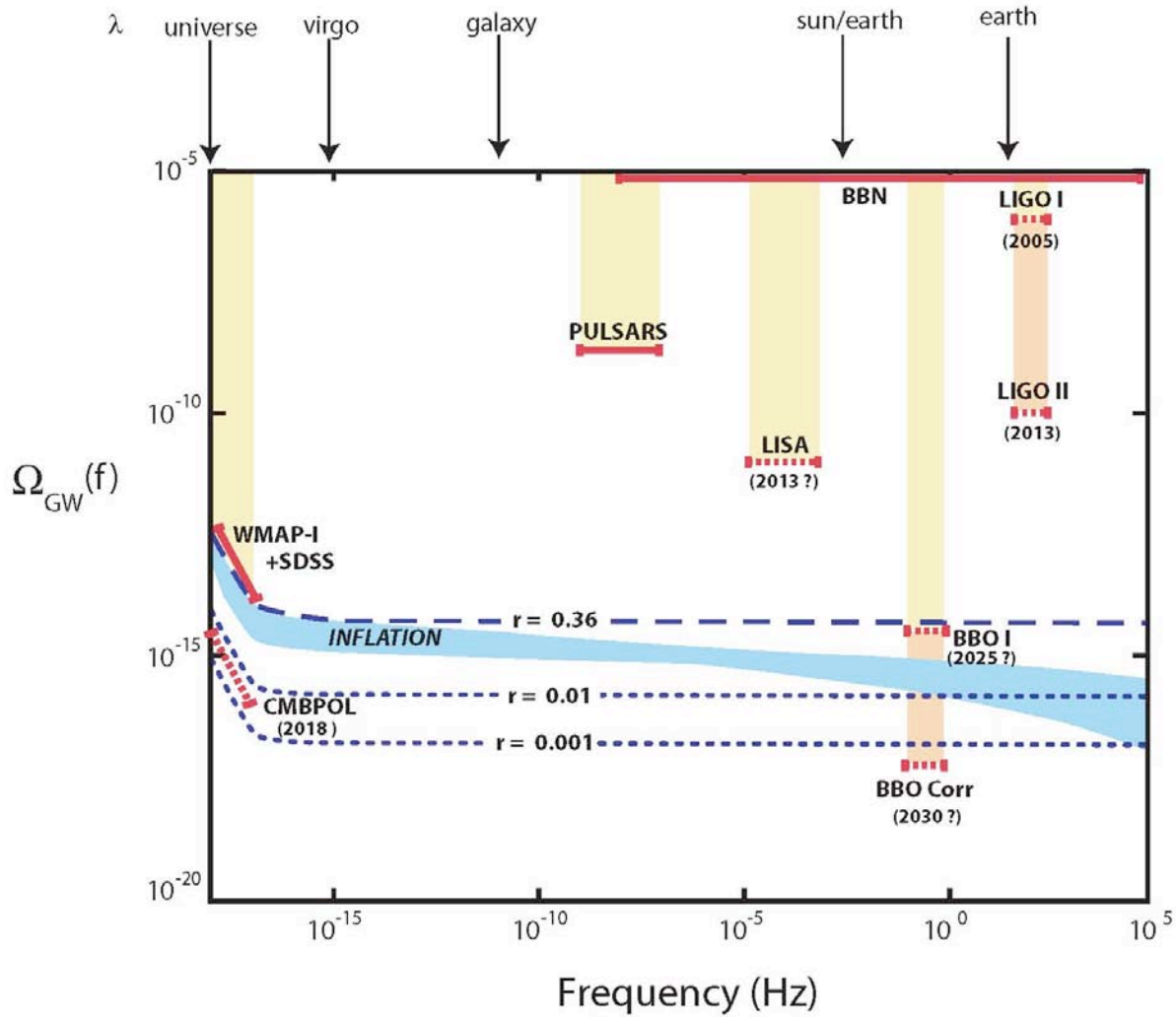

*Theoretical predictions and observational constraints on primordial gravitational waves from inflation are shown in a plot of $\Omega_{GW}(f)$ (the gravitational wave energy density per logarithmic frequency interval, in units of the critical density) versus frequency. The dashed curve (corresponding to tensor-to-scalar ratio r=0.36 is the maximum allowed based on WMAP-1 if the primordial spectrum is perfectly scale invariant ($n_t$=0). The dotted curves are the maximum for $n_t$=0 if r=0.01 or~0.001. Inflation produces a spectrum in which the index changes slowly as a function of frequency: the blue region represents the range predicted for simple inflation models with the minimal number of parameters and tunings. The currently existing experimental constraints shown are due to: big bang nucleosynthesis (BBN), binary pulsars, and WMAP-1 (first year) with SDSS. Also shown are the projections for LIGO (both LIGO-I, after one year running, and LIGO-II); LISA; and BBO (both initial sensitivity, BBO-I, and after cross-correlating receivers, BBO-Corr). Figure courtesy of Latham Boyle and Paul Steinhardt.*

The search for a gravitational wave stochastic background is also being carried out at very much higher frequencies by interferometric gravitational wave detectors. The Laser Interferometer Gravitational-wave Observatory (LIGO) operates 4km long interferometers at two widely separated sites. The interferometers obtain their best sensitivity in a band extending from 70 to 300 Hz. The search for a stochastic background is carried out by measuring noise common to the interferometers using cross-correlation. At initial design sensitivity, which will be attained in 2005, the $\Omega_{GW}(f)$ detectable will be close to $10^{-7}$. With an upgrade planned to be operating by 2013 LIGO will extend its sensitive band in low frequency to 15Hz and improve its limiting sensitivity by a factor of 15. With these changes the limiting value for $\Omega_{GW}(f)$ will become $10^{-10}$, still not at the level anticipated for slow roll inflation but able to detect a variety of other models.

The Laser Interferometer Space Antenna (LISA) is a configuration of three spacecraft placed at the vertices of an equilateral triangle with sides $5 \times 10^6$ km long. The sensitive observing band extends from 0.1mHz to 10mHz. The measurement of a stochastic background would be carried out by measuring the noise of the system using observational modes of the interferometer sensitive to gravitational waves and then subtracting from this (in power) the intrinsic noise of the system in a mode not sensitive to gravitational waves. LISA will need to contend with a foreground of gravitational wave "noise" from the unresolved gravitational wave emission of ordinary white dwarf binaries in our Galaxy. As is true for LIGO, LISA does not have the sensitivity to measure the anticipated level for slow roll inflation.

Big Bang Observer (BBO) is a concept being considered for launch after LISA. BBO is being planned to fill the frequency gap between the ground based interferometers and LISA, the band from 10mHz to 1 Hz. It will use high power interferometry on baselines of $5 \times 10^4$ km in triangular configurations, ultimately in a hexagonal pattern but with three widely separated constellations of spacecraft. The sensitivity projections for a single configuration used in a similar mode to LISA approaches the slow roll inflation prediction for $\Omega_{GW}(f)$. A later phase where cross correlation is done between configurations, much as in the LIGO program, could reach well below the slow roll values. BBO has to contend with the foreground of compact binary coalescences of neutron stars and black holes throughout the entire universe. The temporal signature of these coalescences will be used to remove them from the stochastic background.

# 3 Theory of CMB Polarization and Gravitational Waves

## 3.1 Statistical characterization of CMB anisotropies

To characterize the anisotropies in the CMB, one needs to specify three numbers for every point in the sky: one to give the overall intensity of the radiation (or equivalently the temperature $T$ of the blackbody spectrum) and the other two to specify its polarization properties (assuming only linear polarization is present). Two numbers are needed to characterize the linear polarization, a degree of polarization $P$ and an angle ($\alpha$) between the direction of the polarization and some specified coordinate system on the sky. Rather than $P$ and $\alpha$ it has become standard to use the Stokes parameters $Q$ and $U$, defined by $Q \equiv P \cos 2\alpha$, $U \equiv P \sin 2\alpha$.

In principle, the fourth Stokes parameter $V$ that describes circular polarization is needed as well. However, circular polarization is not expected since CMB polarization is believed to arise only from scattering of the CMB photons and electrons, a mechanism that does not generate circular polarization.

Figure 3.1 shows a simulated CMB map. The polarization map looks almost like a map of vectors (or arrows) with the only exception that the polarization "rods" do not point, that is to say they are not arrows but rods. More precisely, a polarization rod describes the same polarization properties if it is rotated by 180°. For a vector or arrow this would be true only after a 360° rotation.

Maps of vectors can be decomposed into a gradient and a curl part. Similarly, polarization maps can be decomposed into two components usually called $E$ (the analog of the gradient component) and $B$ (the analog of the curl component). That is to say, one can characterize the polarization pattern in a map either by specifying $Q$ and $U$ at every point or by specifying $E$ and $B$. Figure 3.1 shows such a decomposition in a simulated portion of the sky.

As an illustration in figure 3.1, we also show simple cartoon-like examples of polarization patterns that have positive and negative measures of $E$ and $B$ around a certain point. The figure illustrates what is meant by the claim that $B$ patterns are "curl-like". Specifically, $E$ and $B$ have different properties when reflected across a line going through the centers of the patterns. After the reflection, the $E$ patterns are unchanged, while the $B$ patterns change from one to the other, from positive to negative $B$.

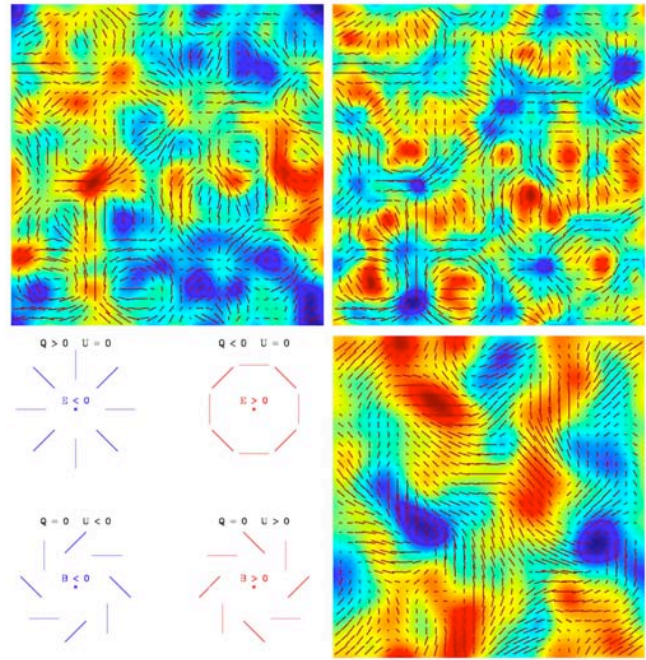

*Figure 3.1: The panels are: (top left) temperature map in the background, with polarization (rods); (top right)* **E** *map of the polarization field shown in the top left panel (background) and polarization coming only from* **E** *(rods); (bottom right)* **B** *map of the polarization field in the top left panel (background) and polarization coming only from* **B** *(rods); (bottom left) sketch of how* **E** *and* **B** *and* **Q** *and* **U** *are defined. The fields are 15 degrees on a side, and the maps have been smoothed with a one-degree beam for clarity. The scales have been adjusted from panel to panel so that the maps always have the same range of colors and the sizes of the rods designating the polarization amplitude are comparable. The scale of the* **B** *map was significantly stretched to correspond to expectations in the CMB. The B modes are predominantly at larger spatial scales.*

Whether a polarization field has an *E* or *B* component is a property of the pattern of polarization rods *around* each point in the map and not *at* the point itself. In this sense, *E* and *B* are not local quantities, they are not a new set of variables to describe the polarization at the point, but characterize the polarization pattern around the point. *The E-B characterization of the polarization field is more than a mathematical curiosity. As we will demonstrate in §3.4, B-type polarization is a powerful way to search for the existence of the stochastic background of gravitational waves predicted by inflationary models.*

The statistics of the CMB perturbations are primarily determined by the two point correlation functions of *T*, *E* and *B*. It is more convenient to describe the correlations in terms of the power spectra, which are effectively the Fourier transforms of the two point correlation functions. Under most circumstances only four power spectra are needed. The first three are the autocorrelation spectra for *T*, *E*, and *B*, which characterize the fluctuation levels in each of the three maps as a function of angular scale. *E* and *B* are shown in figure 2.2. The fourth is the cross correlation spectrum between *T* and *E*, which quantifies the relation between the temperature and *E* polarization maps shown in figure 5.1 as TE. The additional cross correlation spectra between *B* and *E* or *B* and *T* are expected to vanish for cosmological signals, unless the primordial fluctuations that result in the *B* pattern were generated by a mechanism that did not respect parity (or reflection) symmetry. These two cross spectra may thus be a useful tool for monitoring the effects of foregrounds and/or other non-cosmological systematic effects.

### 3.2 The physics of polarization generation

In this section we explain how the polarization anisotropies are generated.

### 3.2.1 The history of Hydrogen atoms

The most abundant element in our Universe is hydrogen and its ionization state has profound consequences for the CMB. Of particular importance for this report is the fact that polarization is generated by the scattering of CMB photons by electrons coming mainly from ionized hydrogen atoms. Thus it is crucial to understand during what periods in the evolution of the Universe hydrogen was ionized, as it is only during these periods that polarization can be generated.

The temperature of the CMB was higher in the past, increasing linearly with redshift, $T \propto (1+z)$. The interaction between the CMB photons and the hydrogen atoms kept hydrogen ionized until a redshift of $z \approx 1300$. After this time there are no longer enough energetic photons in the CMB to keep hydrogen ionized, so it is said to recombine.

When hydrogen is ionized the CMB photons can scatter with electrons, a process called Thomson scattering. The Universe has expanded by a factor of 1300 since recombination. This implies the density of electrons was much higher then than it is now, a billion times higher. As a result Thomson scattering was so frequent before recombination that a typical CMB photon could only travel a very short distance between successive scatterings. The photons and electrons are said to be "tightly coupled" at these early epochs.

The process of recombination happens rather fast. During a span of about 100,000 years most of the electrons in the Universe find protons and form neutral hydrogen atoms. At that point the density of electrons drops significantly, making Thomson scatterings extremely rare. The Universe becomes transparent to CMB photons, a process also called decoupling which happens at a redshift of 1100, 380,000 years after the Big Bang. CMB photons come to us from this epoch, from a spherical shell around us with radius of $4 \times 10^{28}$ cm called the last scattering surface.

Much later in the evolution of the Universe, perhaps 200 million years after the Big Bang, the first stars started shining. The starlight reionizes the hydrogen. The exact time and way that the first stars formed are not yet fully understood. The recent results from WMAP indicate that reionization started pretty early, even as early as $z \sim 20$. On the other hand there are also indications from studies of absorption by neutral hydrogen towards high redshift quasars, that the reionization process was still ongoing as late as a redshift $z \sim 6$, a billion years after the Big Bang. The reionization process may have been very long and complicated.

At this late stage in the evolution of the Universe, the density of hydrogen was much lower than at recombination. Hence, even though hydrogen is fully ionized, only a fraction of the CMB photons scattered with electrons after this reionization. A scattering percentage somewhere around 10% to 20% is the value favored in the latest analysis of WMAP data. Even though only a fraction of the photons scatter after reionization, this still has a dramatic effect on the production of polarization on large angular scales, of order ten degrees. On these scales the polarization produced at decoupling ($z \sim 1100$) is minimal because these scales are much larger than the mean free path at decoupling. Thus, the signal from reionization determines the shape of the power spectrum at large scales.

**3.3 The mechanism that generates polarization**

Polarization is generated by Thomson scattering between photons and electrons, but Thomson scattering is not enough: the radiation incident on the electrons must be anisotropic. Figure 3.2 illustrates how polarization is produced. The mechanism is similar to that operating in the earth's atmosphere. If one looks at the light from the sun that was scattered by the atoms in the atmosphere, it is partially polarized. This is a direct result of both the scatterings and the fact that the radiation field is not isotropic, with light coming mainly from the direction towards the sun.

The need to have both Thomson scattering and anisotropies is what makes the polarization of the CMB relatively small. Before recombination there were in fact too many scatterings. Multiple scatterings have the effect of making the radiation field incident on the electrons isotropic, significantly reducing the polarization that can be generated. On the other hand, immediately after recombination anisotropies in the radiation field can grow but there are no scatterings to generate polarization. It is only after the hydrogen in the Universe reionizes as a result of the radiation generated by the first stars that these growing anisotropies are converted to polarization. At this late stage the density of matter is very low so only a small of fraction of the photons scatter with electrons leading to only a low degree of polarization.

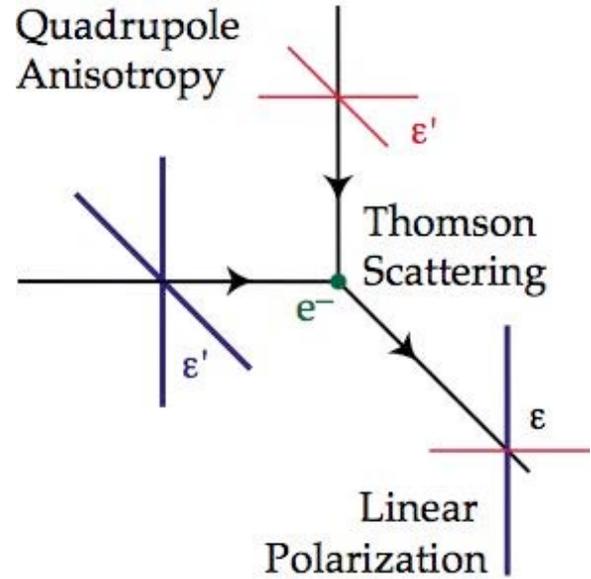

*Figure 3.2: Illustrations of the process by which the scattering of anisotropic and unpolarized radiation can lead to polarization. Unpolarized light of different intensities is incident on the electron. Only the component of the electric field in the up-down (top-bottom) direction is scattered from the radiations coming from the left (top). As a result the amplitude of the two polarizations of the scattered light are different. Figure courtesy Wayne Hu.*

*Polarization can be generated during two epochs, around 380,000 years after the Big Bang when hydrogen atoms become neutral for the first time, and around 200 million years after the Big Bang, when the first stars appear and reionize hydrogen.*

These anisotropies needed to create polarization can have two sources: anisotropies created by a stochastic background of gravitational waves, and anisotropies created by the density perturbations which will eventually grow to form galaxies and other structures we observe around us today.

As photons travel in the space-time perturbed by a gravitational wave their frequency is shifted to the red or blue by an amount that depends on their direction of propagation relative to the direction of propagation of the gravitational wave, the polarization of the gravitational wave and its frequency. The frequency shifts are equivalent to shifts in the temperature of the blackbody spectrum. The temperature fluctuations induced as photons travel between successive scatterings before recombination (or between recombination and reionization) produce an anisotropy in the intensity distribution that through Thomson scattering leads to polarization. The case of the polarization induced by density perturbations is similar. In this case the source of anisotropies are velocity differences in the photon-baryon fluid over a distance comparable to the distance photons can travel between their successive scatterings.

*Polarization is always generated by Thomson scatterings, irrespective of whether the polarization is generated by density perturbations or by gravitational waves.* The only difference is the cause of the anisotropy. This apparently small difference, the source of the anisotropies, has profound consequences. The spatial pattern of the anisotropies created by a gravitational wave has different symmetry properties than the pattern created by a single component of the density field. It lacks reflection symmetry and as a result *the pattern of polarization rods on the sky created by a gravitational wave can have a* B *component.*

### 3.4 The information encoded in the polarization of the CMB

**Gravitational waves and Inflation**: *The study of the anisotropies in the CMB produced by a gravitational wave background left over from the inflationary era is of paramount importance for understanding the physics of inflation. The typical amplitude of gravitational waves in the stochastic background is a remarkable probe of the physics of inflation. It is directly proportional to the rate of expansion of the Universe during inflation, or equivalently to the square of the ratio of the energy scale of inflation to the Planck energy scale.*

Gravitational waves produce both temperature and polarization anisotropies. The standard way to characterize their amplitude is through the so-called tensor to scalar ratio *r*. It gives the ratio between the temperature anisotropies produced by gravitational waves and by density perturbations at *l*=2. A tensor to scalar ratio $r \approx 0.1$ corresponds to an energy scale of inflation, $E_{inf} \sim 2 \times 10^{16}$ GeV, around the expected GUT scale.

The temperature fluctuations due to gravitational waves are produced mainly after recombination and have a power spectrum that is roughly flat up to $l \sim 100$ and then rapidly falls off. By contrast, polarization is produced both around recombination and after reionization. The power spectrum of polarization has two peaks, one at $l \sim 100$ from the polarization generated around recombination and one around $l \sim 6$ from the contribution generated after reionization (see figure 2.2). The amplitude of the peak in the *B* modes at $l \sim 100$ is approximately given by $\Delta T/T = [l(l+1)C_l^{BB}/2\pi]^{1/2} = 0.024 \, (E_{inf}/10^{16} \, \text{GeV})^2$ K.

The shape of the *E* and *B* power spectra created by gravitational waves can be understood as follows. Gravitational waves of wavelength much larger than the distance CMB photons travel between successive scatterings cannot generate an appreciable anisotropy and thus do not lead to polarization. The amplitude of gravitational waves decays when their period is shorter than the age of the Universe at a given epoch. The combination of these two factors selects a scale where the polarization produced by gravitational waves must peak. This scale is around $l \sim 100$ for the contribution coming from recombination, and $l \sim 6$ for the contribution produced after reionization, which occurs when the Universe was much older.

**Reionization:** As we discussed earlier, approximately 200 million years after the Big Bang the first light sources turn on and are able to reionize the hydrogen in the Universe. After this time, a fraction of the CMB photons scatter with electrons, thus increasing the degree of polarization of the CMB on large angular scales. A detection of this large scale signal will allow a precise measurement of the fraction of CMB photons that scattered during this period (the optical depth to the surface of last scattering) which can be translated into a measurement of when reionization happened. The height of the reionization peak is proportional to the total optical depth, while the shape and location of the peak have information about when and how reionization happened. The total amount of information that can be extracted is limited because at these large scales there are only a few multipoles that can be measured.

*Nevertheless, a measurement of the optical depth is important because it breaks degeneracies otherwise present in the determinations of several cosmological parameters. In particular, it would drastically reduce the errors in the shape of the primordial power spectrum. The temperature and small-scale polarization data can be explained equally well by models that have different shapes of the primordial power spectrum and different optical depths to compensate. Thus a definite measurement of the optical depth from the large-scale polarization will favor one model over the other, providing invaluable information to constrain inflationary models.*

**The properties of the primordial seeds:** It is customary to assume that the power spectrum of primordial fluctuations is a simple power law, perhaps with a logarithmically varying spectral index. Although this is well motivated in standard theories of inflation, it must be verified. So far the best constraints on the power spectrum of primordial fluctuations come from comparing the amplitude of temperature perturbations at different angular scales. At least some of the changes produced by differences in the primordial spectrum can be mimicked by changes in other cosmological parameters. If both temperature and polarization are measured, there are three independent measures of the level of fluctuations on each scale ($T$, $E$ and their cross correlation). Moreover, the shapes of the temperature and polarization power spectra depend differently on cosmological parameters. Thus, the simultaneous measurement of temperature and polarization can be used to separate the early (inflation era) physics and later physics affecting the CMB.

In the simplest inflationary models, perturbations come only in the adiabatic form. That is, we expect no fluctuations in the composition of the primordial soup, no spatial variations in the ratio of dark matter to baryon densities, for example. However, in more complicated inflationary models or in other classes of early Universe "scenarios," composition fluctuations usually called isocurvature perturbations can arise. In a model with photons, baryons, neutrinos and cold dark matter present, there are actually four isocurvature modes in addition to the adiabatic one. The simultaneous measurement of temperature and polarization will allow experiments to put constraints on small admixtures of these components that would be impossible to detect otherwise.

Finally, the Gaussianity of the primordial seeds is also an important window to understanding the physics responsible for their generation. The measurement of both $T$ and $E$ produced by density perturbations doubles the amount of information about the primordial seeds, significantly improving the ability of CMB observations to constrain any departures from Gaussianity.

**Consistency checks:** Many aspects of the polarization perturbations are accurately predicted once temperature is measured. As a result polarization can be used to make several consistency checks on the way recombination occurred. In particular, changes in fundamental constants such as the fine structure constant, the gravitational constant, or the speed at which the Universe was expanding during recombination should be severely constrained once polarization is accurately measured. Also constrained would be any presence of ionizing photons in addition to those from the CMB itself.

# 4 Astrophysical Disturbances in Measuring the CMB Polarization: Gravitational Lensing and Polarized Foreground Emission

Measurements of the CMB polarization will be disturbed by matter intervening between the last scattering surface and the observer. The perturbations occur both in the propagation of the radiation through the spatially varying gravitational potentials that generate gravitational lensing and by the direct emission of polarized radiation from electrons and dust. Because the amplitude of the CMB polarization signals is small, foreground emission will present a bigger problem for polarization observations than for measurements of temperature anisotropies, as suggested in figure 2.2. While correction for polarized foreground emission will be a challenge, there are well established and proven techniques for making such corrections. These are based on the spatial distribution and frequency dependence of foregrounds that allow them to be distinguished from the CMB signals. Direct measurements and modeling of foreground emission should conservatively allow us to reduce its effect by a factor of 10, allowing us to measure CMB polarization at levels ~1/10 of the overall foreground emission level.

While the perturbations introduced by gravitational lensing and by emission from electrons and dust present problems for CMB observations, they are of considerable scientific interest in their own right. In this section we focus primarily on foregrounds as a source of noise, postponing discussion of the rich science yield of the foreground measurements we propose to later sections.

## 4.1 Weak Lensing

CMB photons are affected by a number of astrophysical processes after they leave the surface of last scattering at z ~ 1100. As they propagate from the last scattering surface they are gravitationally deflected by mass concentrations, the large-scale structure of the Universe. The typical deflection is around two arc minutes. Gravitational lensing changes both temperature and polarization anisotropies but has a particularly profound effect on the pattern of CMB polarization. Even a polarization pattern that did not have any *B* component at recombination will acquire a *B* component as a result of gravitational lensing. The effect is simple to understand. Consider a polarization map imprinted at the last scattering surface, and assume it just contains an *E* pattern. The effect of lensing is to shift around the position of the different polarization rods by random amounts of order one arc minute. This random shifting modifies the pattern, and because it is random it will naturally distort the map in a way that creates both *E* and *B* components. In Figure 2.2 we show the power spectrum of the *B* component generated by lensing of the *E* mode. It is clear from the figure that for a given low level of the gravitational wave background, the lensing signal would be larger on almost all scales, except at small *l*. The lensing *B* mode, however, does not have the reionization signature at low *l*, because the power is actually coming from "aliasing" of the small-scale polarization power rather than from a rearrangement of the original large-scale power.

The ultimate limitation on detecting a stochastic background of gravitational waves via the *B*-mode polarization they produce comes from the spurious *B* modes generated by lensing. The lensing distortions to the temperature and polarization maps also make them non-Gaussian. Methods have been developed to use this non-Gaussianity to measure the projected mass density, which can be used to correct this contamination, at least partially. *With the simplest "lensing cleaning" methods proposed so far, the lowest energy scale that seems measurable is* $E_{inf}$ = 2 x $10^{15}$ GeV. *At least in principle, however, it appears possible to go even lower with more sophisticated techniques.*

The lensing effect is not only a nuisance for detecting gravitational waves, it is interesting in its own right as a constraint on the large-scale structure that is doing the lensing. *The last scattering surface is at such high redshift that observations of the intervening lensing of the CMB may eventually provide one of the deepest probes for large-scale structure.* The amplitude of structure on scales of order 2 to 1000 Mpc at redshifts from z ~10 to z ~0 may eventually be constrained with this technique. Lensing will not only allow a measurement of the level of fluctuations but may lead to actual reconstructed maps of the projected mass density that can be correlated with maps produced by CMB experiments and other probes.

*CMB lensing maps, especially when combined with other probes of lensing towards the same direction of the sky, can also provide constraints on various cosmological parameters. In particular, the mass of cosmic neutrinos could be constrained to m < 0.05 eV, comparable with the amplitudes of the mass differences measured using neutrino oscillations.* Measurements of the lensing $B$ modes could also be used to study the growth of large-scale structure through cosmic time. These results in turn could constrain both the equation of state of dark energy and its sound speed.

### 4.2 $E$–$B$ Mixing: systematic effects

There are a variety of other systematic effects, all of which lead to mixing between $E$ and $B$ modes. In finite patches of sky, the separation of the two modes cannot be done perfectly. To do so is analogous to trying to decompose a vector field into its gradient and curl parts when it is measured on a finite part of a plane. Vector fields that are gradients of a scalar with zero Laplacian will have both zero curl and zero divergence. In fact, in a finite patch one can construct a basis for the polarization field in which the basis vectors can be split into three categories. There are pure $E$, pure $B$ and a third category of modes that are ambiguous, including contributions from both $E$ and $B$. The number of ambiguous modes is proportional to the number of pixels along the boundary of the patch. Aliasing due to pixelization also mixes $E$ and $B$, and the power that is aliased from sub-pixel scales leaks into both $E$ and $B$. This is particularly important in searches for $B$ modes because $E$ mode polarization is expected to be much larger than in the $B$ modes and because the $E$ polarization power spectrum is relatively very blue.

### 4.3 Foreground Emission

Emissions from astronomical foregrounds have affected measurements of temperature, or total power, fluctuations in the CMB, but have not dominated them. As indicated schematically in figure 2.2, foregrounds will present bigger problems in CMB polarization measurements. First, some foreground sources have a higher fractional polarization than expected for the CMB signals. A second and larger problem is that we know much less about the amplitude and scale of the polarization of foreground emission, especially from our own Galaxy, than we do about total power emission. A better understanding of polarized foregrounds may be needed even to plan optimum CMB polarization searches: for instance, the nature of foregrounds may have a strong bearing on the best observing frequencies to select. Better characterization of foreground polarized emission from the Galaxy and from a variety of extragalactic sources is therefore a key milestone in the roadmap we propose.

Three physical mechanisms are known to produce foreground contamination at microwave frequencies: synchrotron and free-free emission are major contaminants at frequencies below 50 GHz, while above approximately 100 GHz, dust emission is the major contaminant. There is also evidence for what is referred to as "anomalous emission," at centimeter wavelength, possibly from spinning dust grains, but very little is known about its origin, amplitude or frequency spectrum. Radiation from extragalactic objects is usually referred to as point source contamination and is significant primarily on small angular scales. Diffuse Galactic emission from the Milky Way causes fluctuation mainly on large angular scales. Except for free-free emission, these mechanisms are known to produce polarized radiation, although polarization fraction can vary widely, depending both on astrophysical conditions and on the observation wavelength.

Information about inflationary gravitational waves and reionization is mostly contained in large angular scale polarized fluctuations. The large patches of sky required to understand these phenomena will include areas where foreground emission from the Galaxy is significant and the resulting contamination of the polarization signal is unavoidable. On the other hand, small angular scale measurements can be made in selected

"clean" areas of the sky, where foregrounds are particularly small.

### 4.3.1 Galactic Synchrotron Emission

At this point we know very little about the polarized contribution of Galactic synchrotron emission at CMB frequencies. Recent measurements in the radio reveal that the polarized component of the synchrotron emission of our Galaxy shows significant structure and that it depends weakly on Galactic latitude. Analyses of synchrotron radiation from near the Galactic plane seem to suggest that synchrotron radiation will have little effect on the CMB polarization at frequencies above 100 GHz, at least on small angular scales. This conclusion, however, is based on an enormous extrapolation in frequency from 2.4 GHz, where the measurements were made, to the frequencies of most relevance to CMB measurements. And, although Faraday depolarization is substantial at low frequencies and insignificant at 100 GHz, spatially varying Faraday depolarization could still alias large-scale structure to smaller scales. Furthermore, synchrotron emission properties near the Galactic plane are likely to be substantially different from those at high latitudes, which argues for studies of foregrounds off the Galactic plane.

Uncertainties in our understanding of polarized Galactic synchrotron emission are reflected in figure 4.1. The figure displays the spectrum of the polarized foreground emission fluctuations as well as the anticipated amplitude of the *B* and *E* signals. There is a clearly defined "sweet spot" in frequency, which we expect to lie near the minimum in temperature foregrounds at ~70 GHz.

### 4.3.2 Dust Emission

Optical measurements of the absorption of Galactic starlight by intervening dust have revealed that dust can be polarizing. The dust grains are aspherical and are aligned orthogonally to the magnetic field of the Galaxy. When the warmed dust re-emits, the radiation is polarized perpendicular to the Galactic plane. The polarized dust emission depends on various astrophysical properties such as the dust's intrinsic dielectric properties, the geometry of the magnetic field, and coherence. Some of these properties in turn depend on complex astrophysical processes like star formation and turbulence. Because of the complex combination of factors and the lack of data, modeling dust contributions to polarization is difficult. Another problem with the existing data on dust polarization is that the measurements are almost exclusively on small scales near the Galactic plane, whereas the greatest need is on large scales far off the Galactic plane.

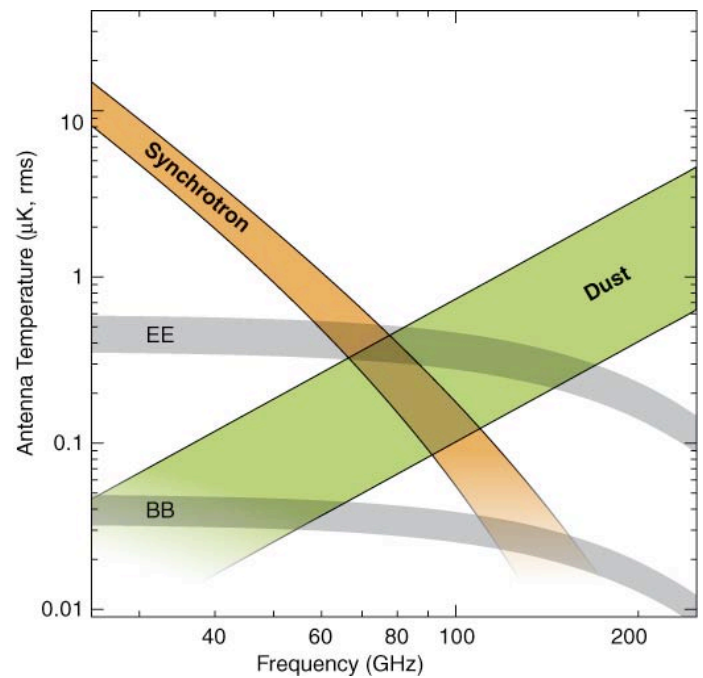

*Figure 4.1: The rms fluctuations in the polarized CMB and foreground signals as a function of frequency. For each emission component, the band represents the rms signal expected from the large-scale emission (2<l<20), consistent with the models used in figure 4.2. The orange band is the synchrotron emission, green is the dust emission, the upper dark band shows the **EE** portion of the CMB signal, and the lower dark band shows the **BB** portion of the CMB, assuming **r**=0.01.*

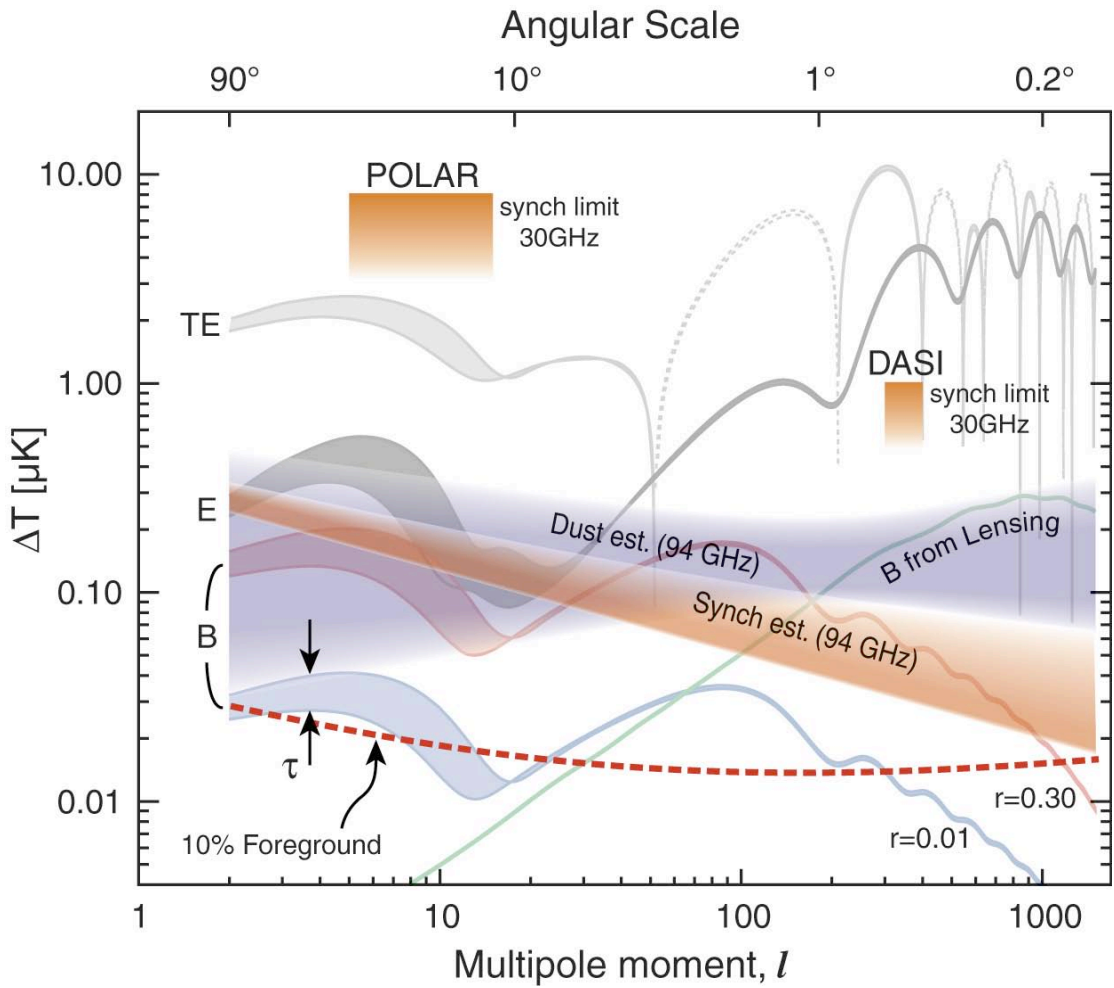

*Figure 4.2: Current estimates of the polarized Galactic foreground signals due to synchrotron emission from cosmic ray electrons and to thermal emission from interstellar dust grains. We show the foreground signals expected at a fixed frequency, 94 GHz. Also shown are the limits on polarized synchrotron emission reported by the POLAR and DASI experiments. These observational limits have been extrapolated to 94 GHz, and plotted in orange with an estimated error band. The thermal dust signal, plotted in blue, was estimated by scaling the unpolarized dust model of Finkbeiner, Davis and Schlegel, then applying a uniform 5% polarization. The large error band for dust emission reflects considerable uncertainty in both the polarization percentage and the coherence of polarization angle as a function of angular scale. We expect to be able to clean the foreground from the polarization maps to the level shown by the dashed red line. Over a substantial range of l, this lies below the B mode signal for r > 0.01.*

Estimates of the foreground emission from Galactic dust and electrons in synchrotron motion are in both figure 4.1 and 4.2. It is widely accepted that we will be able to clean the foregrounds to at least a factor of ten below these levels. This process will involve subtraction of much of the foreground signals using their characteristic spatial and spectral signatures. The dashed red line in figure 4.2 indicates our estimate of the residual foreground contamination, after careful correction for foregrounds. These are the fundamental limits to the search for *B* modes proposed in this document and are the limits we refer to when we speak of planning a CMB mission to work "down to the limits set by astrophysical foregrounds."

### 4.3.3 Extragalactic Point Sources

Extragalactic radio sources affect primarily large $l$ measurements. If they are randomly distributed,

their rms contribution scales simply as $l$. At observing frequencies below ~50 GHz where synchrotron emission dominates, temperature fluctuations are expected to scale as $\nu^{-2.7}$, for a typical synchrotron spectral index of –0.7. Careful modeling, involving an extrapolation in frequency to the 50–150 GHz range, shows that fluctuations produced by unclustered sources fall below the amplitude of CMB temperature fluctuations at all values of $l$ up to about 1000. Similar modeling has begun for polarized signals. The NRAO VLA Sky Survey (NVSS) provided the flux density and Stokes parameters for almost $2 \times 10^6$ discrete radio sources covering 80% of the sky at 1.4 GHz. These measurements were combined with the higher frequency but lower resolution Green Bank 4.85 GHz Survey (GB6) to make a prediction of the polarization properties of extragalactic radio sources at CMB frequencies. These results, which required significant extrapolation of both source fluxes and polarization properties from 5 GHz to the much higher frequencies used in CMB measurements, indicate that radio sources could severely constrain the detectability of high-$l$ $B$-mode polarization at all frequencies up to 100 GHz. Additional observations of extragalactic radio sources are needed to improve the modeling and extrapolations that have been made. In addition, there are classes of radio sources with inverted spectra that may not be fairly represented in the low frequency NVSS and GB6 catalogs. Their numbers, fluxes and polarization properties are poorly known. Nor can we avoid problems from emission from extragalactic sources by moving to higher frequencies: above ~100 GHz, dust re-emission dominates in the spectra of most extragalactic sources, and the high-frequency polarization properties of dusty galaxies are even less well constrained than the properties of ordinary, synchrotron dominated, radio sources.

**4.4 Current and Future Foreground Measurements**

An increased effort over the next five to ten years to measure and model polarized foregrounds will be required if CMB polarization is to be accurately measured. Without these data, it will be difficult to plan optimum CMB polarization observations, or to carry them out. Characterization of Galactic emission is particularly important, and is crucial for the measurement of the fundamental, low-$l$ $B$ modes. Ideally, we would have full maps of polarized Galactic emission available. This is unlikely, so the removal of Galactic foregrounds may have to be done on a statistical basis. To do so, we need information on the polarization level as a function of both angular scale and frequency. Measurements over a wide frequency range, say 30–300 GHz, will be required to characterize the various foreground emission components. We are just beginning to be able to gather such data. Currently, there are a number of projects underway or in the planning stages. Support for these should be maintained.

Excellent, all-sky, low frequency, data will soon be available from the WMAP satellite. The 22–94 GHz data will provide information on Galactic synchrotron emission (and on the spectral index of the emission) on large scales extending down to ten degrees. Several ground-based measurements at similar frequencies will complement these data with higher angular resolution and sensitivity over restricted regions of sky. They will also look for a possible additional source of CMB polarization contamination due to spinning dust grains. By 2009, significant new data covering almost the full range of frequencies of interest to CMB studies, over a wide range of angular scales (but not the whole sky), should be available. At about the same time, high sensitivity, all-sky measurements at 30–70 GHz will become available from the LFI instrument on Planck.

High frequency observations to investigate dust emission have been made and more are planned for sub-orbital and orbital platforms. Most recently, the Archeops balloon mission measured Galactic dust polarization on scales of 15 arc minutes to several degrees at sub-millimeter wavelengths. Archeops has shown that radiation from diffuse Galactic dust is polarized (as expected) generally at the 3–5% level, but in some regions the polarization is as high as 10%. The polarization is coherent on large scales and is aligned orthogonal to the Galactic plane. While the measurements were made near the Galactic plane, they point to the need for more detailed information at higher latitudes with higher sensitivity and more resolution. Such measurements could be made by a series of sub-orbital missions. Combining the data sets will provide the inputs needed to test and refine models of foreground emission.

The Planck satellite has a planned launch date of 2007. The HFI instrument on Planck will have polarization sensitive bolometers (PSBs) operating at 100, 143, 217, and 353 GHz. With all-sky coverage at 5 arc minute resolution, the 353 GHz band will provide an unparalleled data set for studying polarized dust. However, there are a number of reasons to promote vigorous parallel efforts. Smaller sub-orbital missions with narrower goals will return data more quickly. They could study smaller patches of sky with higher sensitivity and spectral resolution. These data would make the extrapolation of existing data up to CMB frequency bands more secure.

Such complementary missions would also mitigate the impact of a possible delay in Planck's launch on the design of the next generation polarization mission. Given a 2007 launch, Planck data would not be fully available to the scientific community until the end of 2009 at the earliest.

An increased understanding of extragalactic point source contamination will only be possible with high-resolution observations of sources at frequencies approaching those to be used in searches for CMB polarization. Wide area surveys are underway at 15 and 18 GHz to identify sources for follow-up at higher frequencies; accompanying polarization measurements are required and are being discussed. There are also small ground based programs to measure polarization of a sample of rising spectrum sources and dusty galaxies. The 100-meter Green Bank Telescope could be used to survey and measure the polarization of 3000 low frequency selected sources at 30 GHz over the next several years. Combining the 30 GHz survey measurements with the existing low-frequency data on a smaller number of selected sources will provide a significantly more complete picture of foreground effects on CMB polarization.

### 4.5 Program for Controlling Foregrounds

Proper characterization of polarized foregrounds is key to the design, execution and analysis of future CMB polarization measurements. We therefore strongly **recommend a systematic program to characterize astrophysical foregrounds, especially from the Galaxy, over a wide range of frequencies.**

Elements of such a program, in order of priority, include:

∗   A series of suborbital missions to measure dust polarization. Both large and small angular scales should be probed at millimeter and sub-millimeter wavelengths where the dust signal is large and easily detected and characterized.

∗   Continued support for theoretical modeling and interpretation of polarized foreground measurements.

∗   Continued support for ground-based efforts to produce 3–15 GHz large-scale maps of the polarized Galactic foreground, to guide extrapolation of polarized synchrotron emission to higher frequencies at large angular scales.

∗   A program to measure the polarization of several thousand selected radio sources at frequencies from 30–100 GHz, and of a statistically significant number of sub-millimeter sources in the ~90 and ~350 GHz atmospheric windows.

# 5. Current Polarization Measurements and Near-Term Program

There is already a great deal of effort going into addressing the science embedded in the CMB polarization. In just the last few months, several new results have been presented. These include the final results from DASI, initial results from CBI and CAPMAP, and polarization from the reionization epoch measured by WMAP. The measured *TE* correlation signals from WMAP show that the *EE* spectrum is going to be close to that anticipated. Confirmation of the *TE* spectrum, a measurement of the *EE* spectrum, and upper limits for the *BB* spectrum have been presented by the BOOMERanG collaboration. Results of the direct measurement of the *EE* power spectrum from WMAP are expected soon. These initial experiments have receivers numbering in the 5–15 range; their results are plotted in figure 5.1. The measurements serve to reinforce our picture of the early Universe, and, as already noted, they have suggested that the Universe was reionized at an earlier epoch than previously thought. They are not yet of sufficient sensitivity to be able to detect the signature either of gravitational lensing or of gravitational waves.

The sister experiments DASI and CBI are interferometers at 30 GHz. The elegance of the interferometric approach is that the detectors are directly sensitive to particular Fourier modes on the sky, without scanning. They can also directly decouple *E* and *B* modes by appropriate weighting of the real and imaginary parts of the phase and amplitude signals they measure. The two experiments operate with different baselines and hence angular scales: DASI has its greatest sensitivity in the region of the second polarization peak, while CBI is sensitive in the region of the fourth peak.

For the future, though, it is unlikely that this approach can be extended with adequate sensitivity to see the *B* modes, and 30 GHz is probably too low a frequency to be able to separate the *B* modes from Galactic synchrotron emission even in the very cleanest patches of the sky. From studies within their data sets, the level of understanding and control of systematic errors is at roughly several $\mu K^2$.

CAPMAP used four correlation polarimeters at 90 GHz in its first season for its result. It is now operating with twelve polarimeters at 90 GHz and four at 44 GHz and should obtain an order of magnitude more data in the current season. It is a single dish experiment that scans a small region of the sky near the North Celestial Pole. The size of the primary (7m) gives CAPMAP excellent sensitivity up to $l$=1500, covering the region where the *E*-mode power is expected to peak. Systematic control is again at the several $\mu K^2$ level.

In addition to these results, there were important upper limits from POLAR, PIQUE, and COMPASS, and the Archeops experiment reported valuable results on the polarization anisotropy for Galactic dust. Upper limits from this current round of experiments on the *B*-mode power are now at the level of 5–10 $\mu K^2$ (95% confidence limit).

CBI, CAPMAP, and WMAP are HEMT-based experiments. BOOMERanG, MAXIPOL and B2K are bolometer-based experiments to measure CMB polarization. BOOMERanG uses PSBs at 145 GHz, and polarizing grids at other frequencies. Modulation is accomplished by means of the scan and sky rotation. MAXIPOL uses a fixed wire grid to define a polarization direction and a rotating half-wave plate to enable polarization modulation and extraction of both Stokes parameters, this primarily at 140 GHz.

Planck, which has a mixed HEMT and bolometer focal plane, is scheduled to be launched in 2007; it has a factor of five improvement in temperature sensitivity over WMAP and will similarly map the entire sky. Planck was not initially designed with polarization sensitivity as a primary goal. We assume here that the Planck team can understand their residual systematic uncertainties (and foregrounds) sufficiently well to allow them to extract science at the level of 300 nK, or about a factor of 10 poorer than the requirements laid out in §6 for CMBPOL, which will be designed to measure $r < 0.01$.

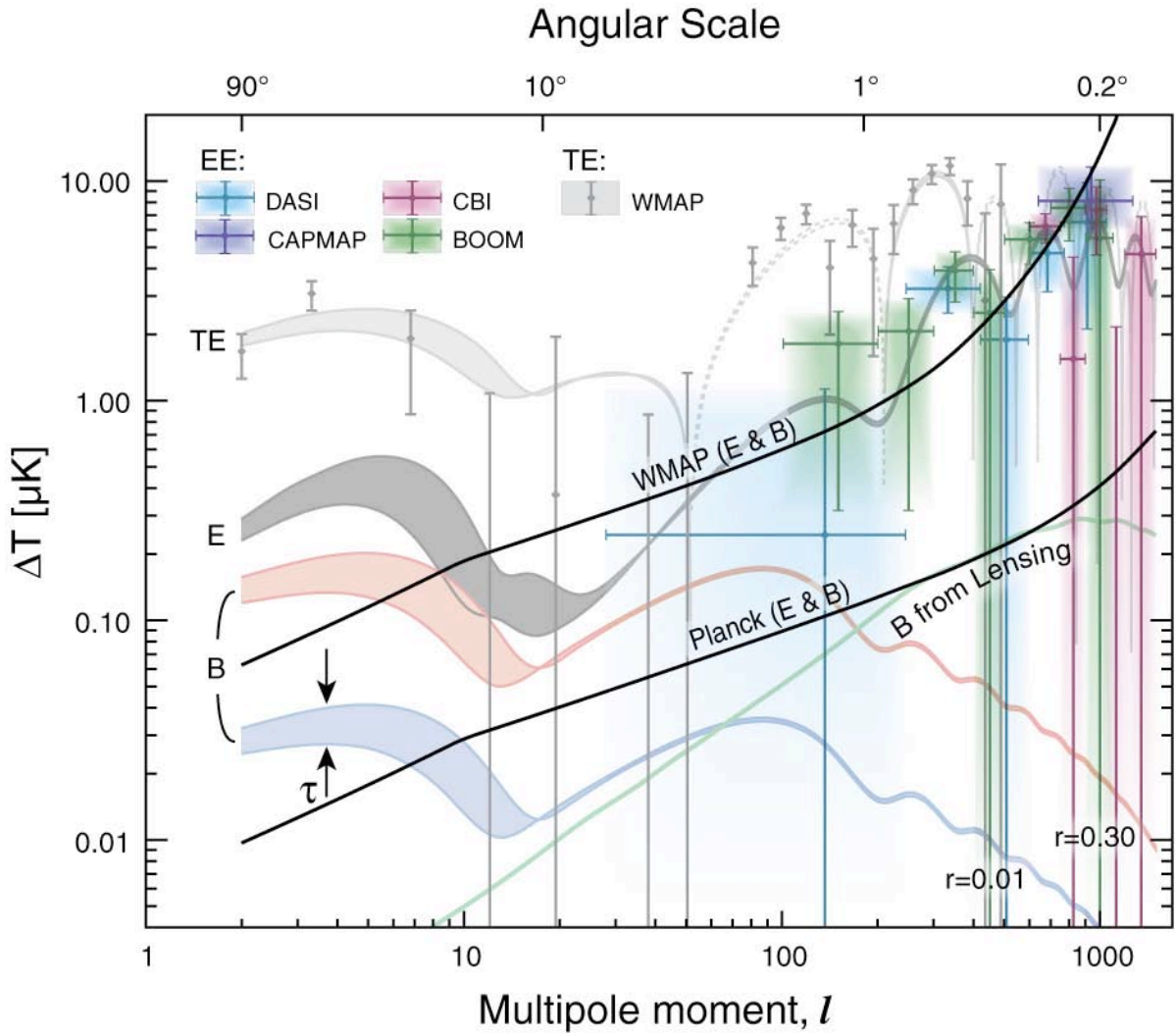

*Figure 5.1: Current measurements of the polarized CMB signal. The **TE** measurements (grey) are from the first-year WMAP data. The **E** measurements (colored) are from the ground-based experiments CAPMAP, CBI, and DASI, and from the balloon-based experiment BOOMERanG. These results are all consistent with the signal predicted by the ΛCDM model and they demonstrate excellent technical progress in our ability to measure CMB polarization. The black curves indicate the one-sigma sensitivity estimates for WMAP and Planck but without correction for foreground emission. The WMAP estimates are based on measured noise properties of the instrument and assume 8 years of operation. The WMAP satellite should measure the **E**-mode signal at low and intermediate l, and may detect a **B**-mode signal if the gravitational wave amplitude is high. Additionally, it will produce sensitive full-sky maps of the polarized synchrotron emission. The Planck estimates are based on noise measurements from the test-bed High Frequency Instrument and assume 1.2 years of operation. Planck will obtain precise measurements of the **E**-mode signal, and can potentially detect a **B**-mode signal from gravitational waves. It will also produce high quality maps of the polarized dust emission from its high frequency channels.*

Between now and the era of CMBPOL, a healthy program of ground- and balloon-based experiments is both planned and needed. For one thing, there is science that can be explored with such programs that will likely not be done even with CMBPOL. This includes a detailed study of the lensing signal at high multipoles. Also, sensitive searches for new astrophysical effects can be made from the ground with long integration times on a small percentage of the sky, achieving better than 100 nK sensitivities per square-degree pixel. This permits studies at the same depth as CMBPOL but enough years before launch to inform that mission.

In addition, there are important systematic issues that need addressing soon, well before a new satellite mission. These include developing means of polarization modulation, finding effective scan strategies, and deep studies and searches to characterize and perhaps discover new astrophysical foregrounds. Even the testing of algorithms for separating pure $E$ modes, pure $B$ modes, and ambiguous polarization modes, should be done with real data sets with adequate sensitivity.

We envision then a program of experiments over the coming 5–8 years to address the science and systematic issues which use both HEMT (coherent) and bolometric detector technology. A discussion of the relative merits of the different types of detectors is given in §§7 and 10. The experiments will involve, progressively, tens, then hundreds, and finally thousands of detectors.

Experiments with tens of detectors are already underway. The sister experiments QUaD and BICEP observe from the South Pole using PSBs at 100 and 150 GHz. QUaD, with a 4' beam, is optimized for detecting the gravitational lensing, while BICEP, at about 40', is going after the gravitational wave signature. These are important experiments for demonstrating modulation techniques and for understanding scan strategies and other systematic effects with bolometric detectors. Again at this number of detectors, MBI is testing the idea of using bolometers configured as an interferometer; and PAPPA is a balloon effort using waveguide-coupled bolometers from GSFC. These latter two experiments have beams in the range of one degree.

It will take hundreds of detectors to be able incisively to study gravitational lensing and be able to make a significant advance in the search for gravitational waves. Several programs approaching this scale have been proposed or recently funded. The two funded experiments both employ transition edge (TES) bolometers operating at 90, 150, and 220 GHz. These are:

• ClOVER, the lone European effort on this scale, with a 15' beam observing from the South Pole (described below)

• EBEX, a balloon-borne experiment funded by NASA, with a 4' beam.

We recognize that other programs using both bolometers and coherent detectors may soon be planned or proposed. This generation of experiment needs to (and can) address systematic issues well below the µK level, especially including the effects of synchronous offsets in the scan strategy, means of modulation of the polarized signal, optical imperfections which can fake or corrupt a polarization signal, immunity to $E$ power leaking to $B$ power, and the means to identify and clean foregrounds. One of the major recommendations of the Task Force is specifically to carry out this ground-based and balloon-borne program to develop the technology and to learn how to do these difficult experiments. Not only will this enable the planning and design of CMBPOL, but it may well produce significant scientific results along the way.

**European ground based programs.** There has been a long and vibrant record of CMB-related research in Europe both independent of and in collaboration with the US. A prime example is the collaboration between the US, Canada, Italy, and the UK on the BOOMERanG experiment. Current collaborations include the QUaD experiment between the US and Cardiff University, as well as the Planck mission. Groups in Canada also have significant collaborations with CMB researchers in the US, for both instrument building (e.g. WMAP, BOOMERanG) and data analysis (e.g. BOOMERanG, CBI, ACBAR). In most cases, the detectors are developed and supplied by the US.

The UK has recently committed significant funds to develop two new ground-based instruments independent of US participation. ClOVER is a series of three independent telescopes operating from 90 to 220 GHz. Each telescope has four co-pointed optical assemblies feeding an array of detectors. *Q/U* polarization is determined by a cross-correlation receiver with waveguide phase switching. They plan to field the instrument to Dome C in Antarctica in 2007. The second effort, BRAIN, is an interferometer with the goal of measuring the *B*-mode signal at low-*l*. BRAIN uses many of the same receiver elements as ClOVER, including detectors and phase switches. The funding for these two experiments includes significant support for TES detector and SQUID development. This follows a major investment in detector development for the SCUBA2 sub-millimeter receiver for the JCMT in Hawaii.

The detector and SQUID development in Europe will complement the virtual monopoly Cardiff University has on millimeter and sub-millimeter filter technology. Many of the bolometer-based CMB experiments for the last 15 years have used filters from Cardiff. They are now developing sophisticated polarizing grids and half-wave plates for use on polarization experiments. One of the reasons they are able to maintain this advantage is the recognition by the UK funding agency that filters are a resource that must be maintained for multiple experiments. The "rolling" grant model for funding has supplied the stable support required to keep this effort alive.

# 6 Requirements: Observation Strategy and Control of Systematics

## 6.1 Overview

Measuring gravitational waves via the CMB polarization will be challenging. We expect that a convincing detection will require a space mission optimized for detecting polarization – and rejecting systematic effects – on large angular scales where the cosmological *B*-mode signal is strongest. In this regime, the sensitivity to measuring gravitational waves should ideally be limited only by confusion from the astrophysical foregrounds. One of the primary short-term goals of the research program defined by our roadmap is to identify the critical scientific requirements for a space mission and the derived instrumental performance requirements. Key elements of the research program focus on assessing:

- The degree to which astrophysical foregrounds, including polarized Galactic emission and gravitational lensing, will compromise the detection of gravitational waves. New data on these foreground signals is imminent and will provide a solid handle on the scope of the foreground challenge. However, it is important to stress that ground- and balloon-based efforts to measure these foregrounds are an important and integral part of the roadmap.

- The degree to which systematic errors can and must be controlled and verified to reach the required accuracy. This includes an assessment of which measurements can be done from ground and sub-orbital platforms and which *must* be done from space.

- The choice of detector, polarization modulation, and other experimental technologies required to meet sensitivity, systematic error, and calibration error budgets.

## 6.2 Sensitivity

CMB photon statistics and practical constraints on focal plane area and mission lifetime limit the ultimate sensitivity of a CMB polarization mission, even for an instrument with no sources of internally generated noise, e.g., noiseless detectors and cold optics. The total instrument sensitivity sets a limit on the ability to measure CMB polarization. The actual sensitivity of the instrument will depend on design choices; however, an analysis of the fundamental limits for an idealized instrument serves as a starting point for any practical instrument design.

To estimate a limiting case, we consider a telescope with a maximum practical diameter of 2.5 m. (For comparison, the WMAP primary diameter is 1.6 m and the sub-millimeter Herschel primary is 3.5 m.) We assume Gaussian illumination with an edge taper of –20 dB and assume that the detectors are arranged in the focal plane with a spacing of $1.27\,f\lambda$, where $f$ is the focal ratio and $\lambda$ is the wavelength of the radiation. In this case, single-mode detector pixels are spaced as closely as possible and are matched to the telescope. We assume a 25% fractional bandwidth, a 4°-diameter diffraction-limited telescope field of view, and an overall optical efficiency of 40%. This idealized focal plane has about 2,000 pixels at a fiducial frequency of 150 GHz. The only noise contribution we consider is photon noise from the CMB itself. The result is that a 150 GHz (single-frequency) mission would have a resolution of 7.5 arc minutes (2σ width) and a sensitivity of $w_p^{-1/2} =$ 2.5 μK-arc minute in one year of observation of the full sky. The noise level shown in figure 6.1 was based on background-limited detectors distributed over the three frequency bands 150 GHz, 200 GHz and 300 GHz. Figure 6.1 shows that such a mission would allow us to reach the confusion limit presented by the gravitational lensing foreground.

This idealized sensitivity estimate serves as a guide to the sorts of scientific results that are possible in principle. It also helps to define a multi-dimensional sensitivity trade space that interacts with the systematic error budget and mission complexity in the following ways.

- **Optics:** The photon noise from the optics can be reduced by cooling them. Also, the diameter of the telescope can be traded against the usable focal plane area. This affects the trade-offs amongst angular resolution, far sidelobe pickup, beam symmetry, cross-polarization pickup, cryogenic heat loads, feasibility, number of detectors, and hence cost. A mission based solely on corrugated feeds is also possible.

- **Detectors:** The parameters to trade off include the operating temperatures of the bath and the sensor, the characteristic response time of the detectors, the spectral bandwidth of the system, feasibility, and cost.

The pixel spacing of a practical focal plane may have a density several times lower than the ideal case, thereby reducing the sensitivity. The instrument will also undoubtedly have multiple frequency bands, which might be accomplished by interleaving pixels at different bands or by frequency multiplexing.

### 6.3 Angular resolution and sky coverage

There is little doubt that the most promising regime in which to detect gravitational waves is on large angular scales and that a space mission should measure polarization over a large fraction of the full sky. It is less clear how much angular resolution a space mission *must* have in order to meet the high level scientific requirements. The primary advantage of high angular resolution is that it gives one a better ability to remove the foreground signal induced by gravitational lensing of the CMB. This, in turn, lowers the confusion limit set by this effect. However, angular resolution tends to drive mission cost and complexity, so the angular resolution requirements must be well formulated.

A key requirement of our roadmap is to look for break points in science return, control of systematic errors, and cost as a function of angular resolution. Two extreme cases are exemplified by a low angular resolution mission ($\sim 1°$, $l \lesssim 200$) and a high angular resolution mission ($\sim 5'$, $l \lesssim 2500$) as shown in figure 6.1. These mission concepts would have significantly different optical designs and would probably have different observational approaches. The systematic effects associated with each would also be different.

Understanding this trade-off involves a tight interplay between the instrument and the spacecraft design, as informed by mission simulations and data analysis techniques. It will also be important to understand what can be reasonably achieved from ground- and balloon-based experiments in the coming years. *Such data will provide a valuable complement to space-based data and may alleviate the pressure to make a space mission overly complex and risky by pushing to higher resolution.*

### 6.4 Frequency coverage

The choice of observing frequencies is dictated by the necessity to remove Galactic and extragalactic foreground signals from the data. As noted in §4, the two major foreground effects for a space-based polarization mission are diffuse emission from the interstellar medium of our Galaxy and the conversion of primary *E*-mode polarization into *B*-mode signal due to gravitational lensing. The latter signal has the same frequency spectrum as the CMB itself, since lensing does not alter the frequency of the lensed photons. Thus the choice of observing frequencies is dictated essentially by the need to remove the diffuse Galactic emission.

It has been known for some time that the *unpolarized* Galactic signal has a local minimum at ~70 GHz where the radio emission from synchrotron and free-free sources falls below the thermal emission from interstellar dust grains. Ideally, measurements should bracket this minimum on both sides to enable a robust characterization of both the radio component and the thermal dust component. For example, the HEMT-based WMAP mission had 5 observing frequencies from 22 to 94 GHz, which allowed for a very good characterization of the radio component and a fair characterization of the dust. Most bolometric instruments have operated from 100 GHz and up, enabling a characterization of the dust.

It is likely that the minimum of the *polarized* Galactic emission also lies near 70 GHz, though current uncertainties in the polarization fraction of the synchrotron and dust components make this uncertain to 20 or 30%. In order to maximize the power of the data to reject polarized foregrounds, it will be important to bracket this minimum on both sides. But, while multiple frequency channels are crucial, the details of the frequency coverage depend strongly on the foreground properties. If foreground spectral indices or polarization patterns vary significantly across the sky, it may be advantageous to decrease the separation between channels, since channels far from the minimum near ~70 GHz may have limited utility. The details of the frequency selection are best made after the initial round of polarization measurements discussed in the previous section have been more fully understood. The forthcoming WMAP polarization data will yield a great deal of information about polarized synchrotron emission. Information on polarized dust will be coming from a variety of near-term bolometric experiments, including further results from Archeops.

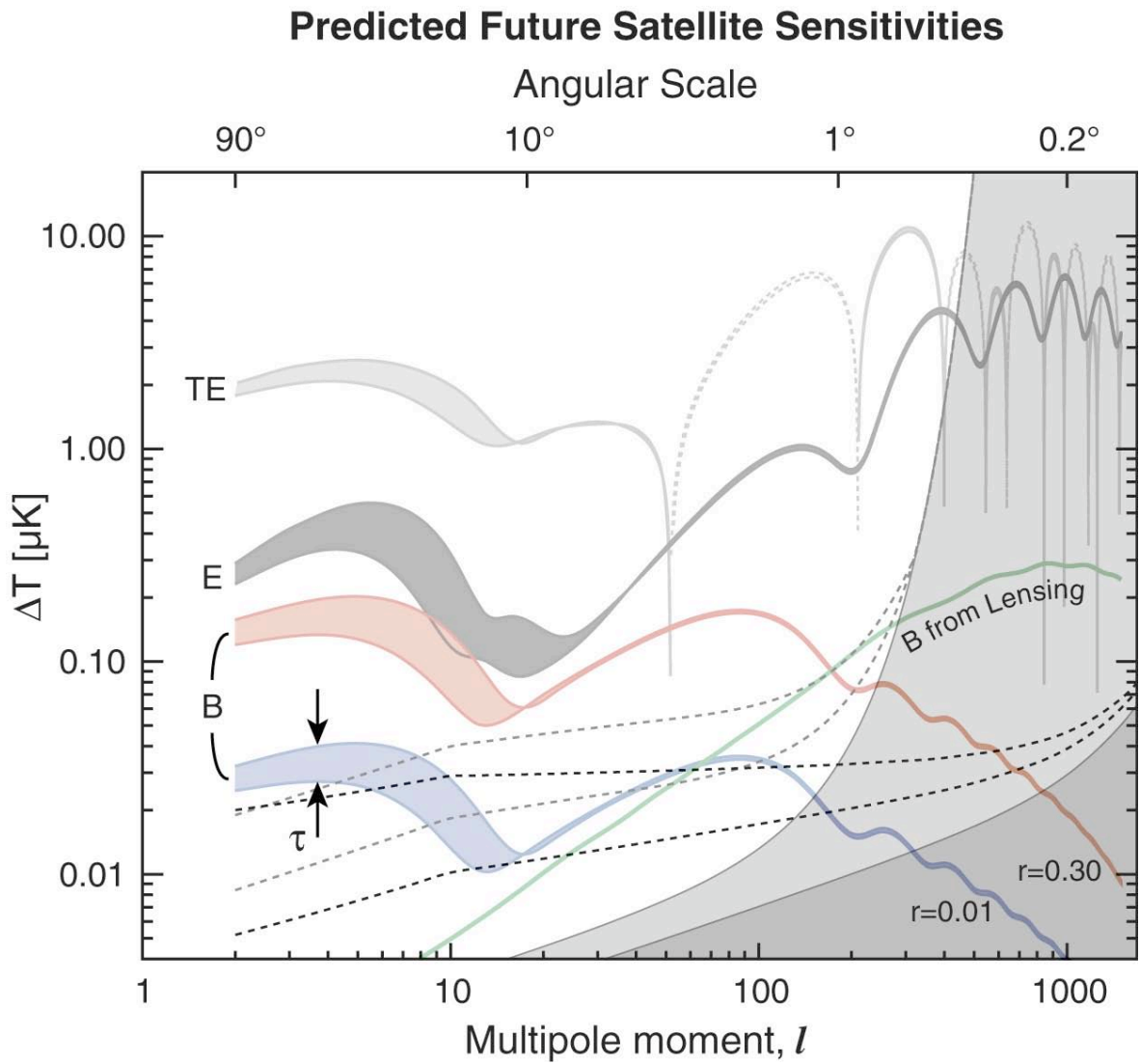

*Figure 6.1: Estimates of the sensitivity that will be achieved by a space mission of the type called for in our roadmap (light grey), and for a more ambitious mission with higher angular resolution (dark grey). The light grey band assumes a 1000-pixel instrument with a 1 degree resolution beam, the dark grey band assumes a 2000-pixel instrument with a 0.1 degree beam. In each case, the detectors are assumed to operate at the background noise limit and to observe for one year. For each mission, the dashed lines indicate two estimates of the residual sensitivity after using the multi-frequency data to subtract foreground signals. The actual sensitivity will depend on many factors that are poorly known at this time, including the amplitude and complexity of the foreground signal, and the instrument's frequency coverage and sensitivity per band.*

## 6.5 Observing strategies and experiment modeling

Designing a space mission to measure gravitational waves via CMB polarization will require a comprehensive study to identify the critical scientific requirements for the mission and their derived instrumental and mission performance requirements. Key elements of the design process must focus on assessing:

- The level and complexity of the foreground signals that will be present, as a function both of frequency and of angular scale on the sky. This will largely determine the range and density of frequency coverage required, and will contribute to the angular resolution requirements. For example, the most thorough methods proposed for subtracting the gravitational lensing foreground require relatively high angular resolution.

- The degree to which systematic errors can and must be controlled and verified to reach a foreground-limited measurement of the sky polarization. The suppression of instrumental systematic effects is strongly coupled to the scan strategy used to observe the sky, so the design of the mission, per se, must take into account the design of the instrument hardware. Also, since some instrumental effects are frequency dependent, the systematic error requirements may impact the range of frequencies chosen for the mission.

- The extent and uniformity of sky coverage for the mission. Since the $B$-mode signal from gravitational waves is expected on large angular scales ($l \leqslant 100$), the expectation is that a space mission will need to cover the full sky and that near uniformity of the coverage will be desirable. The ability to achieve this is strongly dependent on the sky scan strategy and the orbital parameters of the mission.

- The choice of detector, polarization modulation, and other experimental technologies required to meet sensitivity, systematic error, and calibration error budgets.

Cross-cutting all of these considerations is an assessment of the trade-off between scientific return, mission complexity, and cost. A leading question is whether a mission requires high angular resolution, which is typically a strong cost driver for missions. Another issue is whether the data rates anticipated will require some form of lossy compression or other on-board processing. Available down-link rates may also drive the choice of orbit.

## 6.6 Removal of Foreground Emission

Foreground signals are discussed in some detail in §4. A CMB mission designed to measure sky polarization to the limits set by the astrophysical foregrounds requires careful consideration of foreground removal techniques. A variety of methods have been used or proposed for removing foreground signals from temperature (intensity) data and many of these methods should carry over to polarization data as well. All of these methods rely either on combining multi-frequency data, or on exploiting various differences in spatial morphology between the foreground signals and the primordial CMB signal. This, in turn, places requirements on the extent and density of frequency coverage required of a mission, and on the angular resolution and sky coverage required.

Unfortunately, relatively little is known about the nature of the polarized foreground signals at this time, so it is premature to attempt detailed planning on this basis now. However, as discussed in §4 and §11, there will be a wealth of new foreground data coming in the next few years to help guide this process. The amplitude and spectral complexity of the diffuse Galactic foreground signals will tell us a great deal about the specific frequency coverage required for a polarization mission. Also, the amplitude of the Galactic foreground signal relative to the gravitational lensing signal will help determine the degree to which the gravitational lensing signal needs to be removed. Specifically, if our ability to remove the Galactic foreground signal leaves uncertainties that are still above the lensing signal in the angular range of interest, then the requirements for cleaning the lensing signal can be relaxed. This, in turn, has an impact on the angular resolution required of a mission since high resolution is needed to most efficiently clean the lensing signal.

## 6.7 Control of systematic errors

The gravitational-wave CMB polarization signals are small compared to the temperature fluctuations, and therefore one critical step in the roadmap is to achieve the necessary raw sensitivity. As shown in §7, the required sensitivity can be achieved by increasing the number of detectors in the focal plane. The second critical step is to evolve experimental technique to the point where the systematic errors can be kept below the statistical error.

The measurement of the (quite small) temperature anisotropies in the CMB has required excellent control of systematic errors from instrumental effects. The ability to control these effects was developed over several decades of instrument building and observational experience. Many of the lessons learned from those efforts are directly applicable to CMB polarization measurements. However, the measurement of polarization also introduces new instrumental effects. The smaller signal levels of $B$-mode polarization mean that such effects must be controlled at a correspondingly lower level than has been required so far. In this section, we review the most important of these instrumental effects and outline how instruments can be designed to overcome them.

To measure $B$-mode polarization spectra well, an instrument must be capable of making maps of the Stokes parameters $T$, $Q$, and $U$ with very high fidelity: the signals should not be mixed or biased in any way. The goal is to design instruments that limit spurious $B$-mode signals to well below the target signal level. In this report, we propose a target of T/S = $r$ = 0.01 (limited by foreground sources), i.e. ~30 nK rms signals. Our goal is to keep each systematic effect a factor of ten below this level. With this in mind, we can estimate quantitatively the required rejection level for various systematics.

Instrument-induced effects include mixing of the Stokes parameters by the optics, drifts in detector calibration and instrument temperatures, and pickup from the Galaxy by telescope sidelobes.

Although keeping these effects below ~3 nK is challenging, they can be controlled in well-designed instruments. Recently the DASI, CBI, and CAPMAP experiments have demonstrated that even observations from the surface of the earth can reject systematic effects well enough to clearly measure $E$-mode polarization at the 1 µK level. These and other ground-based and balloon-borne observations have developed techniques for rejecting radiation from the 300 K earth as well as emission from the atmosphere. They have not yet reached any fundamental limit.

Instrumental effects can be controlled at even lower levels in space where the power illuminating the instrument is a factor of 100 less than on the earth and the thermal environment can be exceedingly stable. For example, during its first year of operation, WMAP has demonstrated rejection of spin-synchronous spurious temperature signals below the 180 nK level. These effects can be modeled and subtracted to well below this level. Experiments designed specifically for CMB polarization measurements can do substantially better than WMAP by exploiting certain advantageous properties of polarization. In particular, the fundamental measurement of the Stokes parameters can be made by differencing signals that follow identical paths through the optical system (and atmosphere for a ground-based system). Mapping requires differencing Stokes parameters from different parts of the sky, but this comparison is made after the polarization signal are already detected in single beams.

Several mission concept studies for Inflation Probe are underway. While it is premature to choose a "best" design now, the systems all include the following features to combat instrument-induced systematic errors: The instruments will simultaneously measure signals from ~ 1000 beams on the sky. As described above, the Stokes parameters are measured for each beam by comparing orthogonal electric field components averaged over each beam. These beams are designed to be highly symmetric so that the sky is sampled in the same manner for the two orthogonal linear polarizations. The polarization in each beam is modulated in some way to recover the Stokes parameters by phase-sensitive detection of signals at the detectors. The polarimeters are designed to be insensitive to temperature (total power) and to respond only to polarization signals (i.e. to have high common mode rejection). Highly interlinked scans of the sky allow faithful reconstruction of sky maps even in the presence of low-level drifts in the instrument gain and offset.

Table 6.1 summarizes the instrument performance goals for the target $r = 0.01$. The table includes the instrumental parameters, their effects on the signal, the performance goals, and plausible methods for achieving the goals. Many of the requirements can be relaxed by clever choices of observation strategies that null the spurious instrumental effect. One of the most important activities of the roadmap recommended in this report is to develop and test instrument designs and observing strategies to achieve these goals.

## Table 6.1: Instrument Performance Goals

| Parameter | Effect | Goal | Method |
|---|---|---|---|
| Cross-Polar Beam response | E → B | < 0.003 | Rotate Instrument, Wave Plate |
| Main lobe ellipticity (0.5° beam) | dT → B | < $10^{-4}$ | Rotate Instrument, Wave Plate |
| Polarized sidelobes (response at Galaxy) | dT → B | < $10^{-6}$ | Baffles/shielding/measure |
| Instrumental polarization | dT → B | < $10^{-4}$ | Rotate Instrument, Wave Plate |
| Polarization angle | E → B | < 0.2 ° | Measure |
| Relative pointing (of differenced samples) | dT → B | < 0.1" | Dual-polarization pixels |
| Relative calibration | dT → B | < $10^{-5}$ | Modulators |
| Relative calibration drift (scan synchronous) | T → B | < $10^{-9}$ | Modulators |
| Lyot Stop Temperature (10% spill, scan synch.) | $dT_{opt}$ → B | $dT_{opt}$ < 30 nK | Measure |
| Cold stage T drifts (scan synch.) | $dT_{CS}$ → B | $dT_{CS}$ < 1 nK | Improve uniformity, measure |

**TABLE 6.1** *Performance goals for a CMB B-mode measurement. The first eight parameters describe instrumental effects that transform various sky signals into false B-mode signals; here we use T to indicate intensity, E to indicate the E-mode polarization signal, and dT to indicate CMB temperature anisotropies. The listed "Goal" is the level at which an individual instrumental effect will begin to cause a 10% contamination (in units of temperature) of an r = 0.01 B-mode signal in the most naïve experimental design. Clever scan strategies and partial correction of known levels of contamination can relax these requirements. See the text for more details.*

### 6.7.1 Specific Instrument Performance Goals

**1. Cross-polar beam response.** Ideally, a linearly polarized detector will have no response to signals from the orthogonal linear polarization. Any sensitivity to such signals is called "cross-polar" response, in contrast to the desired "co-polar" response. The cross-polar response can vary across the sky; that is, it has a beam pattern. A variety of physical causes induce cross-polar response, from detector coupling properties to effects inherent in reflecting optics. Though this response can be measured and corrected for, errors in that correction will lead to a residual systematic error. The primary effect of cross-polar response is to mix $Q$ and $U$ signals, thereby mixing the much larger $E$ mode signals into $B$ modes. Therefore, the uncertainty in the total cross-polar response, integrated over the beam pattern, must be kept to less than $3 \times 10^{-3}$ of the integrated main lobe co-polar response.

**2. Beam ellipticity.** The *Q* and *U* Stokes parameters are each determined by comparison of signals from two linear polarizations. Ideally the beam response on the sky would be identical for each of the two polarizations. To the extent the beams are not identical, temperature anisotropies will induce false polarization signals. For example, in many systems the beam is slightly elliptical, with the pattern rotated for the orthogonal polarization. Differencing such patterns leads to a sensitivity to the quadrupole of the temperature anisotropy at that point in the sky. This effect depends on angular scale, but for an experiment with half degree resolution, the ellipticity must be less than roughly $10^{-4}$. This effect can in principle be corrected if the ellipticities are known, and a higher resolution, high S/N measurement of the temperature anisotropies is available. Observations at a variety of azimuthal angles can also correct this effect.

**3. Polarized Sidelobes.** Sidelobe response to bright sources is an obvious source of contamination. If the sidelobe response is different for the two polarizations, an unpolarized source will induce a polarization signal. The required sidelobe level depends on both the brightness of the source and the angular resolution of the experiment. For a half degree resolution experiment and the Sun as a source, the polarized sidelobes must be less than $10^{-12}$. This is the primary reason WMAP and Planck are both at L2. The three brightest microwave sources in the sky (the Sun, Earth, and Moon) can be kept in one direction, allowing special shielding to keep the sidelobe response in their direction very low. Contamination by the planets, Galactic sources, and the Galactic plane can be subtracted in the data analysis, but sidelobe response to the Galaxy must be kept below $10^{-6}$.

**4. Instrumental Polarization.** Unpolarized radiation entering the instrument can be polarized by the optical elements in its path. For example, reflection from tilted metal surfaces leads naturally to partial polarization. The degree of polarization depends on the surface conductivity of the metal at the photon frequency, and the angle of the reflection. This effect will cause false polarization anisotropies, with a pattern that is correlated with the temperature anisotropies. The situation is complex for focal planes with many detectors at different polar angles viewing different fields, as they will all have different instrumental polarizations. But instrumental polarization is generally extremely stable, so the effect can be measured and corrected. In addition, varying the rotation of the telescope for scans over the same sky will change the relationship between induced polarizations and underlying temperature anisotropies. It is probable that these variations could be used to separate real polarizations from those induced by this instrumental effect, but it is not yet known to what degree this can be done. In the table we cite the level at which instrumental polarization will have to be either controlled or corrected to keep spurious signals below the 3 nK level.

**5. Polarization angle.** Separating the *E* and *B* polarization signal requires precise knowledge of the directions of the polarization pseudo vectors seen in figure 3.1. Errors in the polarization angle will mix *Q* into *U* and thus *E* into *B*. To keep this mixing below the 0.003 level, the error in angle must be kept below 0.003 radians, or 0.2 degrees.

**6. Relative Pointing.** As is the case for the beam ellipticities described above, any relative pointing error in the beam positions for two orthogonal polarizations being compared to measure *Q* or *U* will lead to a sensitivity to gradients in the temperature anisotropy pattern. For the roughly 100 µK/degree gradients in a temperature anisotropy map with half degree resolution, a pointing accuracy of 0.1 arc seconds is required to keep the error signals below 3 nK. This requirement is for relative pointing, e.g. between two detectors in one focal plane pixel, rather than absolute pointing.

**7. Relative Calibration.** Errors in the average relative calibration of the channels in the focal plane will lead to a coupling of temperature anisotropies into polarization signals. For a constant calibration error, the induced polarization signal is (for a given pair of detectors) perfectly correlated with the temperature anisotropies, similar to the effect of instrumental polarization discussed above. It is very likely that the temperature anisotropies, in fact, can be used to calculate the relative calibrations of all the channels to the desired accuracy. However, any residual errors will convert temperature anisotropies into *B*-mode signals. This effect is mitigated by techniques that allow a single detector to measure the Stokes parameters (rather than differencing two detectors). For example, any polarization modulator greatly reduces this effect since the two polarizations are then measured with one detector having a single calibration.

**8. Relative Calibration drift.** For instruments that difference detectors to produce a *Q* or *U* Stokes parameter, small drifts in the relative calibrations of various detectors will cause spurious polarized signals. Drifts that are synchronous with the observing scan will not integrate down with time like noise. They will couple the total power on the detectors (*T*) into polarization signals, and so must be controlled or corrected for at the $10^{-9}$ level. As with the time stationary relative calibration errors, the use of a polarization modulator greatly mitigates this problem.

**9. Temperature drifts of optical elements.** Small scan-synchronous temperature changes in the optics can, through the emission of those optics, induce false polarization signals. To the extent that the emission is polarization-independent there will be no induced polarization signals. However, as noted above, metal surfaces have polarization-dependent emissivity. Here, we assume a difference in emissivities of 0.001. The relevant optics will be warm enough that the observing bands will be on the Rayleigh-Jeans sides of their emission, so the polarization signals induced will be the product of the emissivity difference and the physical temperature change. To keep induced signals below 3 nK, the required optics temperature stability is 3 µK. If the optical system uses a cold stop (such as a Lyot stop) to help define illumination on the large optics, the required temperature stability of that stop is related to the fraction of power spilled onto the stop. Cooling the stop well below 3 K reduces the required stability as the emission moves into the Wien side of the stop's blackbody emission curve. For 10% spill onto a 10 K stop, the required temperature stability of the stop is 30 nK. These optics temperatures can be monitored and corrected for, if necessary.

**10. Cold stage temperature drifts.** These affect the detector performance or response: for instance, consider TES bolometers. To a very good approximation, TES bolometers measure the optical power by measuring the electrical power required to keep the bolometer at the superconducting transition temperature. If the cold bath warms, less power will be required to keep the bolometer at that temperature, even if the optical power remains constant. The corresponding reduction in electrical power will be misinterpreted as a change in optical power. The sensitivity to such cold bath changes is given by the thermal conductance between the cold bath and the bolometer, G. As long as all the bolometers are equally well coupled to the same cold bath, and have the same G, cold stage temperature drifts will appear as common-mode signals and can be removed. However, in systems that difference detector outputs to measure polarization, variations in G from detector to detector will couple cold stage temperature drifts into false polarization signals. For a system with absorbed CMB power of 1 pW in each detector, a baseline G of 10 pW/K, and 10 % variations in G, the required scan-synchronous stability of the cold stage is roughly 1 nK. Here again, a polarization modulator can greatly mitigate this effect since both polarization states are measured with a single detector in a time short compared to any drift. In addition, cold bath temperature variations can be measured and their effects can be accounted for in the detector signals.

While the improvement required from the current state of the art to obtain the goals in table 6.1 is substantial for many of the parameters, few experiments have thus far been designed from the start for polarization measurements. The first generation of such experiments is starting to be deployed. It is clear that the design of polarization-specific instruments and of observation strategies will develop rapidly. The space environment is much more stable than suborbital platforms, and therefore it is likely that the best performance will be obtained there, especially at the largest angular scales. However, ground and balloon-based experiments are capable of very high performance on smaller angular scales, and they will test techniques required for an orbital mission.

**6.7.2 Modulation.** Polarization signal modulation on multiple time scales is one of the most powerful techniques for controlling instrument-induced systematic errors. When a modulation scheme is appropriately selected, the corresponding demodulation recovers the signal of interest while rejecting artifacts introduced by the experiment. One of the major goals of the experiment design process is to identify and analyze modulation approaches that reliably reject artifacts to a level consistent with overall experiment requirements.

Polarization modulators can provide a means of using the same detection chain to rapidly measure different polarization angles at the same point on the sky. This provides a powerful tool for conquering many of the instrumental effects just described (e.g. detector or amplifier gain drifts), at the risk of introducing others (e.g. gain drifts of the modulating element, or drifts in an offset caused by the modulator).

Several types of polarization modulators have been proposed for bolometric focal planes. These include waveguide Faraday rotators, on-chip transmission line switches, and rotating half-wave plates. The closer a modulator is to the sky in the signal path, the more potential systematic effects it can modulate and thus stabilize. For example, all three of these modulation schemes would help stabilize detector gain drifts. However, only the rotating half-wave plate can potentially leave the feed pattern unchanged as the polarization is rotated, thus separating polarization signals from systematic effects related to the feed pattern.

Investigation of such modulation techniques, and characterization of their undesirable properties (e.g. reflections, transmission gain drifts, emission, offsets, frequency dependence and reliability) is a high-priority activity. Equally important, will be work to assess whether any of these modulators are really needed. It is possible that the cleanest and most stable system will use only telescope motion to modulate the signals on the sky.

**6.7.3 Alternative Technologies: Systematics in HEMT Receivers and Bolometric Interferometers**

We have so far focused on systematic effects in systems using standard bolometric imaging arrays. Alternative technologies are being pursued; the two main ones under investigation now are HEMT-based receivers and bolometric interferometry.

HEMT amplifier technologies have now been miniaturized to the point where full ($I$, $Q$, $U$) sensitive receivers can be placed in a small volume behind a feed horn. These systems will be susceptible to a similar set of instrument-induced systematic errors as those listed in Table 6.1, but in many cases with entirely different physical couplings to many of those effects. HEMT based polarization sensitive receivers are discussed in §7 and again in §10.

Interferometry using incoherent detectors (i.e. bolometers) is another technique being considered. It has a different set of instrumental effects than total-power, imaging instruments, because interferometers use a fundamentally different method to measure the polarization. While the pixels in an imaging system measure the power (square of the electric field) focused onto them from a single direction in the sky, interferometers measure the correlations of electric field amplitudes collected by pairs of antennas in an array of antennas. An interferometer-based polarization receiver using bolometers is described in §7.

Interferometers are susceptible to the instrumental effects discussed above, but in different ways from imaging systems. For example, in an interferometer mismatched beams do not lead to conversion of temperature ($T$) into polarization. On the other hand, any coupling ("leakage") between the orthogonal polarizations entering different antennas will produce a spurious correlated signal. This signal is proportional to the total intensity ($T$) so the effect is to mix $T$ into $Q/U$ and $E/B$. This effect can arise from instrumental polarization, such as from imperfect separation of polarizations by the polarizers located after the interferometer antennas. In existing instruments (e.g. DASI) this leakage exists at the ~1% level and is stable over periods of months to 1% of that, or $10^{-4}$ of the temperature anisotropy signal (*dT*), close to the required 3 nK level.

Likewise, the cross-polar beam response of the antennas will cause some leakage of dT into *E/B*. In existing systems this is a more significant effect than the instrumental polarization mentioned above. The corrugated horns antennas/lenses used by DASI to observe the sky directly (without reflective optics) cause an effect at the ~ 4% level that can be reduced to 1% (~ 600 nK) or better by modeling. Bare horns are better by at least a factor of 100. Finally, phase errors in the measurement of the visibilities cause an uncertainty in the knowledge of the interference pattern that samples the sky. For example, a small phase error introduces a small anti-symmetric component to an otherwise symmetric interference pattern. The result is to intermix *E* and *B* polarization. In DASI, for example, this phase error is measured with an uncertainty of 0.4 degrees, or 0.7%. Improvements of a factor of 2 or so are required to reach the 3nK level.

**6.8 Testing to meet requirements and modeling:** For ground and balloon-based experiments, much of the verification of performance occurs in the field, in response to observations. For a space mission, it is critically important to verify the expected level of instrument-induced systematic errors during ground testing and to confirm equivalent in-flight performance.

For example, this is particularly true for the instrument calibration. CMB temperature anisotropy experiments take advantage of the relatively large (3.3 mK) CMB dipole signal for calibration by rapidly scanning large angles on the sky. No such signal is known to exist for polarization, and the rapid scanning of large angles may not be appropriate for a polarization experiment.

Accurate characterization of the antenna beam patterns both during integration and testing and in flight are required to achieve the science goals. Techniques for characterizing the main beam and sidelobes for both single-mode and multi-mode detector systems must be evaluated. Indeed, new high-accuracy measurement techniques may need to be developed to verify the performance of a CMB polarization observatory's optical system.

Control of systematic errors requires a systems level approach to mission design (see §9). An integral part of any serious design process will be the development of mission simulation tools and prototype data processing software. These tools will allow developers to generate mock science data sets for an entire mission and process those data sets through a prototype pipeline to verify that systematic error requirements can be met with a given mission design. At a minimum, these tools should be capable of simulating all of the following:

- A model sky signal that incorporates a range of predicted CMB signals as well as the important astrophysical foreground signals, especially those from our Galaxy.

- Model planet emission tied to a solar system ephemeris.

- A spacecraft orbit ephemeris and scan strategy that samples the sky and produces a flight-like raw data stream.

- Beam response models that include main beam asymmetry, far sidelobe response (which interacts with the solar system and foreground model), and cross-polarization response.

- A polarimeter model that includes detector and modulator systematics, such as offset and gain drifts, thermal susceptibility, and passband models.

As was done in designing the WMAP mission, the flight-like raw data generated by the data simulator should be fed into a prototype data processing pipeline that solves for the instrument calibration and generates scientific data products such as sky maps and power spectra. The output products may then be compared to the known inputs to assess the suitability of the proposed mission concept.

A particular requirement of the simulation and pipeline software that will be new to a CMB polarization mission is the need to process hundreds to thousands of channels in a relatively automated way. To date, CMB data analysis has typically involved a number of operations performed by hand and channel by channel, because current instruments have a small number of individually constructed detectors. This gives each detector channel an individual "personality" (and sometimes even a name) that often requires software tuned to that channel. Fortunately, near-term experience with large format arrays operating in sub-orbital experiments will allow automated data analysis techniques to be tested in the field. Additional data processing and analysis challenges will be discussed in §9.

**6.9 Scan Strategy and Orbit Selection**

The scan strategy for a CMB experiment has a very strong influence on many aspects of the experiment's performance. These include: 1) the extent and uniformity of the sky coverage, 2) the degree to which the beam response is symmetrized by observing each sky pixel at a range of azimuth angles, 3) the structure of the pixel-pixel noise covariance matrix in the presence of 1/f noise, and most importantly, 4) the degree to which polarization signals fixed on the sky are modulated in the data stream to facilitate the separation of sky signals from instrumental artifacts. The ability to measure large angular scale polarization will require careful selection of the scan pattern, the scan speed and the distribution of viewing angles in each sky pixel.

A critically important consideration is the time scale of the sky signal modulation(s) relative to the time scale(s) of the various instabilities in the system. For example, if a detector has a "1/f" knee frequency of 0.1 mHz, it is desirable to modulate the sky signal, either through sky scanning or internal switching, on a time scale faster than 10 seconds. Moreover, it is desirable to modulate the sky signals on a variety of widely spaced time scales. Such multiple modulations typically translate to sensitivity over a wide range of angular scales. For example, the WMAP mission simultaneously observes two spots on the sky separated by ~140° with two feeds, and modulates the signal between the two feeds at a frequency of 2.5 kHz. Additionally, WMAP spins about its symmetry axis once every 2.2 minutes, and precesses about the Sun-Earth line once per hour. Finally, the satellite revolves around the Sun once per year and so repeats a given sky scan each year. This produces a very robust data set that is sensitive to all angular scales from the full sky down to the angular resolution of the instrument.

One technique for separating real polarization from those introduced by the instrument is to compare multiple measurements of one pixel where the angle around the boresight of the instrument is varied. This approach will probably be required in the search for the very faint signal from $B$ modes, even with an instrument optimized for polarimetry. The speed with which this modulation needs to be performed is still an open question. Current and future CMB missions use various scan patterns to provide different levels of sensitivity to the polarization signal, though none of these missions was optimized for polarization sensitivity. The COBE mission achieved an almost ideal distribution of viewing angles at the cost of observing as close as 64° from the Sun. The WMAP mission achieves a good pattern of scan angles while scanning no closer than 87° from the Sun. The scan pattern of the Planck mission gives a rather small range of viewing angles away from the ecliptic poles, where its sensitivity is concentrated.

For an orbital mission, it is important to determine an appropriate scientific metric, such as the noise level in the $B$-mode power spectrum at low $l$ in the presence of 1/f noise, and use it to evaluate various scan strategies. This will require sophisticated mission simulation and data processing capabilities, but, given these tools, one can meaningfully trade off mission performance against mission complexity.

Orbit selection will be tightly coupled to the desired scan pattern and sky coverage, and will have a significant impact on the final mission design. The choice of orbit will in turn affect mission duration, data transmission requirements, thermal/power stability, radiation/magnetic environments, and stray-light rejection.

# 7 Detectors and Focal Plane Instrumentation

For most of its history, progress in CMB observations has been driven by the invention of ever more sensitive mm- and cm-wave continuum detection schemes. Since the 1960s, CMB experimentalists inventing new types of detectors in university and government laboratories supported by DoE, NASA, NIST and NSF have succeeded in improving raw sensitivity by three orders of magnitude, from a temperature resolution of ~100mK to ~100μK in one second of integration. This relatively low-cost effort has enabled spectacular progress via successive generations of experiments, each equipped with one or several of the latest detectors.

This *modus operandi* has now come to an end. Across the entire band of interest (roughly 30 to 300 GHz) detector sensitivity has reached the fundamental limit set by fluctuations in the photon background for ground-based and balloon-borne experiments. Orbital experiments, which operate in the lowest backgrounds and thus provide the greatest opportunity to take advantage of superior detectors, are not far behind. The polarization sensitive bolometric detectors on Planck will achieve sensitivity within a factor ~4 of the fundamental limit due to fluctuations in the CMB itself (~4 BLIP), and within a factor ~2 of the practical limit of sensitivity imposed by such backgrounds for any real detector system. As a result, future progress in the field requires a new approach.

We have argued that, based on our current knowledge of polarized foreground emission, CMBPOL will require at least ten times the sensitivity of Planck to reach the limit set by foreground confusion. Table 7.1 illustrates how this sensitivity might be achieved. Though the necessary number of detectors depends on details of how close to the fundamental CMB background limit they are designed to operate, how many frequency bands are needed, and the total mission integration time, it is clear that CMBPOL will require on the order of a thousand or more detectors. New focal plane architectures – for example, monolithic, multiplexed arrays of a hundred or more detectors – will be required to achieve focal planes of this scale.

**Table 7.1 Parameters Governing Total Mission Sensitivity**

| Mission | WMAP 100GHz | Planck 100GHz | CMBPOL |
|---|---|---|---|
| Duration | 8 years | 1.2 years | 2–8 years |
| Detectors/Band | 8 | 8 | 150–1000 |
| NET/BLIP$_{CMB}$ | ~60 | ~4 | ~2 |
| NET/1°x1° | ≈ 6μK | ≈ 1μK | 20–100nK |

*Notes: (a) NET/BLIP$_{CMB}$ is the ratio of detector sensitivity to fundamental CMB background limit at 100 GHz, (b) NET/1°x1° is the statistical noise equivalent CMB temperature at 100 GHz, assuming total integration time spread evenly over the sky, (c) CMBPOL values are plausible guesses, given to illustrate how CMBPOL might achieve a sensitivity 10 to 50 times greater than Planck.*

The leap from WMAP and Planck to CMBPOL is similar to the revolutionary leap that visible light astronomy made in going from small numbers of photomultiplier tubes to imaging CCDs. Unlike the leap to CCDs, however there are no industrial partners to share the burden of the technology development. The fabrication of large format arrays with a high yield of active pixels places much larger demands on process control than fabrication of individual detectors that can be screened for performance. In addition, the nature of the measurement – high-fidelity polarimetry with an accuracy of a few parts per *billion* of the total background – places additional requirements on the detector system that have yet to be demonstrated, even in systems with small numbers of detectors. These requirements include high common-mode rejection (exacerbating the need for uniformity of pixels in large arrays) and, perhaps, on-chip polarization modulators (complicating the architecture).

For all of these reasons, continued progress in the field requires a concerted and well-coordinated program of detector development of an unprecedented scale. The small efforts that have successfully driven the field for most of its history will continue to play two vital roles: (1) developing the new detector concepts and polarization modulation schemes that will be necessary to achieve polarimetric fidelity at the few times $10^{-9}$ level that is required, and (2) proving these techniques on ground-based and balloon-borne experiments. In parallel, a stable base of funding will be necessary to build and sustain the groups at the national laboratories and NASA Centers (GSFC, JPL, LBNL and NIST) that will ultimately be equipped to produce the large-format, high-yield, flight-qualified arrays that will fly on CMBPOL.

In the next several sections we survey the state of several promising detector technologies, discussing their relative advantages, and also the problems that need to be addressed to make them candidates for a mission to measure the CMB polarization at a level of $r = 0.01$.

**7.1 Practical Limits to Sensitivity: Bolometers and HEMTs**

Fluctuations in the arrival rate of CMB photons impose a fundamental limit of $\sim 30$ $\mu$K$\sqrt{}$(sec) for detection of a single mode of radiation in a fractional bandwidth of 25% from $\sim 30$ to 220 GHz. We will assume throughout this section that each pixel in the focal plane contains two detectors that couple to orthogonal linear polarizations. Averaging the signal from two such ideal detectors measures Stokes $I$ with a sensitivity of NET $\sim (30/\sqrt{2}) \sim 20\mu$K$\sqrt{}$(sec). Differencing the two detectors measures Stokes $Q$ with the same sensitivity, NEQ = NET following the convention that $Q = (Tx - Ty)/2$.

Bolometric detectors can, in principle, approach this sensitivity over the entire frequency range of interest, but at the expense of driving both instrument emission and bolometer NET well below the CMB contributions. When system trades are taken into account, diminishing returns for the cost of decreasing the temperature of optics and of the bolometer heat sink typically lead to an optimized system operating at a factor $\sim 2 - 3$ above the CMB BLIP limit. A realistic goal for the sensitivity of a single bolometric detector is thus $\sim 40 - 60$ $\mu$K$\sqrt{}$(sec). The polarization-sensitive bolometers (PSBs) on Planck fall short of this goal by factors of 1.2 (at 143 GHz) to 1.7 (at 100 GHz). The sensitivity of the Planck detectors is limited by the requirement on the thermal time constant. Next generation bolometers will achieve faster response through the use of antenna-coupling and/or transition-edge superconducting (TES) sensors with higher electro-thermal feedback, and can realistically be

**Table 7.2: Current and Projected Sensitivity[a] of Bolometer and HEMT-based Detection Schemes.**

| Freq. | 2005[b] | | 2010[c] | |
|---|---|---|---|---|
| | Bolometer | HEMT /$\sqrt{2}$ | Bolometer | HEMT /$\sqrt{2}$ |
| [GHz] | [$\mu K_{cmb}\sqrt{s}$] | [$\mu K_{cmb}\sqrt{s}$] | [$\mu K_{cmb}\sqrt{s}$] | [$\mu K_{cmb}\sqrt{s}$] |
| 30 | – | 93 | 57 | 48 |
| 40 | – | 115 | 51 | 51 |
| 60 | – | 175 | 44 | 60 |
| 90 | 67 | 224 | 40 | 75 |
| 120 | – | – | 40 | 93 |
| 150 | 48 | – | 43 | – |
| 220 | 68 | – | 64 | – |
| 350 | 224 | – | 220 | – |

*Notes: (a) All sensitivities are $NET_{cmb}$/feed = $NEQ_{cmb}$/feed. The HEMT NET/feed is divided by $\sqrt{2}$ to provide a fair basis of comparison between the HEMT-based system, which can measure both the Q and U Stokes parameters simultaneously, and the bolometric system, which measures only Q or U at any given time. (b) 2005 sensitivities are (i) Bolometers: average laboratory measured optical performance of Planck flight Polarization Sensitive Bolometers and (ii) HEMTs: best laboratory performance achieved to date. (c) 2010 bolometer sensitivities are projections for antenna-coupled TES devices, assuming 1.5K RJ instrument emission, 25% fraction bandwidth, 40% optical efficiency, a 100 mK heat sink, a 300 mK superconducting transition, and a thermal safety factor of 3. 2010 HEMT sensitivities assume $T_{sys} = 3h\nu/k_B$, 20% fractional bandwidth, and negligible instrument emission.*

expected to achieve ~40 μK√(sec) near 100 GHz, and < 60 μK√(sec) over the range 30 to 220 GHz.

The performance of HEMT-based detectors is fundamentally limited by a quantum noise limit of $T_{sys} \geq h\nu/k_B$; real HEMTS are not expected to achieve performance better than a factor of ~3 times this limit in the foreseeable future. This limit does not significantly degrade the sensitivity of HEMT-based detection systems at low frequencies, but becomes a dominant contribution to the noise at higher frequencies.

Table 7.2 compares the current state of the art sensitivity for both types of detection system with the sensitivity that might be achieved by 2010. In order properly to compare the sensitivity with which the two detection schemes would measure both Q and U, the NEQ of the HEMT-based system has been divided by √2, to account for the fact that the HEMT-based system can be configured to measure both Stokes parameters simultaneously.

Historically, bolometers have been the most sensitive detectors for CMB studies, but have not been competitive with HEMTs at frequencies below 100 GHz. To date, all bolometers used in CMB experiments have coupled to radiation via a thermally suspended absorbing element with dimensions comparable to a wavelength. This coupling becomes impractical at frequencies below ~100 GHz, as the thermal mass of the absorber rapidly dominates the heat capacity of the device.

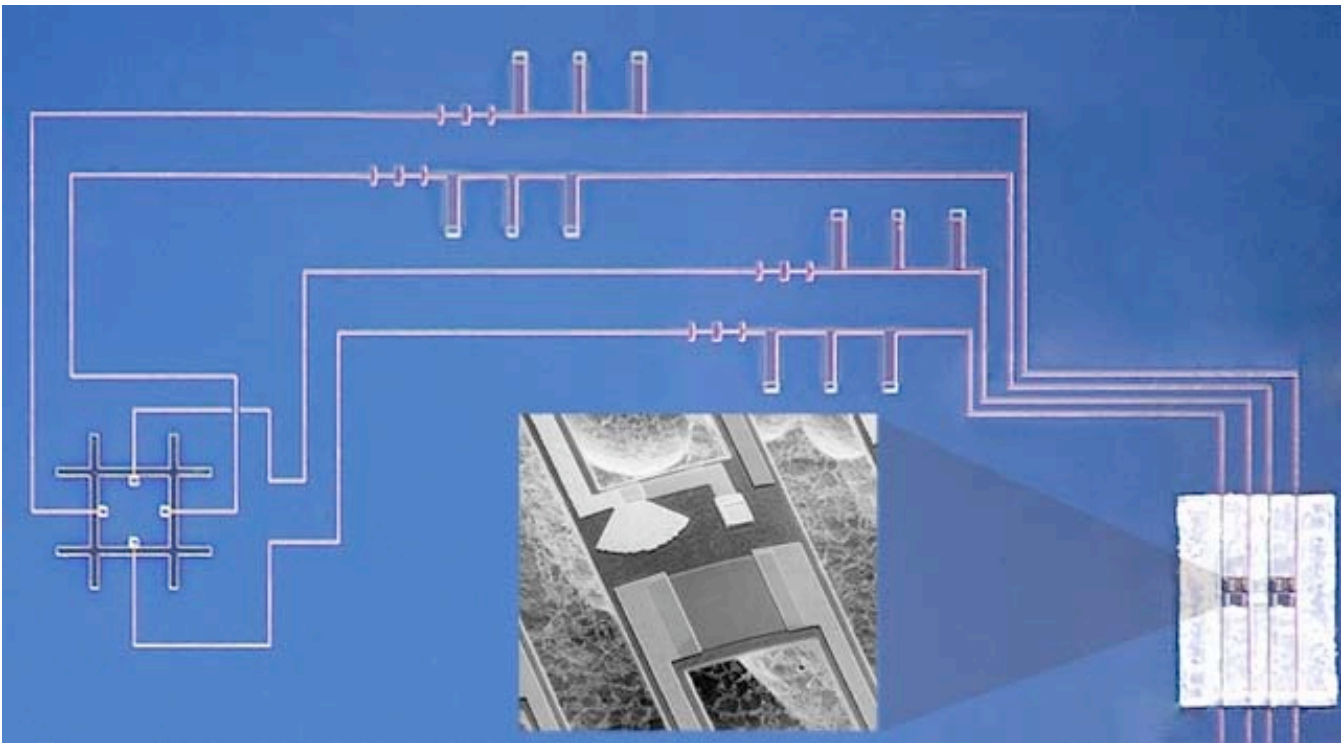

*Figure 7.1: A planar-antenna-coupled bolometer. A dual-polarization antenna is on the left. Each double-slot dipole antenna coherently adds the signal from two slot dipoles to form a relatively symmetric antenna pattern. The slots in this chip are lithographed in a superconducting Nb ground plane. They are ~1 mm long and have a resonant response centered at 220 GHz. Microstrip transmission lines and transmission line filters are used. The filter combination at the top of the photograph includes a low pass filter (left) and a band pass filter (right). The design bandpass is centered at 220 GHz with a 30% bandwidth. These filters have been tested in an end-to-end receiver optical measurement that shows 20% receiver efficiency. The transmission lines terminate in the matched loads on the leg-isolated TES bolometers at the lower right.*

For detectors that are intended to couple to a single polarization of a single spatial mode, the absorber can be replaced by an antenna that is not part of the thermal mass of the detector. Such detectors, an example of which is shown in figure 7.1, have been demonstrated in the laboratory, and will soon be implemented in sub-orbital experiments. The sensitivities for next-generation bolometers given in Table 7.2 assume some sort of antenna coupling. Given this assumption, bolometers achieve comparable or higher sensitivity than HEMTs at frequencies above 40 GHz, thus providing a single technology capable of spanning the entire frequency range of interest.

The BOOMERanG receiver is based on the polarization sensitive, absorber-coupled bolometers. Similar detectors are now in use in QuaD and BICEP, and will be used on Planck. It will be several years, however, before antenna-coupled bolometers are used in real experiments in the field. Currently only the PAPPA experiment is being based on these devices. Given the promise these detectors hold for CMBPOL, we expect other efforts will be proposed in the future.

There are several variants of antenna coupling that are currently in development. These include (i) mounting a dual-polarization planar antenna behind a corrugated feedhorn, (ii) mounting a dual polarization planar antenna behind a lens, and (iii) coupling directly to a planar array of dual polarization antennae on the backside of a silicon wafer. Each of these methods has advantages and disadvantages, and it is only through using each in real experiments that it will become clear which is superior or if a new approach is required.

### 7.2 Bolometers

#### 7.2.1 Polarimetry with Bolometric Detectors

##### 7.2.1.1 Analyzers

A polarimeter consists of a modulator, which switches or rotates the polarization vector, and an analyzer, which detects one or both linear polarizations. The goal for CMBPOL is to extract a *B*-mode polarization signal that may be as small as a few parts per billion of the background. This requires sensitivity to polarization fraction that is unprecedented (to our knowledge in any type of polarimetry) and, correspondingly, a detector system with extremely high common-mode rejection. It is safe to assume that a CMB polarimeter must use a dual polarization analyzer, whereby a single Stokes polarization parameter is obtained by differencing two orthogonal polarization detectors. With a dual analyzer scheme, the two polarization states pass through common atmosphere (if any), optics, feeds (or antennas), filters, and two detectors in close proximity. Thus, common-mode signals are rejected by taking the difference signal between the detectors.

Dual polarization analyzers have been realized in both HEMT-based and bolometer-based systems. An example is shown in figure 7.2.

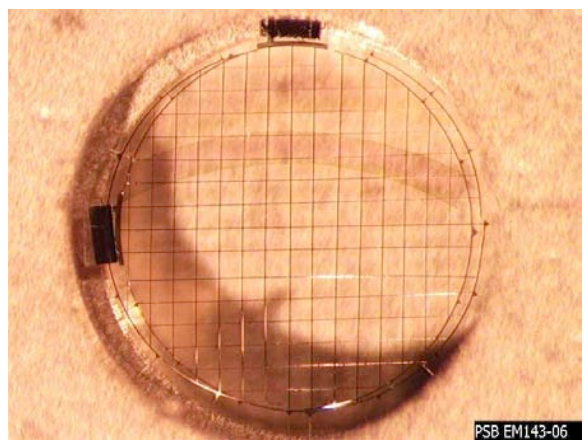

*Figure 7.2: A Polarization Sensitive Bolometer (PSB) pair. The two detectors are oriented to couple to orthogonal linear polarizations and positioned $\lambda/4$ ~500$\mu$ from a backshort. PSBs are a relatively mature technology: 140 GHz PSBs have flown on BOOMERanG (2003); 100 and 143 GHz PSBs are currently being deployed in BICEP and QUaD, and 100, 140, 220 and 350 GHz PSBs have been delivered to Planck for flight in 2007.*

While a PSB is an efficient dual-analyzer, it provides only a single Stokes polarization parameter. A complete measurement of linear polarization, which extracts both Stokes *Q* and *U*, necessitates using two PSBs with one pair rotated 45° from the other. Unfortunately, the PSB pairs thus cannot view the same sky simultaneously, so a complete polarization measurement requires combining observations either with different detectors separated in time, or the same pair rotated by a modulator, also separated in time An ideal analyzer measures Stokes *Q* and *U* simultaneously in a single beam. With antenna-coupled bolometers, such a *QU*-analyzer could consist of an arrangement of standard passive RF components, namely two power dividers and a broadband 180-degree hybrid, and four detectors as shown in figure 7.3. Using this arrangement, all of the linear polarization information can then be obtained in a common beam by taking the pair differences and sums.

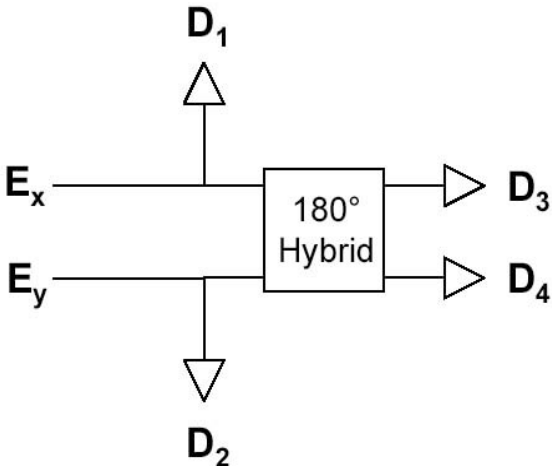

*Figure 7.3:* **QU-analyzer for direct detectors.** *Electromagnetic radiation from a feed or antenna is transmitted to two power splitters. Half the signal from each splitter goes to a 180° hybrid, and half to a detector (D1 and D2). The output channels of the hybrid, which give the sum and difference of the inputs, then pass to detector pairs D3 and D4. The signals in the 4 detectors are a combination of 3 Stokes parameters: $S_1 = E_x^2 = (I + Q)$, $S_2 = E_y^2 = (I - Q)$, $S_3 = (E_x + E_y)^2 = (I + U)$, and $S_4 = (E_x - E_y)^2 = (I - U)$. All of the linear Stokes polarization information can be extracted by forming the pair differences and sums: $Q = 2(S_1 - S_2)$, $U = 2(S_3 - S_4)$, $I = 2(S_1 + S_2) = 2(S_3 + S_4)$.*

**7.2.1.2 Modulators**

The polarization signal must be modulated in order to separate it from the inevitable offset in the output of the polarization analyzer. The simplest way to modulate the signal is to scan the instrument on the sky, and measure spatial differences in polarization. This is the method employed by WMAP and Planck. The challenge of using only this modulation technique is to ensure that the polarization offset is stable over the scan period.

For example, the spin period of Planck is 1 rpm, and that in turn requires instrument stability to 0.016 Hz to recover signals on the largest spatial scales. While faster scan rates are possible, doing so increases the data transmission rate and reduces the required detector speed of response, set by the beam-crossing time. Thus, a detection system suitable for CMBPOL will require either stability on minutes-long time scales or (preferably, since there may be instabilities introduced by other parts of the system) an additional modulation scheme operating on faster time scales.

A classical optical modulator (based, for example, on a rotating waveplate or Faraday rotation in a magnetized ferrite) can be used if the band coverage is limited. Both of these solutions have been developed for funded sub-orbital experiments (Faraday rotation for BICEP, and a magnetically levitated rotating waveplate for EBEX). Both solutions are potentially problematic for use in CMBPOL. The magnetic field strengths required for Faraday rotation make operation in free space impractical over large areas. The devices in use operate in single mode circular waveguide, requiring feedhorn coupling, and dissipate significant amounts of power. A rotating waveplate introduces the risk and cost of moving elements, and the difficult challenge of operating over a broad band.

By modulating the signal on the focal plane, we can avoid the bandwidth problem. Even in a multi-frequency focal plane, modulators may be placed downstream of any filters and dichroics. With direct detectors, an on-focal-plane modulator must be highly efficient, as the modulation occurs pre-amplification. The key to focal plane modulators for direct detectors is the switching component, which must not dissipate significant power at 100 mK, cannot appreciably heat local sections of transmission line and thereby introduce photon noise, and must have stable on and off states so as not to inject noise associated with the switch itself. Possibilities for active components include junctions, transition-edge superconductors, and MEMs-based switches. A prototype polarization switch based on junctions is shown in figure 7.4. This device achieves low power dissipation and is designed for 30% bandwidth, limited by the matching of the switching junctions. A polarimeter based on a MEMs-based switch is shown in figure 7.5.

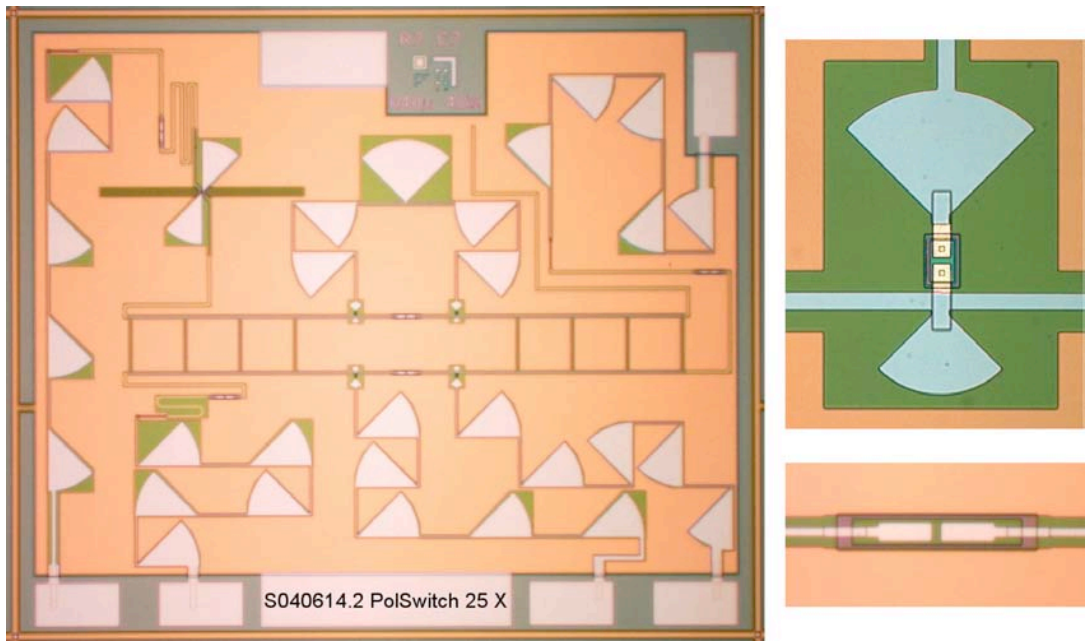

*Figure 7.4: Junction-based polarization switch with two input arms and two output arms. The input arms are 1) a slot antenna, in order to introduce an optical signal, and 2) a termination resistor. The output arms go to junction detectors – these will be replaced by antenna-coupled bolometers in a full device. The signal passes through an arrangement of junction switches (see inset top right) which may be switched on and off by applying bias current. The DC drive current is electrically isolated from the detectors by 4 series capacitor sections in the transmission line (see inset bottom right).*

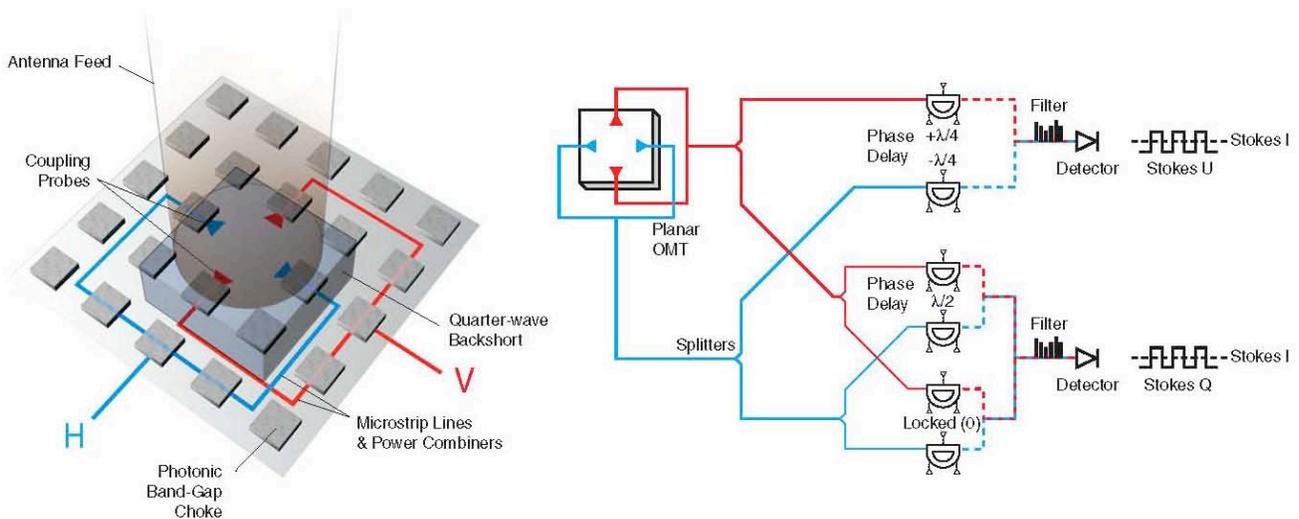

*Figure 7.5: Schematic of a MEMs-switched polarization modulator. MEMs switches synchronously inject half- or quarter-wave phase delay into microstrip transmission lines. After re-combination and square-law detection, the signal has a DC term proportional to Stokes **I** and a modulated term proportional to Stokes **Q** or **U**. This planar "polarimeter-on-a-chip" requires five functional components: a polarizer (ortho-mode transducer) to split the incident electromagnetic wave into orthogonal polarizations, a phase switch to inject half- or quarter-wave phase delays into a single arm, a power divider to recombine the phase-delayed voltages, a filter to eliminate out-of-band power to the detector, and a square-law detector to provide non-linear mixing. Each element can be realized in microstrip. MEMs switched capacitors short out selected lengths of transmission line to provide phase modulation, while transition-edge superconducting bolometers serve as sensitive square-law detectors. The entire polarimeter can be produced using photolithographic techniques, and is fully scalable to kilo-pixel arrays. Figure courtesy of Al Kogut/PAPPA collaboration.*

## 7.2.2 Multiplexed Bolometric Arrays: TES and MKID

Given the large number of pixels in the focal plane, bolometric detectors for CMBPOL will require a cold, low power dissipation multiplexer. There are two thermometer technologies under development that provide this possibility: the superconducting Transition Edge Sensor (TES) detectors and Microwave Kinetic Inductance Detectors (MKID). The TES technology is more mature; and is being widely and rapidly integrated into sub-orbital instrumentation. The MKID technology is still in development, but may offer attractive advantages, especially for an orbital mission.

### 7.2.2.1 TES Detectors

TES is a mature detector technology that has the potential to fulfill all of the requirements of future CMB experiments. TES bolometers cooled to temperatures of 50–300 mK can have a sensitivity that is limited by photon arrival statistics over much of the frequency range of interest. They have two properties that are essential for building large focal-plane arrays (i) They are simple to fabricate using optical photolithography, and (ii) their readout can be "multiplexed" so that a row of detectors can be readout using a single amplifier – this greatly reduces the complexity of the cryogenic wiring.

The TES is a superconducting film biased in the middle of its transition. It is voltage biased, and in this mode it has high stability and linearity due to negative feedback that occurs between the thermal and electrical "circuits" of the bolometer. The signal from a TES is measured using a Superconducting QUantum Interference Device (SQUID) ammeter, which can operate at cryogenic temperature.

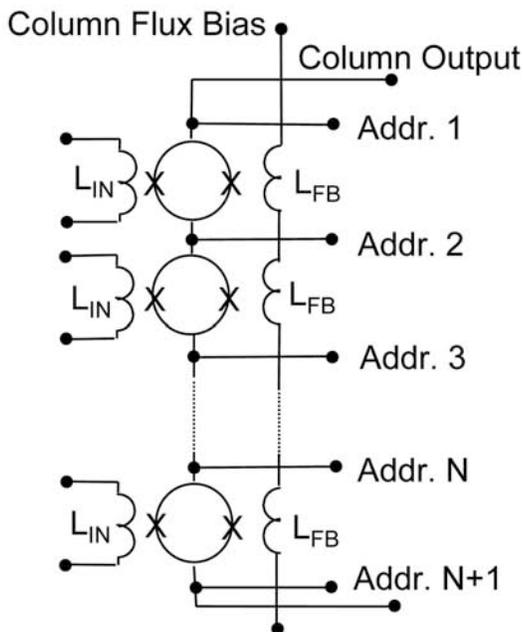
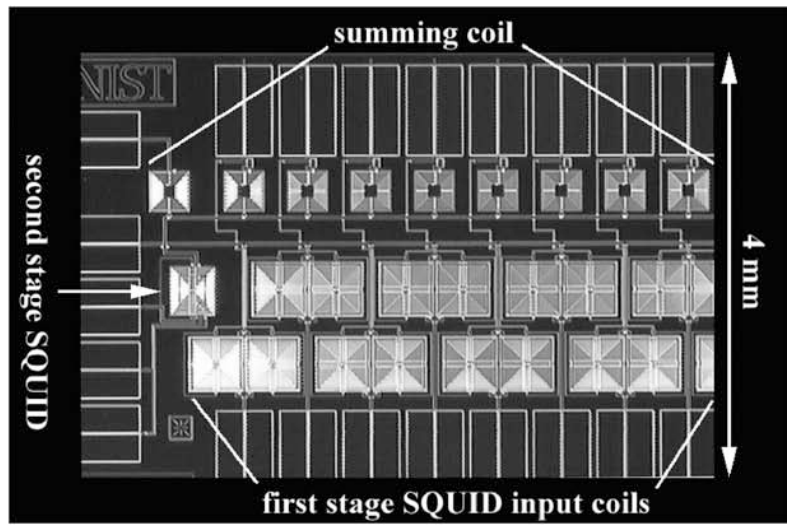

*Figure 7.6: Left: Circuit schematic for an early version of a NIST time-domain SQUID multiplexer, showing its basic functions. Each input inductor $L_{IN}$ is connected in series with a TES detector that is always biased and a bias resistor (neither shown here). The N SQUID loops are connected in series. One SQUID is turned on sequentially using the address lines. Since the other SQUIDs are not biased, they remain in the zero voltage state. The column output is that of the one SQUID that is biased. The feedback flux for each SQUID is stored digitally between cycles and applied during the on state. Right: Photograph of the kilopixel SQUID multiplexer under development for SCUBA2. Figure and photograph courtesy of Kent Irwin.*

There are several readout multiplexer technologies that are reaching maturity, and they can be broadly divided into techniques that divide signals in either time or frequency domains. A time-domain readout multiplexer that uses SQUID switches to sequentially choose the detector that is read with the single output amplifier has been developed at NIST. The time-domain multiplexer can read 32 detectors with a single readout amplifier with no loss in bolometer noise performance or bandwidth. It has been used in an 8-channel system at the Caltech Submillimeter Observatory, and it will be used with arrays of several thousand pixels in several upcoming experiments including SCUBA2 and ACT.

Several groups are independently working on a frequency-domain readout multiplexer. In this scheme, each detector is biased using a sine wave with a unique frequency, the bias signals are amplitude-modulated by the bolometers, and the sum of all the currents is measured using a single SQUID ammeter. This type of multiplexer will be used for 1000 pixel arrays in several upcoming experiments

### 7.2.2.2 MKIDs

The MKID is a relatively new detector for sub-millimeter/millimeter wavelengths. Though less mature than the TES technology, it offers potential advantages that warrant its continued development.

In a kinetic inductance detector, energy absorbed in a superconducting film breaks Cooper pairs, creating excitations from the superconducting state called quasiparticles that modify the surface inductance of the film. In an MKID, the superconducting film is incorporated into a microwave frequency resonant circuit, so that changes in surface inductance translate into changes in resonance frequency and can be sensitively measured. Changes in resonance frequency impart a phase shift to the transmitted on-resonance drive signal, which is detected with a high electron mobility transistor (HEMT). A typical MKID has a resonance frequency of several GHz and a quality factor Q of 100,000–1,000,000.

Like TES detectors, MKIDs can be coupled to planar antennas using microstrip lines, but MKIDs have several advantages over TESs. First, MKIDs lend themselves to a very simple and powerful multiplexed readout. Since the Qs of the resonators are high, a large array of MKIDs, each designed with a unique resonance frequency, may share a common feedline. That single transmission line can carry a comb of microwave signals, each of which interrogates a particular element of the array and is unaffected by the other array elements. Because the excitation signals are in the gigahertz range, there is plenty of bandwidth for each detector, even for arrays of thousands of detectors. A single cryogenic HEMT is used to amplify the signals from the entire array, and the signals are demultiplexed with room temperature electronics. Because complex cryogenic electronics is not needed, the system is very flexible. For example, a single-element readout can be used for testing the entire array during development by tuning through the readout frequencies sequentially. The full multichannel readout system may be used with different arrays simply by swapping cables.

The second advantage of MKIDs is that they detect athermal energy. The fundamental noise is limited by fluctuations in the number of thermal quasiparticles, which, owing to the superconducting energy gap, vanishes exponentially as the temperature is decreased. MKIDs are operated well below the transition temperature, so they are immune to Tc variations across the array and are highly insensitive to variations in substrate temperature.

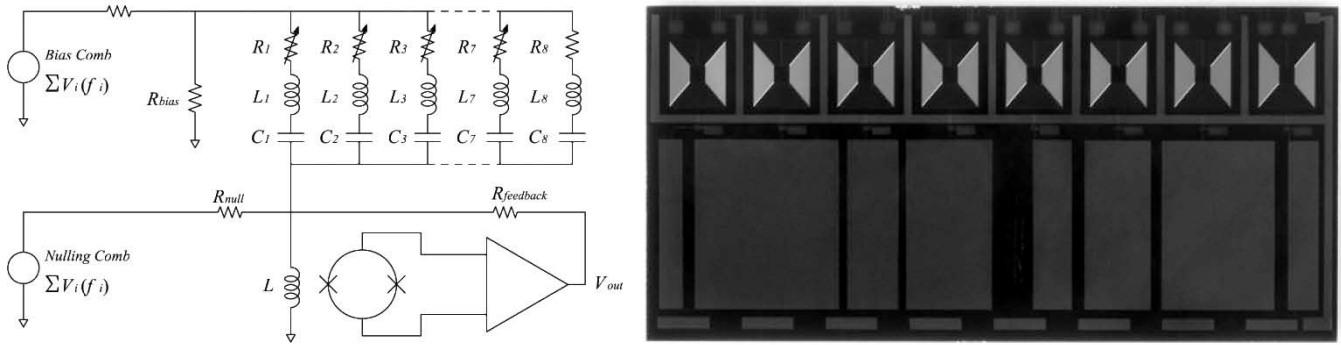

*Figure 7.7: Left: Schematic of a current-summing frequency-multiplexer circuit. The TES devices, represented as variable resistors, are sinusoidally biased each with a different frequency. An LC resonator in series with each TES filters out wideband Johnson noise that would contribute to the other channels. The currents from all muxed TESs are added together at the input inductor of the SQUID. Shunt feedback is used to reduce the input impedance at that point creating a virtual ground. A 180° phased shifted bias signal can be added to the SQUID input to null each of the carriers from the TES partially. The amplitude of this nulling current can be adjusted on longer time scales than those characteristic of astronomical observations. Right: Photograph of a niobium LC filter chip fabricated by TRW (now Northrup-Grumman). The center frequencies vary from 380 kHz to 1 MHz, with an 80 kHz channel spacing. A 32 channel multiplexer can be implemented by using four chips with interdigitated frequencies with a resulting spacing of 20 kHz.*

Finally, arrays of MKIDs are particularly easy to fabricate. Delicate thermal isolation structures are unnecessary, and the sensors themselves are a single superconducting layer, rather than a bilayer, which is difficult to fabricate uniformly across a large wafer. Figure 7.8 shows an antenna-coupled MKID detector, suitable for arraying, *made with only three mask layers*. The antenna is a dual polarization, in-phase combined, slot-antenna array, designed for 350 GHz. The power from each polarization is sent to separate MKIDs, coupled to the same feedline, on the sides of the antenna.

Currently, the sensitivity of MKIDs is limited by noise from the dielectric substrate on which they are fabricated. Dipole moments associated with impurities, defects or surface states in the dielectric couple to the electric field of the MKID and cause resonance frequency noise. The responsiveness of the MKID decreases with absorbed power through the density dependence of the quasiparticle recombination time, whereas the frequency noise remains relatively power independent. The NEP of MKIDs thus increases with loading. The NEP of present devices is adequate for ground-based instruments, which operate at large absorbed power levels and are limited by atmospheric noise. The NEP is about five times the background limit expected for space-based millimeter-wave imaging. There are a variety of methods of lowering the NEP, and photon noise limited performance is likely to be achieved in the next few years as work on MKIDs continues.

### 7.2.2.3 Other detector concepts

In addition to the work that has been mentioned above there are new ideas in receiver technology, which may grow to become candidates for improved polarization observing experiments. Designs exist for small systems that combine traditional microwave techniques with transition-edge hot-electron microbolometers. The result is a single unit comprising a wideband antenna, filter, and a detector that needs no thermal isolation from the bath. Another concept being developed employs mm-wave transistors that can sense the excitation of a single electron. Prototypes of frequency selective bolometers that are stacked in a resonant structure have been tested. With such a device, one spot in the focal plane can be used to sense multiple frequency bands simultaneously without the extra real estate required of beam splitters.

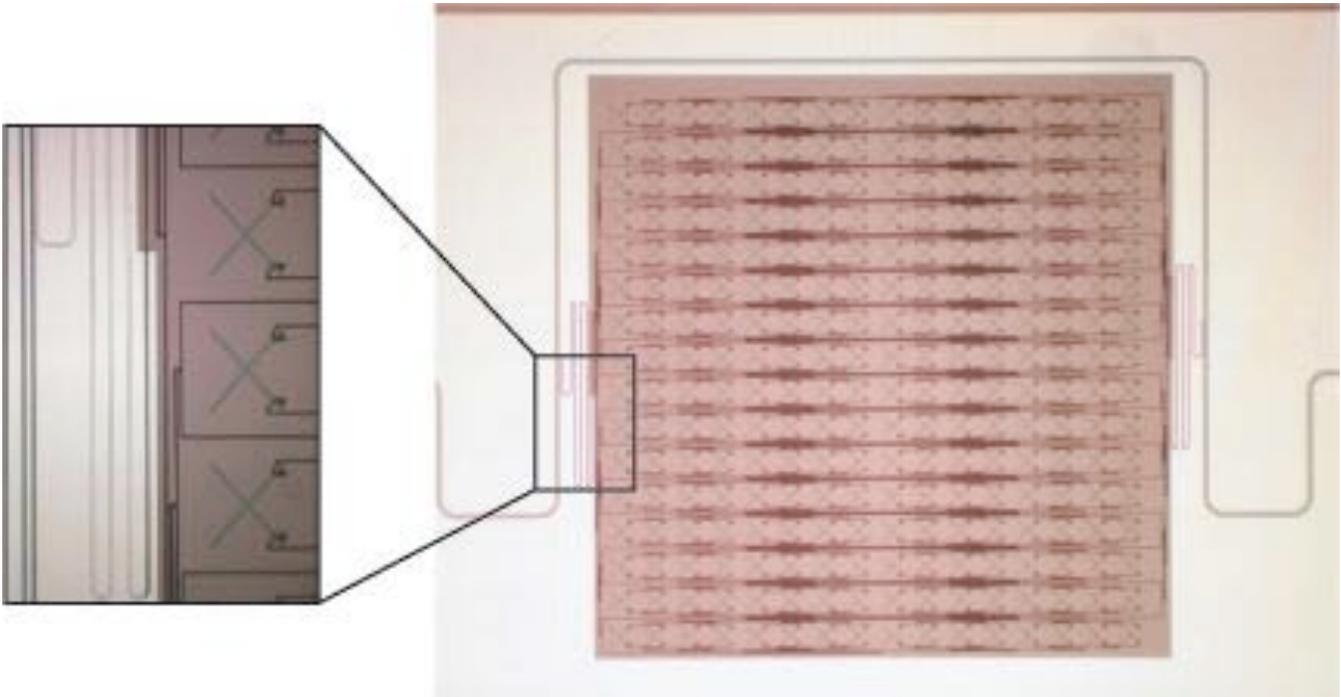

*Figure 7.8: MKIDs coupled to a 350 GHz slot array antenna. The power collected by the antenna is sent to two MKIDs (meanders on either side of the antenna) via microstrip lines.*
*Picture courtesy of Peter Day.*

**7.3 High Electron Mobility Transistors (HEMTs)**

Detectors based on High Electron Mobility Transistors (HEMTs) differ fundamentally from direct detectors. HEMTs are used to amplify the incident signal allowing lossless post-processing prior to detection. The process of amplification adds noise to the signal (the minimum given by the "quantum limit", $T_{sys} \geq h\nu/k_B$), but, because it preserves the phase of the signal, it also enables both *Q* and *U* Stokes parameters to be extracted.

Coherent amplification, or phase preservation of the incident signal through amplification, enables the use of the powerful technique of *correlation*. In correlation polarimetry, the signals are pair-wise multiplied with only the correlated portion providing an output, while the uncorrelated component appears as noise. Each of the first generation of experiments to detect CMB polarization (DASI, CAPMAP, CBI) and/or its temperature correlation (DASI, WMAP) has employed HEMT amplification and either interferometry or correlation polarimetry.

In each of these experiments, the signal is modulated post-amplification using some method of 180° phase modulation of the signals. This modulation and subsequent demodulation reduces systematic offsets by orders of magnitude, providing a relatively systematic-free measurement of the CMB.

When cooled to 20K, InP HEMTs have demonstrated noise of 3–5 times the quantum limit at frequencies below 100 GHz. Cooling below 20K does not significantly improve the noise, and higher temperatures result in only slow degradation of noise.

Arrays of receivers are enabled by monolithic microwave integrated circuit (MMIC) technology in which passive networks are integrated with the active devices on a single chip. InP MMIC amplifiers and phase switches provide state-of-the-art performance from 30–110 GHz. The largest astronomy focal plane arrays to date using MMIC technology are CAPMAP with 16 polarimeters and SEQUOIA, with 32 heterodyne receivers. WMAP with 20 elements uses discrete HEMTs in custom built amplifiers. Planck's Low Frequency Instrument with 20 elements uses a mixture of discrete transistor and MMIC technologies.

MMIC chips can be integrated into multichip modules using semiconductor assembly techniques becoming standard in industry, enabling the development of large arrays of receivers. Using these techniques it is possible to develop completely integrated receiver modules in a small IC style package, quite suitable for large array integration. JPL has demonstrated a small number of compact integrated pseudocorrelation polarimeter modules operating at 40 and 90 GHz. The block diagram for one of these modules is shown in figure 7.9, and photographs are in figure 7.10.

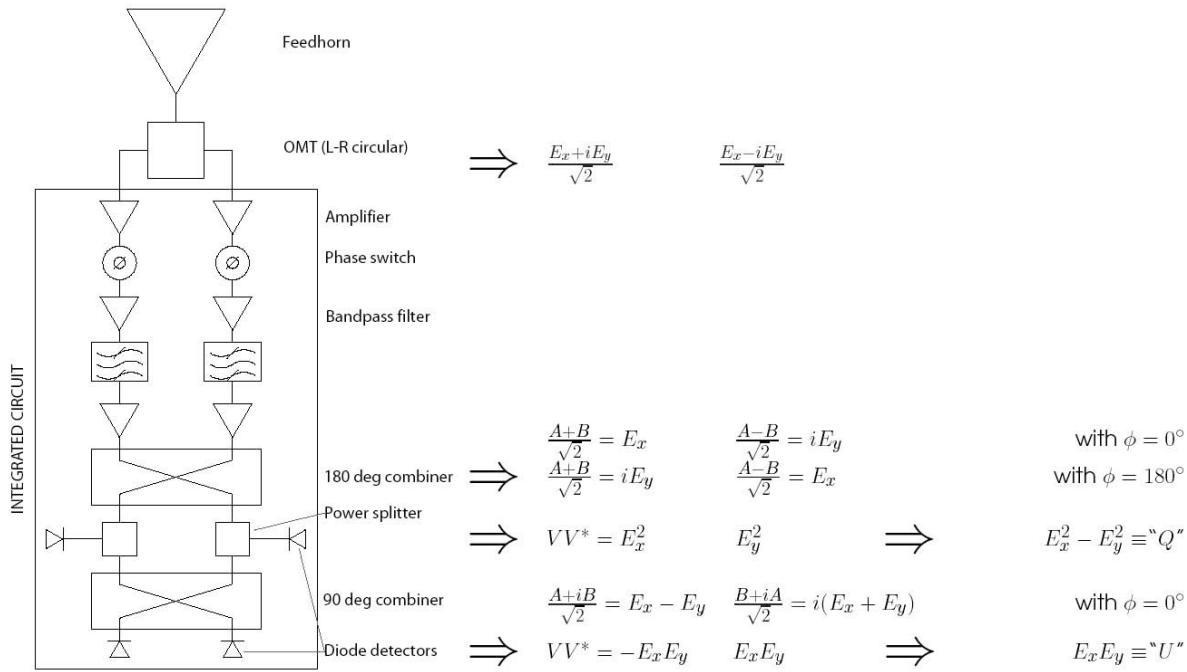

*Figure 7.9: Block diagram of a HEMT polarimeter. The block diagram shows all of the major components of an integrated pseudocorrelation polarimeter and the mathematical functions carried out at each stage. Figure courtesy of Todd Gaier.*

The incoming signal is passed through a left-right circular polarizer, then amplified in two parallel chains, and one leg is phase switched at kHz frequencies. The signals are combined in a 180º coupler, power divided and detected. The detected demodulated signal is the Stokes parameter $Q$. The remaining signals after power division are combined in a 90° coupler and detected. The detected, demodulated output is the Stokes parameter $U$. The simultaneous detection of $Q$ and $U$ with the same beam is another important systematic advantage of coherent detection.

The sensitivity of the 90 GHz MMIC polarimeters is expected to be 350 µK√(sec) for each of the two linear polarization Stokes parameters, when operated from a good terrestrial site. At 40 GHz the sensitivity is expected to be 225 µK√(sec) for each of $Q$ and $U$. For arrays of equal numbers of pixels (i.e. combining the $Q$ and $U$ sensitivities), operating from the ground, HEMT arrays have better sensitivity than existing bolometric systems at frequencies up to about 100 GHz.

Using this approach there is no fundamental limit to the number of receivers that can be fabricated into an array. In large quantities, the cost per element is expected to be less than $500 for the electronics and a comparable amount for the passive feed horn and polarizer. The size of the modules is comparable to the diameter of a feed horn providing an 8º beam, nearly ideal for cross-polar performance. The power dissipated in each

40 GHz module is 20mW, and in each 90 GHz module it is 30mW. For *terrestrial* operation, mechanical coolers with a 20K stage and 10W of capacity can easily cool hundreds of elements, and thousand element arrays are feasible using multiple coolers. Cooling requirements for a 1000 detector *space* mission pose a significant challenge

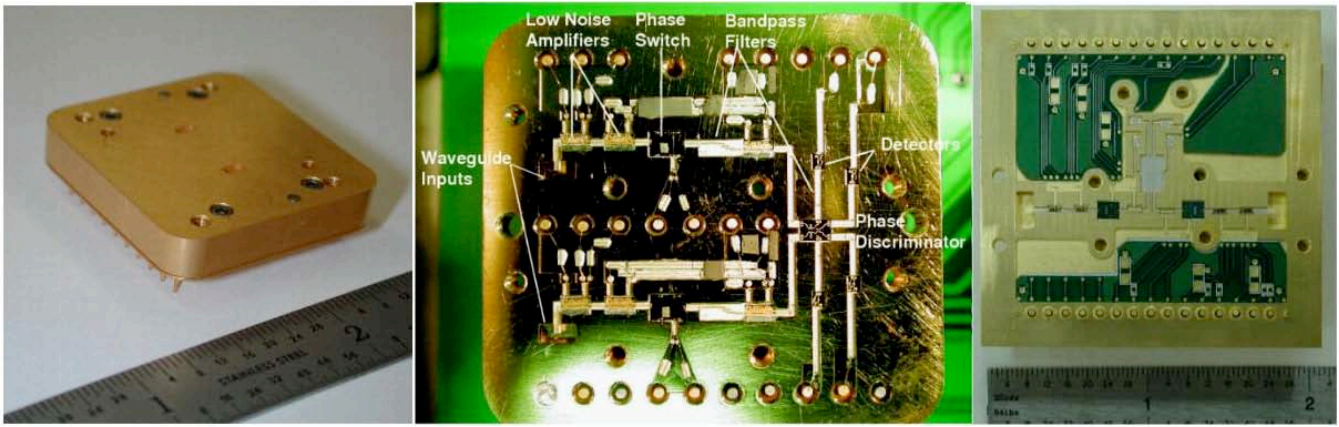

*Figure 7.10: Left: Photograph of a prototype HEMT polarimeter with cover on and input waveguides shown. Center: The complete 90 GHz Q/U module shown with cover off. Right: The 40 GHz Q/U module shown with lid off. Photographs courtesy of Todd Gaier.*

Development of new material systems (such as Sb based devices) in the coming decade may further improve the noise performance and decrease the power dissipation of HEMT devices. With an InAs channel, devices have much greater electron mobility. The theoretical improvements possible with Sb HEMTs are a factor of 2 in noise and a reduction of 2 in power consumption. DARPA has made a substantial investment in room temperature Sb devices.

### 7.4 Interferometric Approaches

An alternate approach to the imaging systems discussed so far is interferometry. Recently two interferometers, DASI and CBI, have detected *E*-mode polarization of the CMB. DASI and CBI use spatial interferometry to measure Fourier components of the Stokes parameters. Interferometers have demonstrated desirable characteristics for high-sensitivity observations of the CMB, primarily in control of systematic effects. (1) They directly measure the power spectrum of the sky, in contrast to differential or total power measurements. Images of the sky can then be created by aperture synthesis.
(2) Interferometers are intrinsically stable since only correlated signals are detected; difficult systematic problems that are inherent in total power and differential measurements are absent in a well-designed interferometer. (3) They can be designed for continuous coverage of the CMB power spectrum with the angular spectral resolution determined by the number of fields imaged. But despite the recent successes of these interferometers for measuring CMB polarization no similar instruments are under development for future polarization observations. The reason for this shift in approach is that current interferometers are limited to just a few receivers (13 for DASI and CBI) and hence have relatively poor sensitivity compared to large imaging arrays of detectors. Furthermore, the most sensitive detectors for measurements across much of the 30 to 300 GHz range relevant to the CMB are cooled bolometers. Bolometers are incoherent detectors, which are not normally used in radio interferometers, which use coherent receivers. (HEMT amplifiers in the case of DASI and CBI.)

Schemes are under development to combine the advantages of interferometry with the sensitivity advantages of large arrays of bolometers. These schemes use adding interferometry as opposed to the multiplying interferometry used for coherent systems so far. Each of the N antennas in the array accepts one electromagnetic mode. The sensitivity is comparable to imaging arrays that accept N modes. To achieve the necessary level of sensitivity, these interferometric arrays will require a comparable number of bolometers as an imaging system (~1000).

One possible scheme appears in figure 7.11. The interferometer is a 2-D array of corrugated antennas, each coupled to an ortho-mode transducer (OMT). The outputs of the OMTs are each phase-modulated at unique low (10Hz – 1kHz) frequencies. The signal amplitudes are added together by a beam combiner (such as a Butler combiner). The *E*-field amplitude from a given antenna is divided by a factor of $\sqrt{(2N)}$, and 2N of these amplitudes, two polarizations from each antenna, appear at the 2N beam combiner outputs. The outputs of the combiner couple to low-temperature bolometers, which act as "square-law" devices, squaring the sum of these amplitudes. The detector signal is then the sum of products of all possible pair-wise combinations of electric field amplitudes. There are $(2N)^2$ such terms for each detector. 2N of these are "self-products" where the amplitudes come from a single antenna in the array. The remaining $2N(2N-1)$ terms are "cross-terms" where the amplitudes come from different antennas. These cross terms represent the pair-wise interference between different antennas. Each detector produces a superposition of signals caused by the beating of these various modulation terms. By phase-sensitive detection at the beat frequencies from these modulations it is possible to isolate the visibility from each baseline.

The beam combination occurs in a guided-wave structure such as waveguide or microstrip transmission lines. Because there are no amplifiers in this system, the losses between the horn inputs and the detectors must be small and requires the use of low-loss waveguide and/or superconducting transmission line circuits.

Many of the necessary technical developments for this type of interferometry are similar to those for imaging arrays of bolometers. Common elements include antenna-coupled bolometers and microstrip modulators and beam combiners.

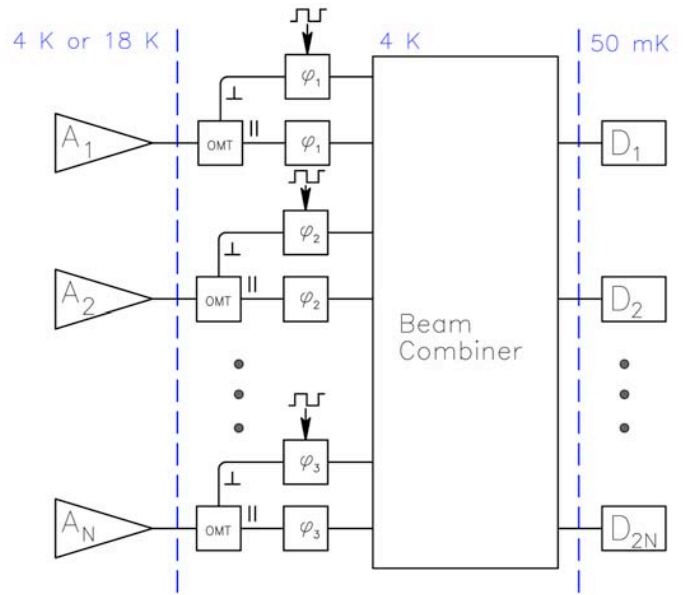

*Figure 7.11: Conceptual design of a bolometric interferometer. A close-packed array of horn antennas (A) observes the sky directly. Each horn output couples to an orthomode transducer. Each OMT output is phase-modulated at a unique frequency and followed by a Butler beam combiner. The detectors are cooled bolometers.*

### 7.5 Sub-Kelvin Coolers

The sensitivity of bolometric detectors depends on the optical background loading on the detector, on the temperature of the bolometer, and on the thermal conductance, G, between the refrigerator cold stage and the bolometer's absorber. For bolometers optimized for observations of the CMB from a cooled telescope in space, the optimum cold stage temperature is well below 300 mK. Reaching this temperature range starting from a liquid 4He bath or a cryocooler requires either dilution refrigeration or adiabatic demagnetization refrigeration. The temperature of the cold stage must be stable to about 10 nK/$\sqrt{(Hz)}$.

The High Frequency instrument on Planck uses a single-shot dilution refrigerator designed to cool a ~2 kg focal plane to 100 mK for 30 months when operated from a 4K heat sink. The Planck cooler has been space-qualified, is now commercially available, and provides continuous cooling. Though the dynamics of the fluid flow through the system leads to significant temperature fluctuations (~100 mK), temperature stability of

~10 nK/√(Hz) has been achieved in Planck through the use of multiple stages of temperature regulation

Adiabatic demagnetization refrigerators (ADRs) provide another option for cooling to 100mK and below. ADRs have been widely used to cool bolometric detectors in demanding applications in remote ground-based sites such as Antarctica (White Dish, Python), scientific balloons (MAXIMA, MAXIPOL, MSAM2), aircraft (HAWC) and space (XQC, XRS). Recently, two-stage ADRs have been developed that can operate from ~4K, allowing them to operate from an unpumped liquid He bath or a mechanical cryocooler. Multi-stage ADRs are under development to operate from a higher starting temperature and provide continuous cooling.

The duty cycle at low temperature is determined by the entropy remaining in the salt pill after initial demagnetization. For typical ADR's (~0.1 kg of paramagnetic salt, 3T superconducting magnet) this hold time is about 24 hours with a 1 µW heat load at 100mK. The recycling takes about 20 minutes. Temperature stability of 50 nK rms has been achieved in a sounding rocket. A new development, the continuous ADR, allows continuous cooling and higher heat loads. Several paramagnetic stages and magnets are arranged in series. Heat passes from the low-temperature to higher-temperature stages in a sequential "bucket-brigade" fashion. The continuous stage always remains at the desired set point. The thermal cycles of the other stages are timed so that when one stage magnetizes (expels heat) the temperature of the following stage is lower and conducts heat away through the heat switch that connects them. The result is an ADR with 100 % duty cycle. The choice of materials, volume, and magnet for the highest-temperature stage can be optimized to operate from a reservoir above 4 K.

For both single-shot and continuous ADR's reliable cryogenic heat switches are required. Gas-gap heat switches, magnetoresistive switches, and superconducting switches, have been used successfully.

ADR's have a number of appealing features relative to the dilution refrigerator used on Planck: (1) operation is all electronic, (2) there are no gases or liquid cryogens to leak, (3) the temperature can be adjusted easily from the reservoir temp (typically 2–5 K) to ~ 0.025 K, (4) the temperature can be easily controlled to high precision.

Although ADRs have been in use for more than 50 years, only in the past 20 years has rapid progress been made toward implementation of ADR-cooled detectors in astronomical observations. Developments in several areas are required before ADRs will be suitable for cooling large arrays of detectors in an orbital environment:

1) Magnet leads: high current, low thermal conductance magnet leads are required to carry several Amperes to the superconducting magnets. These leads form the dominant heat load on the ~4K stage.

2) Temperature stability: 10 nK/√(Hz) has not yet been demonstrated

3) Magnetic shielding: stray magnetic fields from the ADR magnets can interfere with superconducting detectors (i.e. TES) and readouts (SQUIDs). Passive and active shielding methods have been developed but not yet demonstrated in a system with large detector arrays.

There is considerable overlap between the cryogenic needs of the CMB community and of the X-ray community. X-ray micro-calorimeters used for spectroscopy on suborbital (XQC) and future orbital missions (XRS, Constellation-X) use or will use ADRs to cool arrays of micro-calorimeters to about 50 mK. Both communities benefit from NASA's Advanced Cryocooler Technology Development Program, for providing the reservoir temperature to which the ADR is connected. It is for these reasons that we recommend the continued development of sub-Kelvin cryogenics for space applications.

## 8 Optics

All CMB experiments use an optical system to concentrate light from the sky onto the detectors. There is a rich variety of optical systems – interferometers, on- and off-axis reflective single-dish telescopes, and refractive telescopes – each of which gives a set of performance tradeoffs (see figures 8.1 to 8.3). The choice of detector type, e.g. coherent or direct, can strongly affect these tradeoffs.

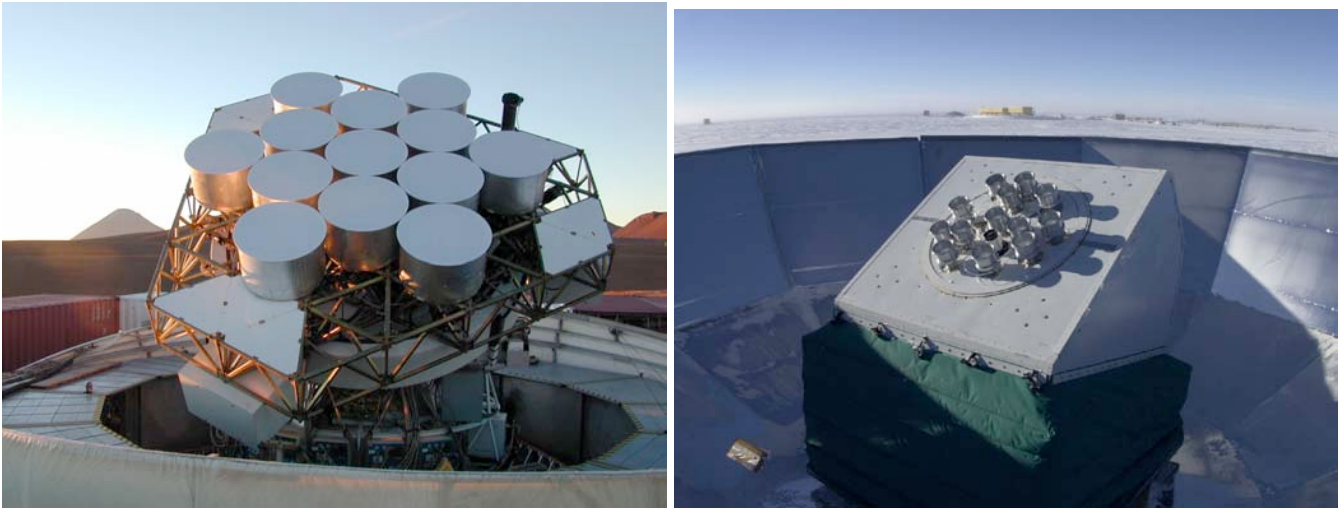

*Figure 8.1: Left: the CBI CMB interferometer in the Atacama desert in Chile. Right: the DASI CMB interferometer at the South Pole. An interferometer forms effective beams by correlating signals between the many antennas.*

Although measurements of temperature anisotropy have given us a broad base of mature optical systems, polarization measurements will make different and more stringent demands on optical systems. To minimize systematic uncertainties, the optical system has to minimize the generation of polarized light from non-polarized light, and the rotation of polarization. CMB polarization signals are formed by differencing two linearly polarized intensities, and differences in the size or shape of the two optical beams will couple to spatial intensity variations to produce false polarization signals. Given the very small polarization signals, optical systems must have stringent rejection of emission away from the main optical beam. For single dish telescopes, large focal plane arrays are required for sensitivity. Telescope designs must balance the requirements of large field-of-view and polarization purity. Both lenses and mirrors can be used in imaging systems. Mirror technology is mature, but low-loss cryogenic lenses with broadband antireflection coatings need to be developed.

Most experiments use a horn or planar antenna to couple a free space wave to a guided wave, which is then coupled to a detector. Horn antenna theory is mature and they have excellent properties. A space mission that employs feeds as optical elements is possible. Planar antennas are under development. Planar antennas can have multi-band response and are an important ingredient for monolithic imaging arrays. Filled bolometer arrays couple free space waves directly to absorbers, but they require quasi-optical components to select frequency bands and polarization

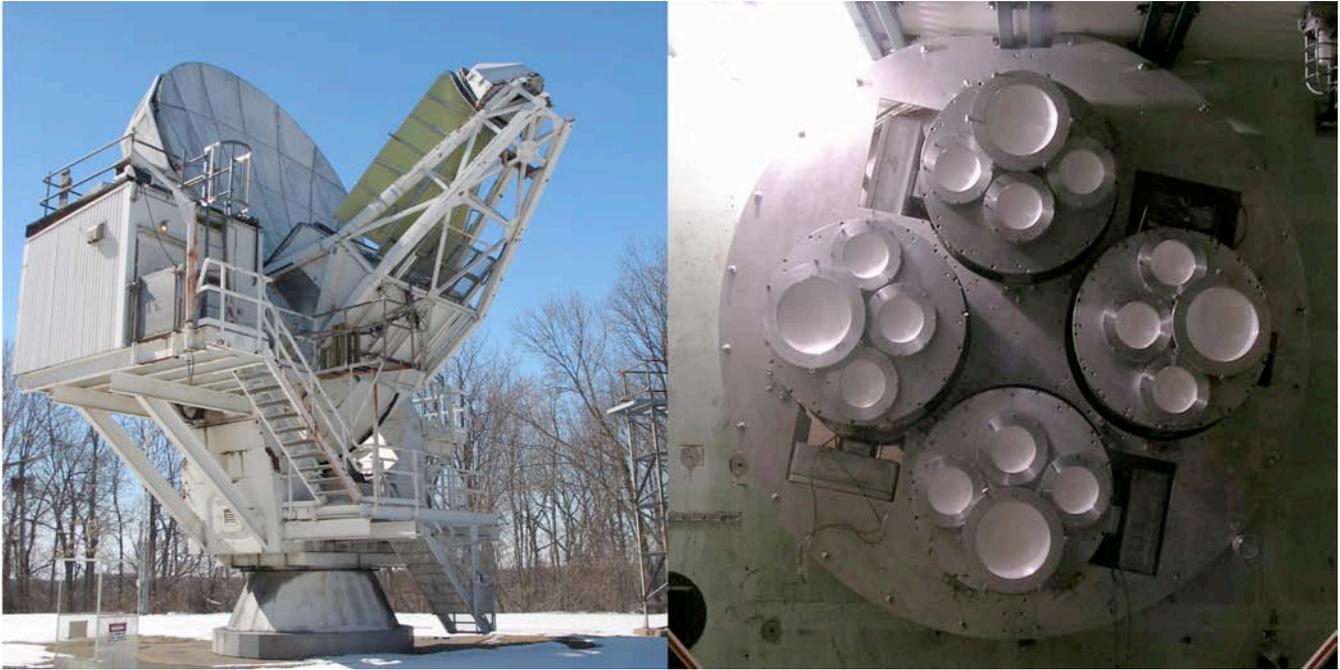

*Figure 8.2: Left: Lucent 7m telescope currently used by the CAPMAP experiment measuring CMB polarization at four arc minute scales. Lucent Laboratories has offered the telescope for continued CMB research. Right: CAPMAP focal plane with its 16 radiometers in 4 cryostats. The useful part of the focal plane is more than 48" in diameter.*

## 8.1 Telescope Design

Telescopes used for observing the CMB are of two basic types, single-aperture imagers and interferometers (see figures 8.1 to 8.3). Single-aperture telescopes scan or chop across the sky to map an extended region. Interferometers, in contrast, track one patch of sky and mosaic several patches to observe an extended region of sky. Interferometers measure in the Fourier domain, and therefore the data are closer to the power spectra that are ultimately desired. Interferometers are not susceptible to some of the systematic uncertainties of single-aperture imagers, but the reverse statement is also true. As CMB experiments move toward instruments with 1000 or more detectors, it becomes technically challenging to build a coherent interferometer due to the large number of receivers and correlations required. Bolometric interferometers are also being developed, but they have analogous technical challenges, as discussed in §6.7.3.

Single-aperture imagers can be further sub-divided into those that use a reflector (mirror) or refractor (lens) for the first elements in the telescope. In the optical regime, reflectors have become the norm, but with CMB measurements a refractive telescope allows a large field-of-view and low generated polarization simultaneously (polarization effects will be defined and discussed below). Large refractive telescopes are technically challenging – large low-loss lenses are difficult to build and the lenses have to be cooled to reduce their emission.

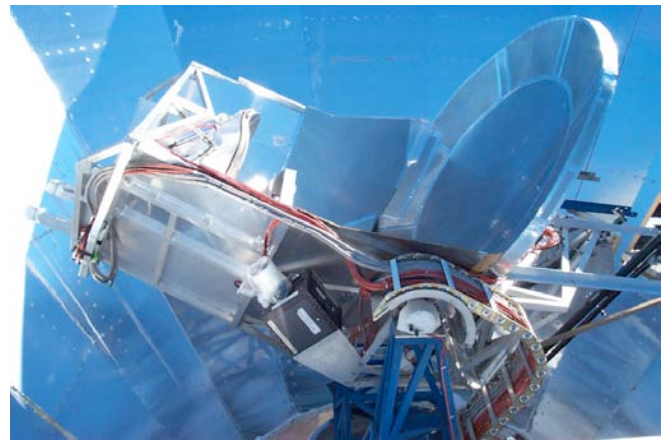

*Figure 8.3: A photograph of the Viper off-axis reflector telescope at the South Pole. Light from the sky is not blocked by the secondary mirror before reflecting from the primary mirror.*

Construction methods for millimeter-wave mirrors are mature. Mirrors up to two meters in diameter are generally machined directly from solid aluminum (see figure 8.3 for an example). Larger dishes are made from machined aluminum panels (see figure 8.2 for an example). For low weight and low temperature expansion coefficients, carbon-fiber mirrors are used.

Reflective telescopes often use refractive reimaging optics, which are usually cooled. In contrast to a large refractive primary aperture, the relatively small reimaging lenses are easier to fabricate. Lens materials with low loss at millimeter wavelengths and suitable broadband antireflection coatings do not exist and need to be developed. Silicon and some forms of plastic are candidate materials for low-loss lenses. A broadband antireflection coating has an index of refraction that gradually changes. This gradient in index could be achieved by using a sandwich of layers with varying index or by fabricating cone or pyramid shapes into the surface.

Many CMB receivers use quasi-optical filters to reject infrared light and to define frequency bands. The theory and technology for these filters is fairly mature, but building filters of the appropriate size for large arrays remains a challenge (see figure 8.4).

Sidelobes refers to the off-axis response of an optical system. Low sidelobe response is especially important for CMB polarization measurements since the Galaxy, the Earth, and nearby structures will have much stronger contrast than that of the CMB polarization. Sidelobes are controlled in two ways. First, a goal of the telescope design is to have low sidelobe response. As mentioned, off-axis telescopes have no blockage of the primary from the secondary, and can have significantly lower sidelobe response than an on-axis telescope. Second, shielding can redirect the sidelobes closer to the main beam(s). It is desirable to have the shields move with the telescope to avoid systematic errors arising from variations in emission and reflection of the shield itself (see figure 8.5).

The design of space-borne CMB telescopes has much in common with that of ground- and balloon-based telescopes, but there are unique advantages and challenges for space instruments. An advantage for space-borne telescopes is that the optical elements and shields can be cooled (passively or actively) which can simplify the telescope design and improve its performance.

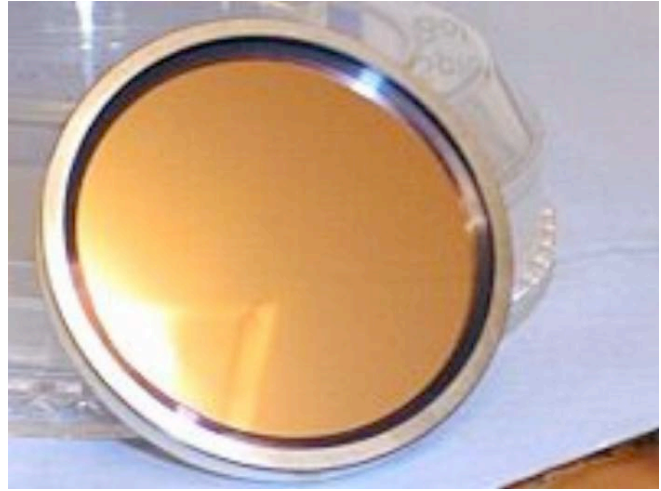

*Figure 8.4: Photograph of a quasi-optical metal-mesh low-pass filter (courtesy of Peter Ade).*

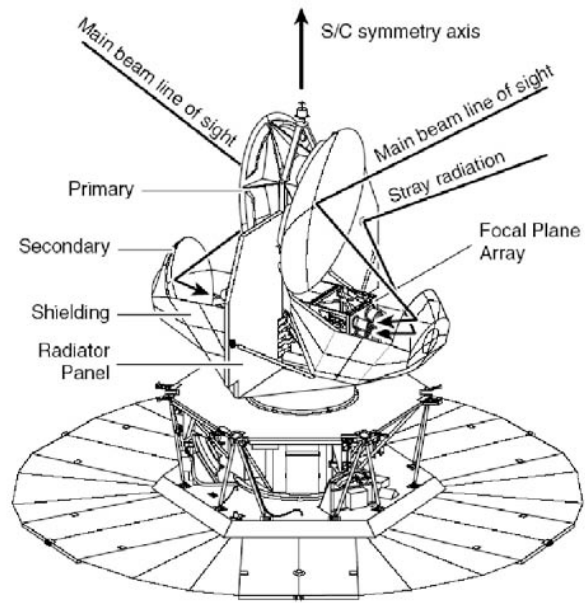

*Figure 8.5: Line drawing of the WMAP space-borne CMB telescope. The two back-to-back telescopes are off-axis 1.3 meter diameter designs. Note the optical shielding of the receiver and secondary, which is done to reduce sidelobe response.*

The weight and size of space-borne telescopes, however, are strongly constrained by launch vehicle requirements.

### 8.2 Coupling to the telescope

The interface between the optical system and the detector system can be done in a number of ways. Single-dish telescopes have a focal plane that can have an array of horn (see figure 8.6) or planar antennas, or the focal plane can be filled directly with the absorbers of bolometers. The antennas limit the solid angle that can be "seen" by the detectors to that subtended by the telescope optics, which is necessary since nearby objects such as the cryostat and ground are hotter than the CMB. For a filled array, a cold aperture stop (also called a Lyot stop) is required to limit the solid angle seen by the detectors. Interferometers currently use horn antennas, but planar antennas would also be suitable.

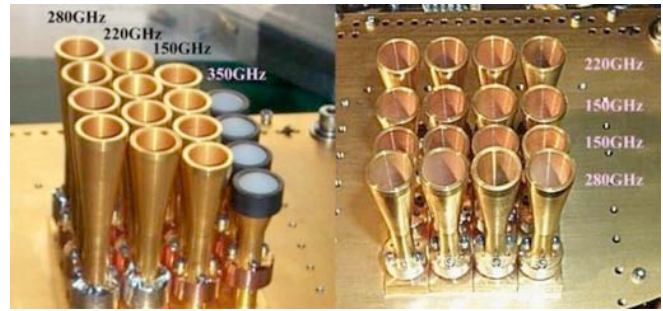

*Figure 8.6: Photograph of a focal-plane array of scalar horn antennas from the ACBAR CMB experiment at the South Pole.*

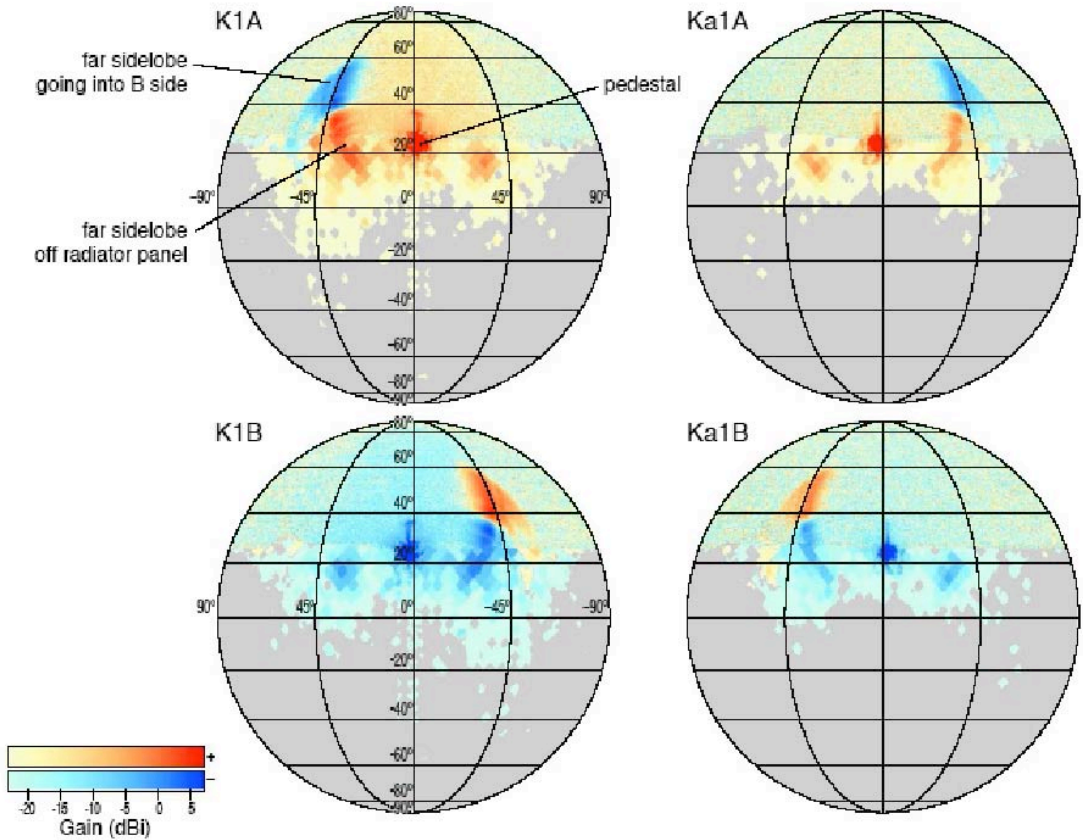

*Figure 8.7: Maps of the sidelobe response of the WMAP space-borne CMB telescope. The measured patterns agree broadly with simulations and ground-based tests.*

### 8.3 Beam and Polarization Purity

The optical system of a CMB telescope can polarize an unpolarized flux (instrumental polarization), rotate polarization (cross polarization), and convert a polarized flux to an unpolarized flux (depolarization). The systematic uncertainties arising from these non-idealities are complex, and are deeply linked to the scan strategy used. Systematic uncertainties due to non-idealities in the optics are discussed in detail in the section on systematic uncertainties (§6). In this section, we discuss the origin of the optical effects.

### 8.3.1 Instrumental polarization

Reflection from a metal sheet at non-normal incidence will polarize an unpolarized flux up to roughly 1%. In a reflecting telescope, the net amount of instrumental polarization induced is given by the degree of symmetry in the optics. The center pixel of an on-axis telescope can have zero instrumental polarization. The instrumental polarization will increase for pixels away from the center of the focal plane. Off-axis telescopes have a relatively strong net instrumental polarization (a fraction of 1%). Off-axis telescopes have two compensating virtues. Firstly, the primary mirror is not blocked by the secondary mirror, and therefore off-axis telescopes have the potential for lower sidelobe response than on-axis telescopes. Secondly, off-axis telescopes have, in general, a larger field-of-view which allows larger focal plane arrays. A large field-of-view for an on-axis telescope requires a large secondary mirror with significant obscuration and associated scattering into sidelobes.

Instrumental polarization can also be induced by lenses. Non-ideal antireflection coatings produce instrumental polarization. Again, symmetry is an important factor. Pixels far from the center of the focal plane will have ray bundles that are not symmetric through all lenses. Broad-band antireflection coatings are not mature and require development.

For an off-axis telescope, the instrumental polarization is largely constant across the beam on the sky. This type of pattern can be referred to as a "monopole" instrumental polarization. Higher order patterns of instrumental polarization are caused by the appropriate asymmetries in the optical system. As mentioned in §6, differencing two beams with different beam shapes gives an instrumental polarization with a polarized beam shape given by the difference of the two intensity beam shapes. Aberrations in an optical system can lead to asymmetries in the beams. The goals of low geometric aberrations and large field-of-view directly conflict, and this presents a challenging optimization problem for CMB telescope design.

### 8.3.2 Cross polarization

Cross-polarization is a rotation of a linear polarization. It is naturally generated by curvature in optical elements and it is related to field distortion. Telescopes can be designed to minimize cross-polar response. The off-axis telescope design of Dragone, for example, has low cross-polar response across the focal plane.

Horn antennas produce a small level of cross polarization (< –40 dB) that has a four-leaf clover "quadrupolar" pattern of alternating positive and negative rotations.

# 9 System Issues

A system issue is any instrument characteristic that emerges after the entire receiver is assembled. While some examples of system issues fall into the category of technical problems that must be diagnosed and fixed for the instrument to be scientifically viable, they can also induce subtle characteristics into the science data during observations. For large CMB experiments, particularly space-borne experiments, system technical problems strongly drive cost risk, because they only present themselves at a high level of integration, at the peak of operational costs. However, even instrument signatures that do not prevent scientific observations can consume enormous resources at the data analysis stage. The issue of handling instrument signatures will be increasingly important as CMB instruments progress from receivers with several detectors, which often have their own idiosyncratic behavior, to large focal planes with thousands of detectors. Any credible development plan for a complex orbital receiver must therefore concentrate on minimizing systems issues by 1) designing defenses for, or entirely avoiding, known systems problems at the initial design phase, e.g. by keeping the receiver as simple and robust as possible; 2) identifying, allocating systems budgets, and testing systems issues at sub-systems level; 3) thoroughly managing interfaces between sub-systems; and 4) increasing the functionality of the sub-systems, effectively moving systems risk to an earlier level of integration. Most importantly however, *systems risk management can only be informed by direct experience*. A complete working knowledge of systems issues is the only true defense against systems risk. This prior knowledge can only be obtained from the end-to-end experience derived from smaller-scale sub-orbital and ground-based pathfinder instruments.

## 9.1 Systems Issues for CMB Receivers

Because a complete taxonomy of receiver systems issues is beyond the scope of this document, we give some general categories of problems, with concrete examples of successful mitigation, while noting that the most important systematic effect may always be the one that escapes the notice of the instrument team. The descriptions given below concentrate on the characteristics of a space-borne experiment, where systems management will be paramount.

### 9.1.1 The detector readout chain

High-sensitivity millimeter-wave focal plane detectors tend to be sensitive to more than just CMB photons. The signal chain from the detector to cold preamplifier electronics to warm readout electronics can be generically sensitive to extraneous sources of interference, such as magnetic fields, conducted and radiated electromagnetic interference, electrical cross-talk between channels, 1/f noise, and even micro-vibrations. Thus while the detector sensitivities required for CMB polarimetry may be demonstrated in a laboratory environment, past experience has shown that many years of hands-on testing in real instruments is necessary before a detector technology is mature enough for a satellite experiment.

The solutions for controlling environmental interference are born of experience, but must be designed into an experiment from the beginning. An example, shown in figure 9.1, is the 4 K Faraday cage for the BICEP receiver, developed for electromagnetic compatibility (EMC) and to minimize susceptibility to electromagnetic interference (EMI). The necessity of such a design, which required developing several new technologies (low-dissipation JFET modules and cryogenic filtered micro-D connectors), only became evident after years of field experience with EMI/EMC problems. Note how difficult it would be to implement such a shield into a detector system retroactively.

The detector readout chain is a complex exercise in systems engineering. The detectors are electrically connected to readout electronics by meters of cable that wind through the cold instrument and spacecraft to the warm readout electronics. On a space project, the individual components of the readout chain must be developed in parallel, which in turn requires defining the interfaces between the systems. The instrument team can only make informed decisions about allocating budgets for noise, capacitance, inductance, shielding, and thermal dissipation based on prior experience. Higher levels of sub-system integration can help defeat these issues. For example, a detector with a high level of functionality (see figure 9.2) is more complex but eliminates interfaces and minimizes systems risk within the instrument. Further integration of multiplexing and pre-amplification on the focal plane itself is now possible (see §7) with superconducting detectors and readouts.

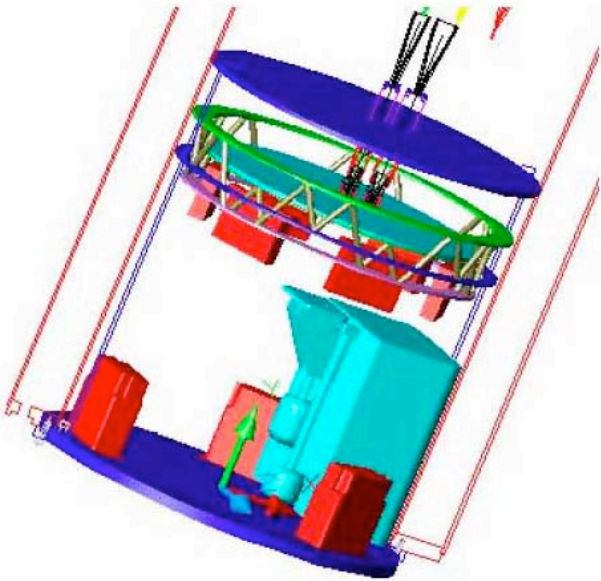

*Figure 9.1: Layout of the BICEP focal plane receiver. The entire 250 mK focal plane, $^3$He cooler, detectors, filters, and low-power JFETs are surrounded by a 4 K Faraday shield, shown in blue. The only penetrations into the Faraday cage are the waveguide holes in the back-to-back entrance feeds, which cannot propagate RF emission due to the waveguide cutoff, and the RF-filtered electrical connectors on the outputs of the 3 JFET preamplifier units (red units at the bottom).*

### 9.1.2 Stray light, filtering, and instrumental emission

Extraordinarily sensitive millimeter-wave detectors not only detect the 2.7 K CMB sky with high signal-to-noise, they can also detect photon emission from the instrument itself. Especially for a space-borne experiment, thermal emission from even a 2K instrument can be significant, as shown in figure 9.3, resulting in reduced sensitivity due to photon noise. The filters used to define the spectral passband of a receiver may also emit, diffract, reflect, scatter and/or polarize stray radiation, leading to excess optical power and spurious optical response. Although numerous instruments have surmounted these issues, surprises still arise at systems level. Unfortunately, even with a small sub-orbital/ground-based receiver, diagnosing these optical systems problems is a time-consuming process involving multiple cool-downs of the instrument.

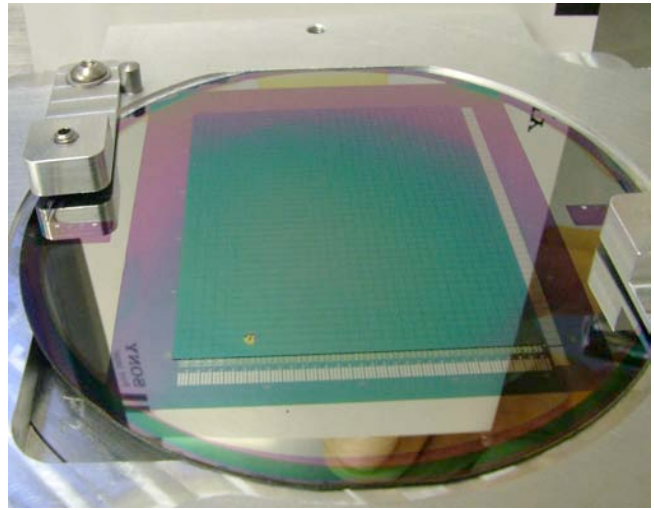

*Figure 9.2: Systems for large format arrays have already been developed for slightly higher frequencies than will be needed for CMBPOL. This picture shows a monolithic array of 32 by 40 TES detectors that are bump-bonded to a similarly sized array of multiplexers. The whole assembly is cooled to 100 mK. It operates at 350 GHz. Such designs are approaching the functionality of a CCD camera, but at millimeter wavelengths. This array was produced at NIST, the Scottish Microelectronics Centre, and Raytheon, and is part of the SCUBA2 camera. First light on the James Clerk Maxwell Telescope is scheduled for 2006. Figure courtesy of NIST.*

Integrated antenna-coupled detectors (see §7) incorporate the functions of band filtering, beam collimation, polarization analysis, and even polarization modulation right at the focal plane. While these functions increase focal plane complexity, they reduce systems risk at a higher level of integration.

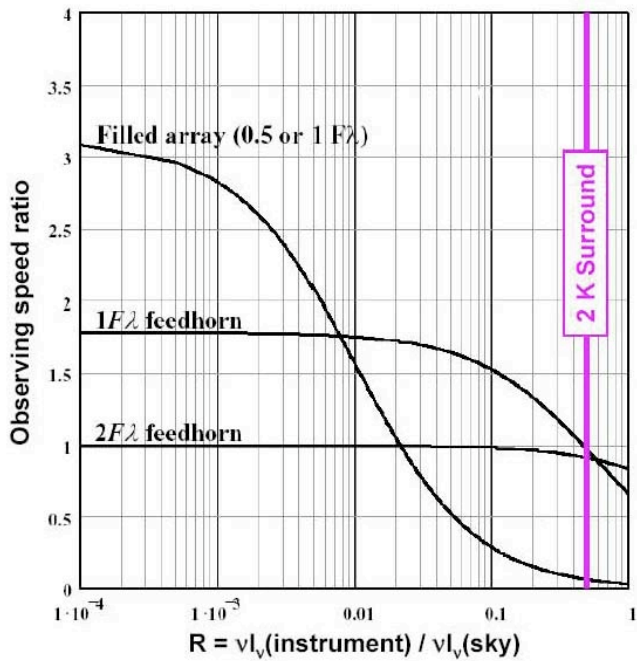

*Figure 9.3: Relative observing speed for background-limited array architectures when viewing both the sky and partial emission from the instrument surround and optics. Instrument brightness from a 2K environment is comparable to the 2.7K CMB sky (about half of the surface brightness at 100GHz) as shown by the purple vertical line in the plot. Photon noise from the instrument decreases the observing speed; therefore, more collimated detector beams (2fλ and 1fλ feedhorns) suffer less degradation than un-collimated bare arrays. Stringent control of the detector field of view is essential for space-borne CMB observations.*

### 9.1.3 The cryogenic chain

A CMB receiver requires a cryogenic focal plane and cooled optics, provided through a chain of cooling technologies. Because the focal plane detectors are generically sensitive to focal plane temperature fluctuations, and thermal fluctuations in the optics that emit detectable thermal photons in the mm-wave band, the driving requirement in systems thermal engineering is stability (see the table in the systematics section, §6).

The Spitzer mission provides a useful working example of the cooling techniques needed in a space-borne experiment – and the associated systems challenges. Spitzer took maximum advantage of passive cooling, first by moving from a low-earth orbit, where the instrument sees both the sun and 2π steradians of 300 K Earth, to a heliocentric orbit. In a heliocentric orbit, the warm portions of the spacecraft, the solar panels, electronics, and telemetry systems, all are located facing the sun, while the receiver section views a large solid angle of cold space. This arrangement allows passive cooling to well below 50 K, greatly reducing the thermal parasitic load into the next cooling stage. Furthermore with a scanning all-sky experiment at L2, such as WMAP or Planck, the radiation input from the sun is nearly constant and thus inherently stable. By launching the telescope warm, and later cooling the optics with the helium vent gas, Spitzer minimizes the cryogenic mass, which must be thermally isolated by a mechanical support system. A warm launch simplifies the mechanical design, since all of the components that must withstand launch vibration are designed and tested at room temperature.

While these techniques greatly reduce launch mass and cost, they also present a systems level challenge: any integrated test of the receiver, telescope, and cryogenics involves operating the entire system in a large cryogenic chamber in order to simulate the passive cooling environment from space. Not only is such a test enormously expensive itself, it occurs at such a high level of integration that fixing any problem in the receiver is very expensive. The practical solution is to avoid full systems testing for measuring performance, and to completely characterize the focal plane, the optics, and the cryogenic chain at sub-system level. This strategy requires a through understanding and designing all the issues involved in integration.

After passive cooling, the next stage in the cooling chain cools the optics directly in front of the detectors, and provides a heat sink for the sub-Kelvin focal-plane cooler. Candidate sub-Kelvin technologies such as dilution refrigerators, ADRs, and $^3$He sorption coolers, can all be designed to operate from a 4 K heat sink. The intermediate 4 K cooler could either be an active cooler or a liquid-helium cryostat. Finally, the lowest temperature cooling stage must provide a stable and continuous operating temperature of 100 – 200 mK. All of the coolers must meet cooling heat lift and stability requirements individually, but the entire system must be designed not only for the allocation of heat loads between the interfaces, but to control any interactions between the systems, including operation in zero gravity.

### 9.1.4 The optical chain

CMB polarization receivers must have unprecedented levels of sidelobe and polarization control. For example, the sidelobe characteristics of a future space-borne experiment must be significantly better than that of Planck, due to polarized off-axis response to the bright Galactic plane. The optical response is not only a function of the telescope, but also the properties of the fore-optics and focal-plane illumination pattern. Unfortunately, an end-to-end characterization of the cooled telescope and cryogenic receiver is a practical impossibility in a satellite experiment. Therefore a program of component-level testing and full electromagnetic analysis of the optical chain is required. Experience with a full system employing the working components (e.g. antenna-coupled detectors, lenses) in understanding such real-world issues is invaluable.

### 9.2 Systems Issues for CMB Data Analysis

The program called for in this report will lead to experiments with more detectors, longer integration times, and smaller beams, and thus larger volumes of data. It is quite likely that future experiments will generate of order 7 to 10 Tb of data per year. While numbers of this magnitude are not startling in the context of modern science, there are several aspects of CMB data processing that are relatively unique to the field.

From the standpoint of computing requirements, the most important is that the *entire* raw data set must be processed and reduced to obtain measurements of the cosmological parameters, including the gravitational wave amplitude. This processing inevitably requires iterative passes through the data and the generation of several copies of the data with different instrumental effects partially or fully corrected as part of the systematic error analysis. Many of the algorithms used in the analysis involve linear algebra computations in which the operation counts scale like $N^\alpha$, where $N$ is a fundamental data set size and $\alpha$ is typically between 1.5 and 3. These factors place a great burden on computer systems: for CPU cycles, for disk storage space, and for the communications fabric that connects them. Moreover, the algorithms needed for processing and analyzing CMB polarization data are not as mature as the ones used for temperature analysis. Polarization codes will need to be developed and refined in step with the experiments they serve.

### 9.2.1 Algorithm Development

This work must focus on a range of issues from experiment-specific systematic error analyses to the more general problems of polarization map-making, power spectrum estimation and cosmological parameter fitting. The most effective way to develop these codes is by running them on high-performance platforms using real data (or high-fidelity simulations of real data). To cultivate this, we call on experimental groups to make their data available to the larger community (for example, via NASA's CMB thematic data center, LAMBDA) and we encourage the support of groups who make CMB data analysis codes available to the community, such as CMBFAST, HEALPix, and MADCAP, to name a few.

There are a number of well-defined tasks in need of further development that will be a part of any CMB experiment's data processing pipeline. These can be roughly categorized by the level of data reduction each stage entails:

1) Managing and configuring the raw telemetry from the experiment. For a satellite mission, this may require some degree of on-board science data processing (e.g. data compression) to keep downlink data rates to manageable levels. This is potentially a serious concern for a mission if downlink times need to be minimized to avoid data contamination. The demands of this "level-0" processing on the ground are relatively modest. Mainly, one requires adequate disk and back-up facilities as well as sound data management policies to protect data from loss and misuse. There will be some effort required to develop data compression and semi-automated data flagging algorithms to efficiently handle the large volumes of data envisioned.

2) Level-1 data analysis entails determining the instrument calibration, characterizing the noise properties of the instrument, and searching for classes of systematic errors that arise in the time domain. Since one is generally breaking new ground with new instruments, this can be the most difficult stage of the processing to define in advance. Sufficient flexibility should be designed in to the data processing pipeline to accommodate lessons learned about the data "in flight".

3) Level-2 products are typically maps of the sky generated from the calibrated time-series data. This step usually entails solving a large linear system for each channel of data and is thus one of the more computationally demanding steps of the processing. In addition, since sky maps are one of the most important legacy products of a CMB experiment, the pixel-to-pixel covariance matrix must also be determined with sufficient fidelity to enable power spectrum estimation from the maps. As a community, we have relatively little experience in generating polarization maps from real data, so much of the development that needs to occur in this area relates to understanding how real experimental properties (noise spectrum, beam response, scan strategy, etc.) impact the fidelity of the final polarization maps. The best way to gain this experience is to work with real data. It would be premature to write a general-purpose map-making code that works with any experiment and runs on any platform. However, developers should strive to produce code that employs general-purpose modules that can easily be shared.

4) Level-3 data are analyzed scientific products such as models of the foreground emission from multi-frequency data, estimates of the CMB angular power spectra (temperature and polarization), and estimates of cosmological parameters from the power spectra. Again, the generation of these products entails solving large linear systems to reduce the multi-channel sky maps (with correlated noise) down to a small number of spectral components and/or cosmological parameters. If an experiment is able to produce sky maps with well-defined pixel-pixel error matrices, the level-3 processing can be relatively insensitive to further details of the experiment. Thus it is often feasible to produce general purpose code modules for level-3 data analysis that are shared via the web. We encourage the support of groups who make such codes available to the community.

5) Simulation products. The ability to generate simulated data at all levels of reduction has been a crucial factor in the success of previous CMB experiments. This applies equally to the planning and analysis stages of an experiment. There should continue to be a vigorous effort to develop instrument simulator tools that can be used to generate mock science data.

The computational efficiency of a code depends on whether the platform can deliver the data to the processors as quickly as they can process it. Different algorithms used in CMB analysis can differ in actual efficiency by more than an order of magnitude, depending on the type of algorithm, the data set size, the platform architecture, and so forth. As data sets grow in size and are distributed over more processors, greater stress is put on the communications network, which further reduces efficiency. A focused computational science research program that targets the implementation of CMB analysis codes on various parallel platform architectures will squeeze the most efficiency out of the data processing pipeline.

**9.2.2 Computational Resources**

The CMB community currently uses several million processor hours of computer time each year, spread over everything from small dedicated clusters to the biggest national flagship super-computers. There is no doubt that we will need these resources and more to support the next generation of experiments, and a plan to meet these needs should be a part of any roadmap towards CMB polarization. Given the iterative nature of CMB data analysis we cannot afford very long wall-clock run times for any single processing pass through the data. A reasonable rule of thumb would be that a single analysis takes no more than a few days, during which time a single 1 GHz processor can perform $O(10^{14})$ operations. For CMB polarization data, we will need to use hundreds or thousands of processors to analyze the data in a timely manner.

Parallel computing resources currently fall into two broad classes: "off-the-shelf" clusters of tens to a few hundred processors, and super-computers of hundreds to tens of thousands of processors built with specialized hardware. The former are inexpensive enough to be purchased at the institution or mission level, and so can provide a dedicated computing resource, but are smaller and may give a lower fraction of peak performance. The latter are expensive enough to be restricted to a small number of national facilities, and so are inevitably shared resources with queuing restrictions, resource budgets, and annual allocation processes, but they provide much greater computing capability, significant economies of scale, and a potentially higher fraction of peak performance.

Both classes of parallel computing are currently being used for CMB data analysis. Of the former, WMAP uses a cluster of 6 32-processor SGI Origin 300 machines at NASA's Goddard Space Flight Center, while Planck's plans include two 256-processor clusters at its Data Processing Centers in Paris and Trieste. These systems provide on-demand resources to the projects and they simplify the procedures required to ensure data security since access to these systems is limited. Of the latter, more than a dozen CMB experiments currently use the 6,600-processor IBM SP3 at the DoE's National Energy Research Scientific Computing Center (NERSC) at Berkeley Lab. In 2004, NERSC estimates that more than 1,000,000 processor-hours will be devoted to CMB data analysis. The NSF's National Center for Supercomputing Applications at the University of Illinois at Urbana-Champaign provides some time on its 2,500 processor PowerEdge/Xeon cluster to local researchers, while the 10,000-processor Project Columbia at NASA's Ames Research Center will soon be another resource.

National supercomputing facilities provide exceptional resources, with both the capacity and the capability to support large collaborations, along with high-quality administration, user support and long-term system upgrade plans. However, for an experiment to rely on such resources there must be guarantees both of their general availability over the lifetime of the experiment, and of their specific, timely, availability once the data start arriving. We encourage experimental groups to avail themselves of this resource when it makes sense for them, and we encourage the agencies that support these centers to give CMB data analysis a high priority when making time allocations.

## 10 Roadmap for CMB Polarization Research: Timeline and Estimated Costs

The goal of the field is to measure the CMB with increasing precision and accuracy in order to constrain the physics of the earliest moments of the Universe. Great strides have already been made. Over the course of the next decade, we will continue to refine measurements of the polarization and of the fine-scale temperature anisotropies in the CMB. In this section, we focus on the roadmap for polarization research, with the understanding that observations targeting fine-scale temperature anisotropies are both fundamental to the advancement of the field, and will serve as important proving grounds for technologies and techniques relevant to the improvement of polarization measurements.

Improved CMB polarization results will significantly constrain properties of cosmological models and will also reveal the optimal way to extract information from the CMB in the presence of foreground emission. They will also make clear the best experimental techniques with which to make definitive and lasting measurements of the CMB polarization. The ultimate goal is to measure the $B$-mode polarization to the limits set by our ability to model and correct for astrophysical foreground emission, which is expected to require a dedicated satellite mission. *A phased program leading to the launch of such a satellite is the major recommendation of this Task Force.* A roadmap for that phased program is laid out in the remainder of this section.

### 10.1 Primary Goal of the Roadmap: CMBPOL

In this section we present a roadmap that builds toward a space mission, which we will call CMBPOL. We are confident that in following this path, excellent science will result and technological advances with wide applicability will be made. However, we cannot yet predict where the necessary technological breakthroughs will come or the best way to measure $B$-mode polarization. The timeline is planned so that by 2011 a coherent and technically well-motivated proposal can be made for the CMBPOL mission.

The roadmap is directed toward the ultimate goal of a space-borne CMB polarization mission, and predicated on requests for proposals for CMBPOL in 2011, a mission start in 2012, and launch in 2018. The major challenges in the technical and scientific formulation of CMBPOL are listed below.

### Detector Technology

Given our current state of knowledge of contaminating polarized foregrounds, we assume that CMBPOL must cover the entire frequency range 30 – 300 GHz. A single detector technology covering this band, while not strictly essential, will greatly reduce mission cost, risk, and complexity.

There remain technical challenges to be met before either bolometers or transistors (e.g., HEMTs and MMICs) can be used for a space mission. For bolometers, the challenges fall into two categories: expanding the frequency range down to approximately 30 GHz, depending on what is learned about the foregrounds at low frequencies, and devising an unambiguous method for detecting polarization from the difference of total power measurements. To make bolometers successful, new experimental techniques may be needed to modulate polarization. For MMICs, the challenges also fall into two categories: expanding the frequency range to possibly as high as 300 GHz, again depending on the nature of the foregrounds, and finding a way to use less power. To make MMICs successful for space applications, there must be advances in both high frequency transistors and in the development of high capacity cooling systems. *The challenges facing coherent detector schemes have led us to assume bolometric detectors when modeling the CMBPOL mission in this Report (see recommendations below).*

### Polarized Foreground Emission

Knowledge of astrophysical polarized foregrounds is central to the planning of CMBPOL. We must anticipate that a space mission will ultimately be limited by foreground subtraction, even after using all of the spatial and spectral information available. Thus foregrounds set not only the requirements for frequency coverage, but also the target for system sensitivity, both in the size of the focal plane array and the mission life.

Knowledge of the polarized foregrounds must come from a dedicated program from ongoing space experiments, and ground-based and balloon-borne receivers. Of course the role of these experiments extends beyond simply understanding foregrounds. They will shed further scientific light on reionization, *E*-mode polarization, CMB lensing, and small-scale temperature anisotropies. They will also provide significantly improved upper limits, and possibly even detections, of gravitational wave *B*-mode polarization.

**Receiver Systematics**

The successful operation of complete receiver systems is a vitally important step leading to CMBPOL. Future polarimeters will operate by combining the outputs of arrays of detectors to produce a power spectrum; it is thus crucial that the entire system work with well-behaved noise properties. Techniques for measuring the polarization signal, particularly in the case of bolometric detectors, are likewise vital to a successful CMBPOL mission. Such techniques are already in development, but must be fielded and shown to work in real systems. Only with such systems can we quantitatively test the control of systematic errors, and prove out signal and scan modulation strategies specifically adapted to studies of polarization.

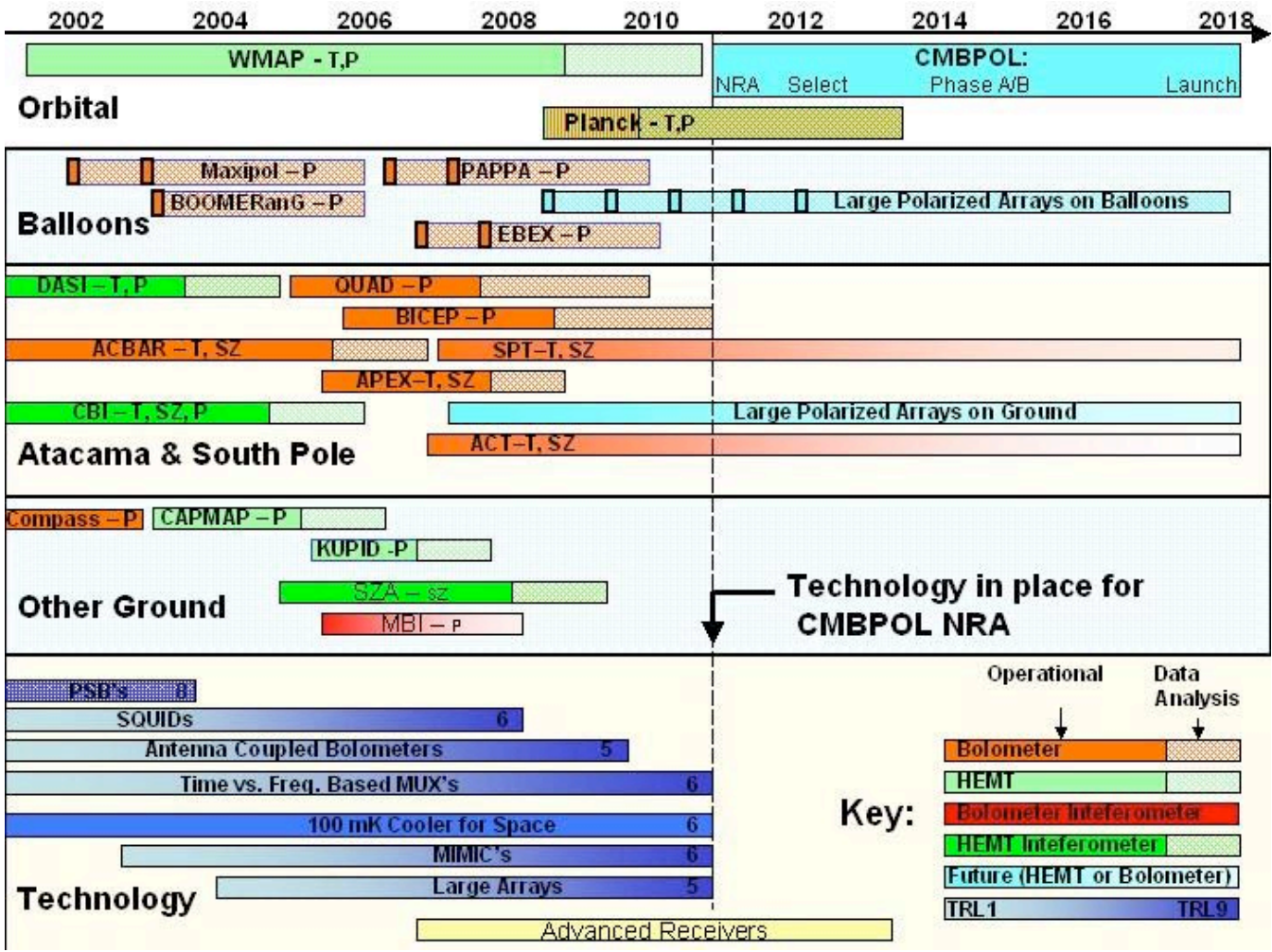

*Figure 10.1: Schematic timeline of research programs observing CMB small-scale temperature fluctuations, CMB polarization, and the Sunyaev-Zel'dovich effect. The projects included in technology development are needed for ground-based, balloon and space observations. Technology development, first at the component level, and then at the systems integration level, must be in place by 2011 for the CMBPOL mission opportunity.*
-71-

## 10.2 Key Elements of the Roadmap

The timeline of currently funded receiver development efforts, anticipated new receiver efforts, and major technology developments is shown in figure 10.1. The essential elements in the roadmap leading up to a space mission are shown in figure 10.2. The following are the key programmatic elements of the roadmap.

**Ongoing and future space missions**

WMAP is in orbit, taking polarization data in five frequency bands between 22 and 94 GHz. It is funded to observe through 2009, at which time the polarization sensitivity would reach the level shown in figure 5.1. This sensitivity would enable WMAP to obtain improved measurements of the reionization optical depth and the polarized synchrotron emission, both important in the design of CMBPOL.

Planck, with an anticipated launch in late 2007, is expected to achieve five times better polarization sensitivity than WMAP. We hope for polarization results by 2011. Though both missions have the capability to measure polarization, neither was optimized for it. Data from both missions will provide information about foreground emission essential for planning CMBPOL.

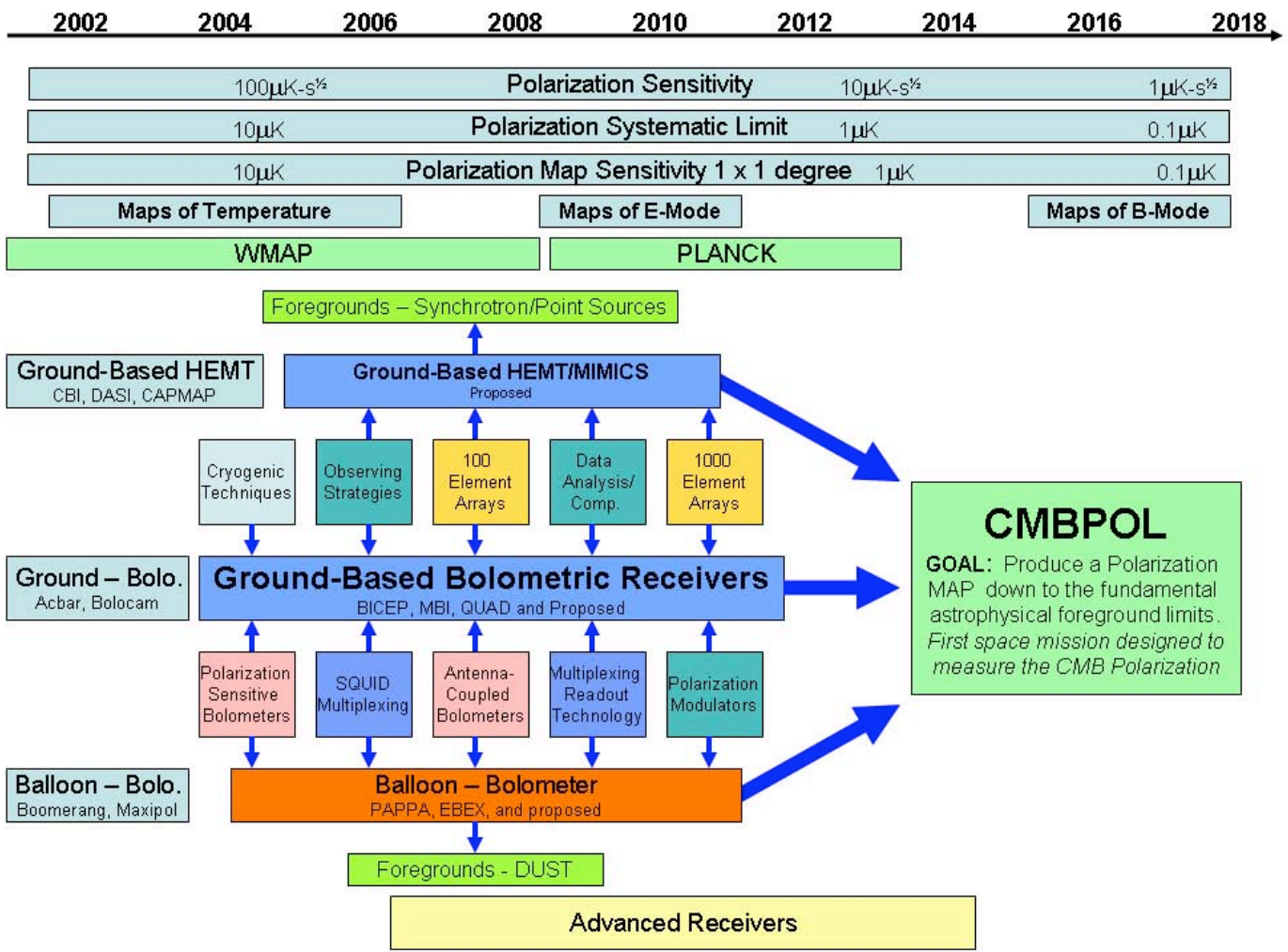

*Figure 10.2: The elements of the CMB polarization research timeline showing the range of improvements in sensitivity expected and the activities that need to be carried out to realize a space-borne CMB polarization measurement in the next decade.*

**Support of Detector Technology**

A significant effort was made by the Task Force to establish a rationale for a recommendation on the continued development of HEMT and bolometer detectors for the future of CMB research. The properties of the detectors and future advances that are anticipated are described in §7. Our findings are:

Bolometers have a clear advantage over MMICs for a potential space mission to measure the *B* modes to a level of $r = 0.01$. Nevertheless, MMICs cannot be ruled out as a possibility. For example, it may become clear that a mission with only $r = 0.05$ sensitivity is required in which case the technological challenge is reduced. The advantages of bolometers come in three areas.

1) The necessary sensitivity per device at many of the frequencies that will be important to the mission has already been demonstrated. For the MMICs to reach the same level will take, experts estimate, another five years.

2) The cooling requirements for bolometers are based on systems that have been demonstrated in space, such as on Spitzer. A technology for MMICs has been demonstrated for Planck, though a MMIC based mission will require considerable power. For example, the need to lift 10W at 20K is being considered. Such a spacecraft will take 4.5 kW (19m$^2$ of solar panels) and a 25 m$^2$ radiator at 45K. Additionally, the requisite thermal stability has not been demonstrated.

3) For the CMB (30–300 GHz), bolometers cover a wider frequency range than do MMICs. This will be especially important for identifying and removing contaminating foreground emission from the CMB signal. Bolometers still need to be developed for 30 GHz but no technological hurdles have been identified by anyone in the field. For competitive power consumption and sensitivity, MMICs are expected to top out at 120–150 GHz.

Our recommendations are the following:

1) Baseline bolometers for a space mission emphasizing the need to develop new sensors and architectures that minimize systematic effects.

2) Ensure that NASA continues to fund MMICs at a nominal level. They are a core technology with wide applicability and may prove ultimately to be a suitable choice for a space mission. They are critical for ground-based experiments that will be essential for designing the next space mission.

**Bolometers for the CMBPOL mission**

The most sensitive CMB temperature receivers to date, by a large margin, have been designed around bolometers. However, significant adaptations will be required for polarimetry. While there are no known fundamental problems for achieving either the frequency range or the stability for the total power differencing, a system achieving the demanding requirements of CMBPOL must be demonstrated prior to its use in space.

Bolometers have been used for many years to make high-fidelity polarization measurements at infrared to sub-millimeter wavelengths. Their use for studies of CMB polarization has begun relatively recently. Antenna coupled or similar bolometers hold particular promise because in principle they offer a simple and scalable method of selecting just one polarization of the incident field. They have been under development for a number of years, though efforts have intensified recently. Bolometers may have to operate at frequencies as low as 30 GHz to perform polarized foreground subtraction. The physical limitations of traditional absorbing structures for bolometers suggest that antenna coupling may be the best approach to reaching these low frequencies.

CMBPOL will require many detectors, but the necessary number is within the capabilities of multiplexing technologies now in development. At sub-millimeter wavelengths bolometer arrays with of order 1000 elements already exist and are being tested in the lab for SCUBA2, an instrument that will ultimately field 10,000 detectors. CMB research groups are currently developing large-format arrays to measure temperature anisotropy for SPT and ACT. These devices will be built, characterized, and fielded on balloons and the ground prior to their application to CMBPOL.

**Polarization Modulators**

To measure the polarization, two power signals must be subtracted to approximately one part in a hundred million in the final map. Any scheme to detect the polarization requires a way to modulate the polarization. In space this may be as simple as a rapidly spinning satellite, though many researchers believe that another level of modulation is required.

An active differencing and modulation scheme can provide this additional level of signal modulation, and may perhaps adapt techniques already applied successfully to bolometer-based astronomical polarimetry in the sub-millimeter band, although these all have limitations when applied to a broad spectral band. The development of modulation schemes requires ingenuity and the testing of multiple schemes, and no viable technique for space has yet been demonstrated. We anticipate that the design, development, and testing will carry on until 2011.

**Ground-based and balloon-borne program**

It is not possible to overemphasize the importance of supporting a vigorous ground- and balloon-based program to measure the CMB polarization. Many space missions (COBE, Spitzer, HST) were tested first with balloon payloads, and we expect balloon-based experiments to be important for testing concepts relevant to CMBPOL, as noted in earlier parts of this report. Furthermore, we expect that significant experience for CMBPOL will be gained from ground-based measurements. In particular, NASA should support ground-based experiments to test CMBPOL technology. There is ample precedent for this in the support of experiments that led to COBE and WMAP (see the sidebar on history accompanying §1 of this report).

**Balloon-borne program**

Existing balloon-based experiments for measuring the polarization include BOOMERanG and MAXIPOL. Both had their origins in temperature anisotropy experiments and were adapted for polarization. A new generation of balloon-based experiments, designed from the start to measure polarization, is being funded. They so far include PAPPA and EBEX.

**Coherent detectors**

Per measuring device, MMICs offer sensitivities competitive with bolometers at frequencies of 100 GHz and below. We strongly endorse continued, substantial support of ground-based measurements of CMB polarization using coherent (HEMT or MMIC) detectors as well as continued investment in new types of HEMT detectors.

At least two ground-based experiments to measure the CMB polarization are currently operating, CAPMAP and CBI. Additional efforts incorporating coherent detectors are anticipated. There are a number of compelling reasons for supporting research on coherent detectors in addition to the CMB science already coming from experiments incorporating coherent detectors. These include:

(1) There is a long history of CMB polarimetry with coherent systems. Hence, the measurement process is understood, and most if not all of the systematic errors are understood and can be modeled with high accuracy.

(2) The scientific impact of ground-based polarization measurements made with arrays of 100–1000 MMICs will be large and will inform future measurements.

(3) Coherent systems are already developed for 40 GHz and will unveil the polarized foreground emission long before 40 GHz bolometric systems are operational.

(4) Experimental techniques developed for coherent systems will be directly applicable to bolometric systems; examples are novel feed designs and orthomode transducers.

(6) Technology developments in Sb transistors and MMICs may be leveraged from efforts in the military and commercial sectors, so this development can be cost effective.

**Bolometric receivers**

Existing, funded bolometer experiments include BOOMERanG, BICEP, EBEX, MAXIPOL, MBI, and PAPPA, and QUaD. Such experiments and platforms will be the test beds for CMBPOL. Not only should important science result from them, but they will identify promising observing strategies and chart the foreground emission, especially from Galactic dust, in limited regions of the sky studied with high sensitivity.

We anticipate multiple generations of receivers with tests of different bolometric systems on a variety of platforms over the next seven years. Unlike the case for coherent receivers, there is not one clearly "best" method. We urge broad and sustained support from NASA, NSF, DoE and NIST to develop the technical and scientific base for CMBPOL. This is the basis for our recommendation on detector development, leading to arrays of thousands of polarization sensitive detectors and adequate support of facilities to produce them.

**Supporting observations of foreground emission**

Foreground emission will be a much more important concern for CMBPOL than it has been for WMAP or will be for Planck. Over much of the sky, the *B*-mode signal will be comparable or smaller than the uncorrected foreground emission (see figures 4.1 and 4.2). To plan CMBPOL, we must understand foreground emission, which is one of the reasons for including support for ground- and balloon-based precursor experiments in our major recommendation. WMAP will provide full sky maps at frequencies down to 22 GHz. To understand the emission mechanisms, polarization measurements near 5 and 10 GHz will be important for separating any potential spinning or magnetized dust component from the synchrotron component. The KUPID experiment covering 12–18 GHz is a first step towards this. At high frequencies, individual experiments will measure the foreground emission and Planck will give an all-sky measurement at 353 GHz (its highest frequency polarized channel). Deep pointed measurements at higher frequencies will be important for a clear understanding of the mechanisms of dust polarization. Lastly, little is known about the polarization properties of ensembles of extragalactic sources. More measurements are needed. Recent examples of the utility of such dedicated observations are the application of the WHAM and Finkbeiner-Davis-Schlegel maps to the WMAP analysis.

**Data analysis**

The CMB community currently uses several million processor hours of computer time each year, spread over everything from small, dedicated clusters to the biggest national flagship supercomputers. There is no doubt that we will need these resources and more to support the next generation of experiments, and a plan to meet these needs should be a part of any road map towards CMB polarization. National supercomputing facilities, such as DoE's NERSC at LBNL provide exceptionally useful resources, with both the capacity and the capability to support large collaborations, along with high-quality administration, user support, and long-term system upgrade plans. However, for an experiment to rely on such resources there must be guarantees both of their general availability over the lifetime of the experiment, and of their specific, timely, availability once the data start arriving. We encourage experimental groups to avail themselves of this resource when it makes sense, and we encourage the agencies supporting these centers to give CMB data analysis a high priority when making time allocations.

As discussed in the prior section, there needs to be a concerted effort on the part of the CMB data analysis community to develop algorithms for all stages of the data processing pipeline. This development must focus on a range of issues from experiment-specific systematic error analysis to the more general problems of polarization map-making, power spectrum estimation and parameter fitting. The most effective way to develop these algorithms is by working with real data, and/or with high-fidelity simulations of real data. To cultivate this approach, we call on experimental groups to make data and algorithms available to the larger community (for example, via NASA's thematic CMB data center, LAMBDA). We also encourage the support of groups who make CMB data analysis codes (such as CMBFAST, HEALPix, and MADCAP, to name a few) available to the community. These efforts have certainly benefited in the past and into the present from well-funded satellite data analysis programs (COBE, WMAP, and Planck) as well as the related LTSA, ADP, and ATP programs at NASA. We strongly encourage the continued support of these programs. We also call for a focused computational science research program that targets the implementation of CMB analysis codes on various parallel platform architectures to squeeze the most efficiency out of the data processing pipeline.

**Theory**

There is also a need for fundamental theoretical research on the CMB and foregrounds. While the basic paradigm of inflation is clear, there are many details that need attention before a satellite mission is undertaken, spanning the range from fundamental physics research to CMB phenomenology and foreground modeling.

**European programs**

The Task Force has concentrated its report on ongoing and planned activities in the United States related to a CMB polarization mission. In §5 we touched on a variety of ground-based efforts being planned in Europe, and we anticipate there will be future efforts to develop CMB satellite experiments.

**10.3 CMB research funding roadmap**

The Task Force has examined this program in terms of a variety of factors including cost. We have compiled an estimated funding profile based on the priorities detailed in this report, considering both past funding as a baseline as well as future resources. The profile shown in figure 10.3 is designed to:

– Augment core capabilities at the major detector facilities.

– Maintain robust and innovative university-based detector programs.

– Expand the very successful ground-based and balloon-based programs while maintaining existing efforts through to scientific completion.

– Maintain theoretical research to guide planned future observations and interpretation.

– Create a CMBPOL funding line for mission planning to address mission-specific design challenges and to develop mission-specific technologies.

Each of these goals is critical to the successful completion of the program we describe. Our analysis is the result of a significant prioritization exercise where many different plans and programs were presented.

To put the funding profile in context we collected data on the funding associated with CMB related research encompassing all of the categories detailed above. We did not include funding from private sources, mission operations and analysis costs for WMAP and Planck, and non-detector mission hardware costs for Planck. We also did not include essential logistical support provided by NSF's Office of Polar Programs and NASA's National Scientific Balloon Facility. Our baseline runs from 2003 to 2005 and future estimates are made in 2005 dollars.

The level of funding to implement the roadmap, as shown in figure 10.3, is quite comparable to the level that exists today. This is possible because large programs currently in place, such as the satellites and major ground-based facilities, will taper off in future years. The program therefore requires shifting these funds from existing efforts into new efforts that support the roadmap. The roadmap does not require an overall increase in the funding to the field as a whole. However, implementing the roadmap as proposed does require close coordination between DoE, NASA, NIST and NSF in order to optimize limited resources.

**Base Detector Technology**

The funding of detector development is central to realizing the new scientific capabilities planned in the roadmap. However, we have serious short- and long-term issues that need to be addressed. For the past five to seven years, detector development at major facilities at NIST, NASA, and DoE labs has relied heavily on funding derived from large related programs. These include mission hardware, such as Planck and Herschel, and to a lesser extent Code R technology funding from NASA. While satellite funding does not directly support detector technology used in non-mission experiments, it creates and maintains a core capability that can be leveraged to pioneer new detectors. In 2005, the support base provided by the Planck and Herschel missions will ramp to zero. In addition, NASA Code R funding was abruptly terminated in 2003. This represents a crisis to our program that must be immediately addressed. Our considered estimate is that $7M/year in base technology funding will be required to maintain the capabilities and continue the technology development program, leading to the most promising detector technologies by 2011. We note this is specifically base funding and does not include the cost of fabricating focal plane arrays for specific ground-based and sub-orbital experiments once the technology is in hand. Our estimate of $7M/year is commensurate with the level that existed prior to the demise of Code R.

The development of CMB detector technology will have broad applicability in other areas of science. For instance, the core technologies used in the CMB detectors (as described in §7) all have viable applications in space-borne far-infrared, UV and X-ray astronomy, in cryogenic dark-matter detection, in sub-millimeter cameras, and in ground-based instrumentation for optical photon counting and spectroscopy.

**Ground-Based Receiver Development**

CMB research requires a healthy level of interaction and feedback between detector development, receiver/instrument development, data analysis, and theory. Our program provides support for both receivers and assembled instruments to test the detectors, test schemes for systems integration, probe systematic effects, and test observing strategies and data collection techniques. Based on past activities in these areas we estimate a need for $10M/year to support ground-based receiver development. This program first of all requires completing and maintaining programs such as the South Pole Telescope and the Atacama Cosmology Telescope. However, as SPT and ACT funding tapers down, the overall level of ground-based funding must be maintained through 2011 to support a new generation of polarimeters with higher sensitivity. Such instruments will be needed to probe the power spectrum more deeply, to gather vital information about foregrounds, and to test in the field the future technologies needed for CMBPOL.

**Balloon-borne Receiver Development**

We anticipate that balloon-borne instruments will continue to play a vital role in CMB research. Compared to observations from the ground, balloon observations offer low atmospheric emission, permitting us to use a wider range of frequencies to explore foregrounds, and allowing an experiment to cover large regions of sky to explore lower multipoles than can be realistically measured from the ground. We foresee a modest increase in sub-orbital funding from the current level, increasing up to $3.5M/year.

**Satellite Mission Planning and Technology**

The program to launch a complex cryogenic satellite mission in 2018, six years after selection in 2012, requires a high level of mission planning as well as the advance development of mission-specific technologies. We anticipate that the goal of defining the detector and receiver elements will be completed by 2011. However, these technologies must then make the transition from demonstrations on the ground and balloons to reliable space-qualified systems. These critical technologies include not only the detectors themselves, but also the polarization modulators.

A space mission also requires the design and development of technologies that cannot be readily developed on a ground-based or sub-orbital platform. In particular, the cooling technologies for CMBPOL will be specific to space. These include passive thermal design, taking maximum advantage of the space environment, which may necessitate a multi-layer, deployed sunshield. An active cooling system or a long-duration cryostat providing a 2 K base temperature, operating from 30–50 K, must be developed from component technologies. Development of space-qualified, continuous, and highly stable coolers to 0.1 K will also require a dedicated funding line.

Significant mission planning will be necessary to understand experiment-specific challenges such as systematic error simulation and analysis, optical design, power requirements, data rates, and orbital and scanning parameters. NASA has already made an initial step in funding mission concept studies for CMBPOL in the Beyond Einstein program, and we recommend that this funding line be gradually increased up to mission selection.

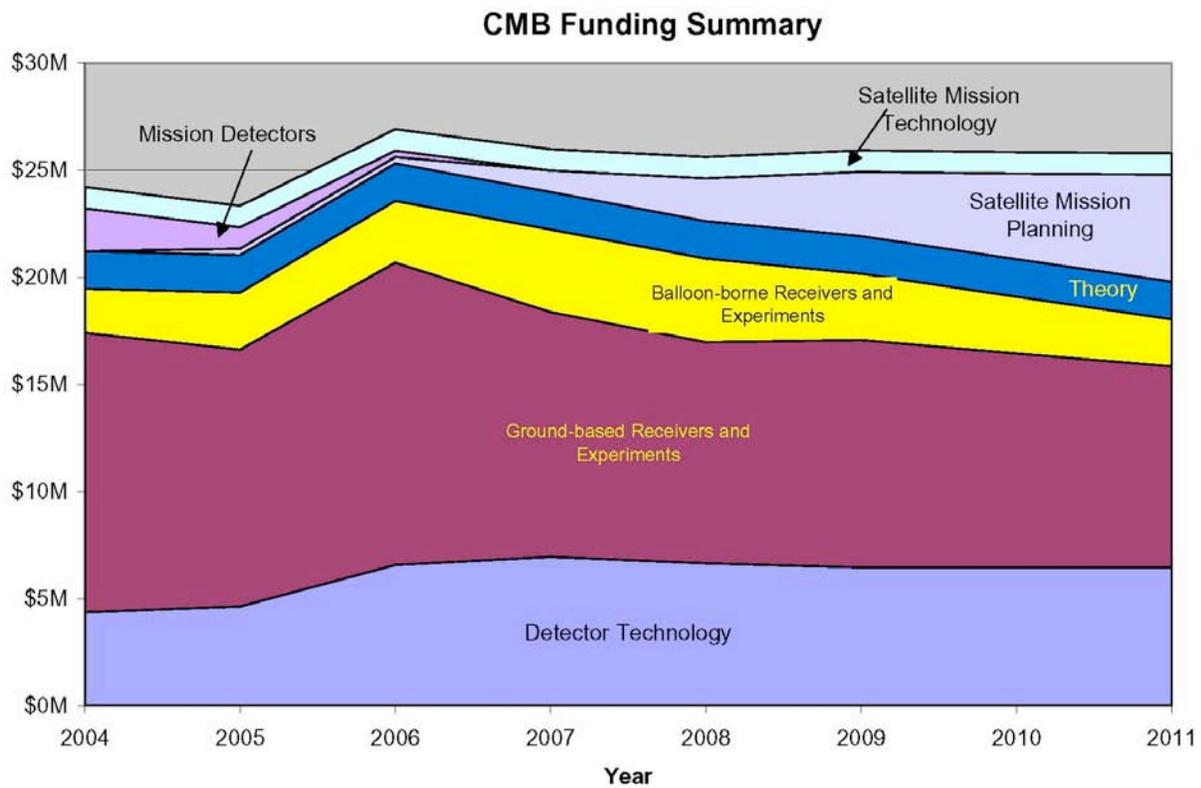

*Figure 10.3: Past (2004–2005) and projected (2006–2011) funding levels for CMB research. The plot is made in 2005 dollars.*

# 11 Small-scale temperature anisotropy and Sunyaev-Zel'dovich effect measurements.

## 11.1 Introduction

The *B*-mode polarization of the CMB is just one of the observables that will increase our understanding of the physics of the birth and evolution of the Universe. Considerable and complementary insight will be gained from measurements of the small-scale anisotropy in the CMB ($l > 800$, or angular scales smaller than a quarter of a degree) and measurements of the Sunyaev-Zel'dovich effect (SZE) in hundreds to thousands of galactic clusters. In this section we give an overview of these effects and then indicate how they complement the polarization measurements.

There are three parameters in the standard cosmological model that pertain to inflation and related theories. They are the tensor to scalar ratio, $r$, the scalar spectral index, $n_s$, and the running of the scalar spectral index, $\alpha = \dfrac{dn_s}{d\ln(k)}$, where $k$ is the wavevector measured in inverse megaparsecs. Our current knowledge of these parameters is based on small angular scale measurements of temperature fluctuations combined with large angular scale measurements. Indeed, in the first release of the WMAP data, the satellite measurements were combined with the CBI and ACBAR ground-based measurements to improve the parameter determination. The reason why different experimental results are needed is quite simple. Over a limited range of $l$ (or $k$) the parameters are degenerate with each other. In other words they may be traded off against each other to produce essentially the same observed spectrum. As the angular range over which the parameters are determined increases, they may be separated better. For example, $n_s$ represents the overall slope in figure 11.1. A factor of three increase in the range, say from $\Delta l = 1000$ to $\Delta l = 3000$, can lead to as much as a factor of three decrease in the uncertainty in $n_s$.

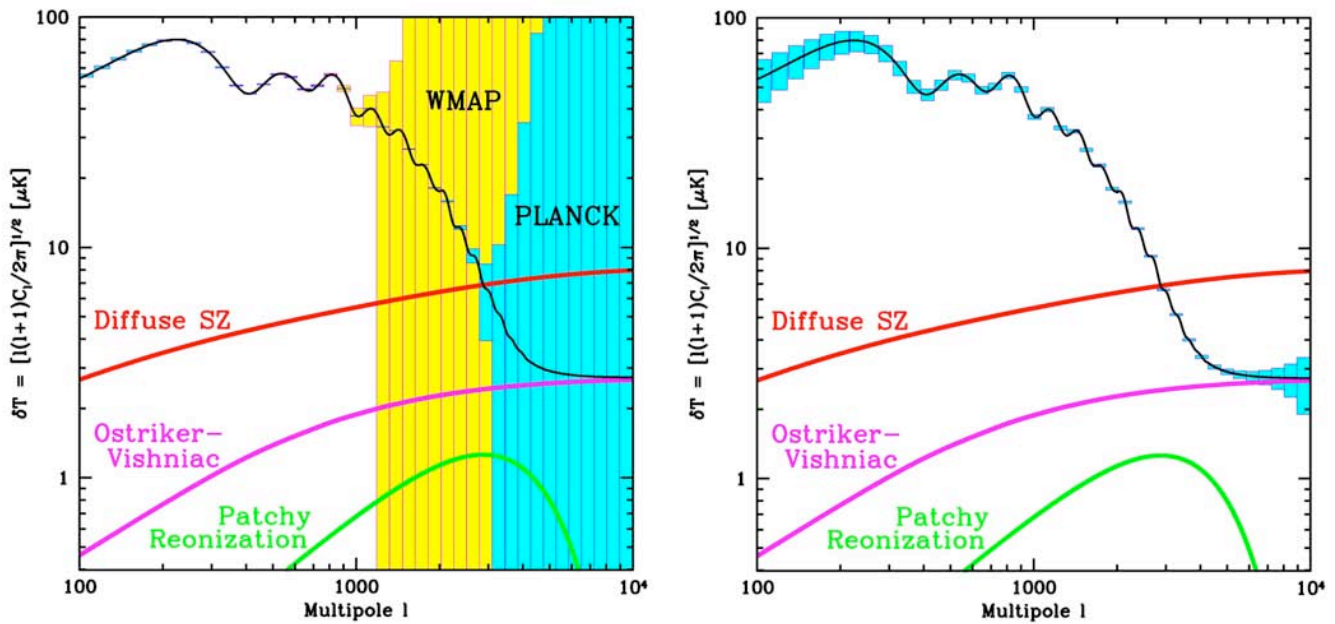

*Figure 11.1: The predicted angular power spectrum of the CMB from the quadrupole to arc minute scales. The angular scale is related to multipole as $\theta \approx \pi/l$. The linear regime covers the range $2 < l < 2000$. Secondary effects, dominant at $l > 3000$, are distinguishable by their spectral and spatial distributions. In the panel on the left, the boxes correspond to error bars for WMAP (measured), and Planck (anticipated). The panel on the right shows an estimate of the statistical error for the bolometer arrays on ACT and SPT. Figure courtesy of de Oliveira-Costa.*

In the SZE, energetic electrons at temperatures near $10^3$ eV in clusters of galaxies inverse Compton scatter CMB photons, altering the CMB spectrum. As a result, a cluster appears colder than the CMB at a frequency below about 217 GHz and hotter than the CMB at higher frequencies. The magnitude of the effect is large, of order 1 mK at 30 GHz in the most massive clusters.

The great utility of the SZE as a probe of cosmic structure is twofold. First, as one moves to higher and higher redshifts, the number of clusters decreases exponentially. The exponential cutoff means that the number versus redshift is a sensitive probe of the scale of the Universe as a function of distance. In turn this tells us about the equation of state of the Universe, as explained below. Secondly, the magnitude of the SZE is nearly independent of redshift.

**11.2 The Small Scale Anisotropy.**

To a good approximation we can divide the temperature anisotropy into a linear regime at angular scales larger than 5 arc minutes (multipoles $l<2000$), and a non-linear regime at smaller angular scales. In the linear regime, the microwave background fluctuations are dominated by primary anisotropies, described by small linear perturbations in the primordial plasma's density and velocity at the surface of last scattering. At smaller scales, the primary signal is suppressed because photons diffuse out of the gravitational potential wells (the so called "Silk damping") and the fluctuations average out over the line of sight. Here, the dominant contributions to the anisotropy are produced by nonlinear effects at much more recent epochs. Understanding the transition from linear to non-linear structure formation is crucial to our understanding of the cosmological parameters and of how the first structures formed.

The three major contributors to the power spectrum at small angular scales are (1) the diffuse SZE, due to the scattering of microwave photons from hot electrons in dense regions (not necessarily clusters of galaxies); (2) the Ostriker-Vishniac (OV) effect, from the scattering of microwave photons from free electrons after the reionization of the Universe (this is similar to the kinetic SZE [KSZ] discussed below, although the OV effect is diffuse); and (3) the Rees-Sciama effect and gravitational lensing, arising from variations in mass density along the path of the microwave photons, as discussed in §3. Figure 11.1 displays various contributions to the microwave background power spectrum for both large and small angular scales.

When combined, these different effects tell how cosmic structure formed as a function of redshift. In turn, given the standard model of cosmology, the growth rate of structure is a direct function of the equation of state of the dark energy, determined by the parameter *w*, the ratio of the cosmological pressure to its energy density, and of the mass of the neutrino.

Thus the small-scale anisotropy not only provides a key component for the tests of inflation through the parameters $n_s$ and $\alpha$, but it also tells us about the formation of the first objects, the dark energy, and the neutrino mass.

**11.3 Simulations and Analysis of the Sky at Arc Minute Resolution.**

Based on the standard model of cosmology, we can simulate the millimeter-wave sky at arc minute resolution to view some of these effects. Figure 11.2 shows the predictions for three frequency bands centered on 150 GHz, 217 GHz, and 265 GHz over a two square degree region of sky. At each frequency, the four major components – the primary CMB fluctuations, the SZE, the OV and KSZ effects, and extragalactic point sources – are separately shown along with their sum. The components have a variety of frequency and spatial characteristics that allow them to be separated.

The central goal of the analysis of the observed sky images is to extract these four components from the measurements, and to distinguish them clearly from foreground emission and instrumental systematic errors. Once the components are separated, the various effects are combined in one grand cosmological analysis. For example, the diffuse SZE is expected to dominate maps at 150 GHz at $l \approx 2500$, though the primary anisotropy should be visible in 217 GHz maps up to $l \approx 3000$.

The analysis of the high resolution sub-millimeter sky meshes well with trends in astrophysical simulation. While primary anisotropies dominate at degree scales and above, non-linear cosmological effects will be important at small scales. Thus, high fidelity and high resolution simulations of the sub-millimeter sky are essential. These simulations are expected to become as important to cosmology as the large Monte-Carlo programs are to high-energy physics.

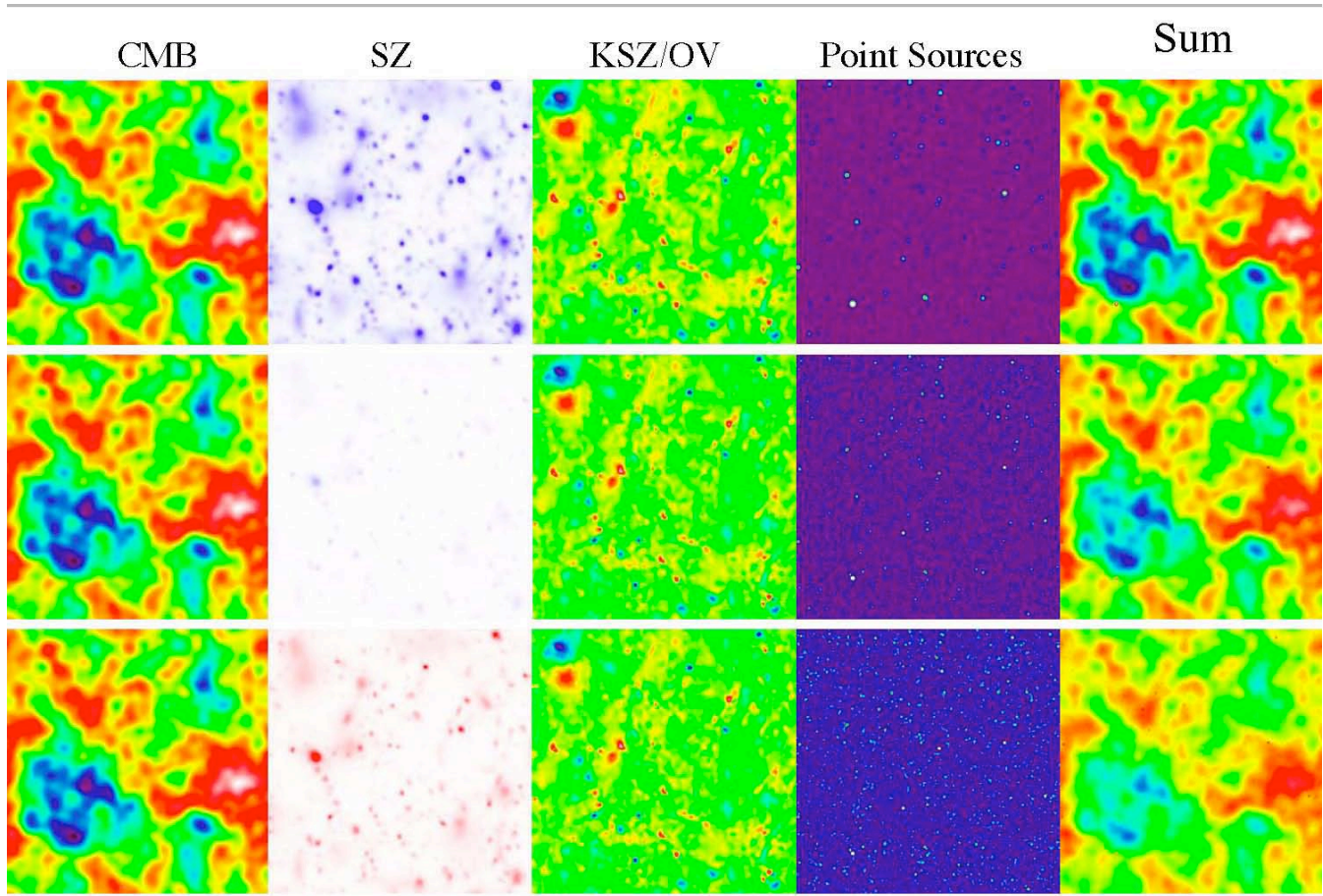

*Figure 11.2: Components of the sub-millimeter sky. From top to bottom, the panels correspond to 150, 217, and 265 GHz. Each panel shows a 1.4 by 1.4 degree patch of the sky. The various components that comprise the sub-mm sky are shown separately. The color scale has blue as negative and red as positive. Thus, at 150 GHz the SZE shows a decrement, at 217 GHz there is almost no signal, and at 265 GHz there is an increment.*

*Figure courtesy of U. Seljak, J. Burwell, and K. Huffenberger.*

There are a number of anticipated correlations between the diffuse CMB-related signals and galaxies as measured with optical telescopes. For example, the same matter distribution that leads to weak lensing of galaxy images also distorts the CMB. This lensing of the CMB is measurable. When combined with the optical measurements it yields a picture of the matter distribution to high redshift.

**11.4 Program to Measure the Small Scale Anisotropy.**

There is an active and growing program to measure the CMB at small angular scales. Currently, experiments with tens of detectors such as ACBAR, CBI and the VSA are unveiling the arc minute CMB sky. Though we do not yet have maps as detailed as the simulations above, future experiments may match this resolution and sensitivity.

Not only is the science from the small-scale anisotropy inextricably intertwined with that for the CMB polarization measurements, but the technology is as well. The program to measure the small-scale anisotropy requires that arrays of thousands of detectors be built and matched to dedicated telescopes. The numbers of detectors is simply dictated by the sensitivity levels required to clean the signal of foreground emission and to clearly identify the constitutive components of the sub-millimeter sky. Medium scale telescopes are required to reach arc minute scales. For example, the diffraction limit of a 6m telescope with a typical feed at 100 GHz is approximately 2 arc minutes, corresponding to $l$=6000. As the effects one endeavors to measure are at roughly a part in a billion of the temperature of the local environment, special purpose, carefully shielded, telescopes are required.

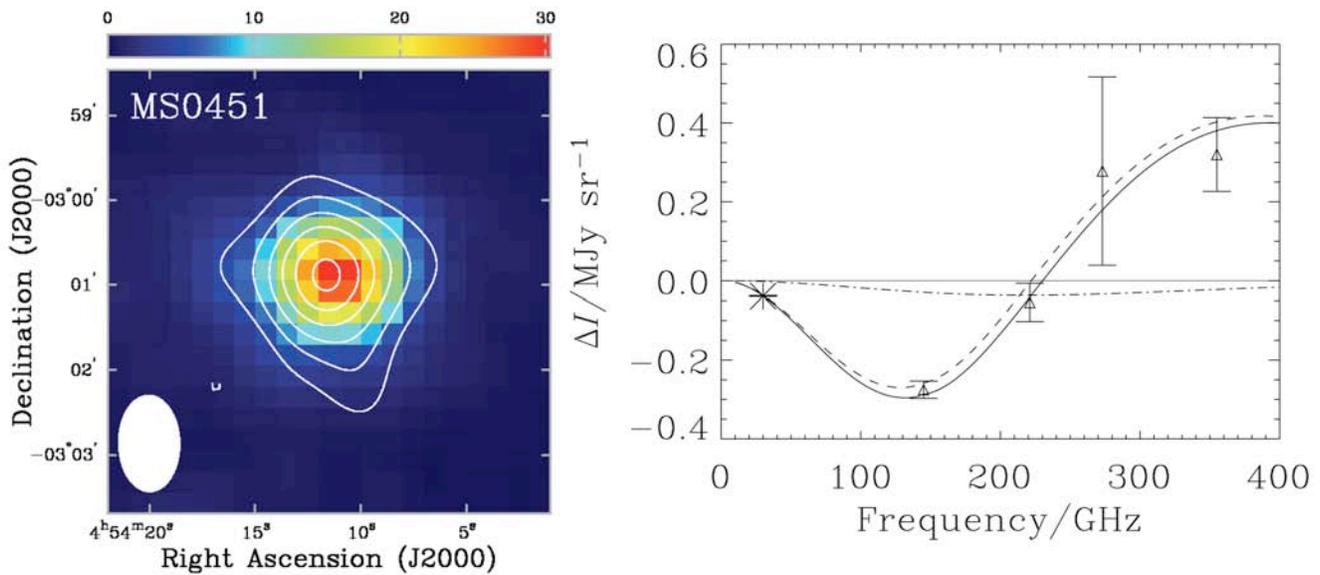

*Figure 11.3: Current experiments measure both the spatial distribution and the spectrum of the SZE. Left: A measurement by the BIMA team shows the SZE as contours plotted over the X-ray emission from the hot gas in the cluster. Right: The triangles show measurements from the SuZIE experiment of the spectrum of the SZE; the star is the 30 GHz measurement taken from the picture on the left. The thermal SZE has a distinctive spectrum, allowing clusters to be distinguished from CMB fluctuations. The dot-dash line is an extra component to the SZE caused by the extra motion of the cluster in addition to the expansion of the Universe (the Kinetic SZE).*

There are two telescopes under construction that are designed specifically to measure the small-scale structure in the CMB. One is the 10m South Pole Telescope (SPT) at the South Pole. The other is the 6m Atacama Cosmology Telescope (ACT) located in northern Chile, near the future site of the Atacama Large Millimeter Array (ALMA). Both telescopes are surrounded by large ground screens and scan the sky by moving the entire structure, to avoid systematics. Special purpose "cameras" with arrays of thousands of detectors specifically designed to couple to the telescope are under construction. These systems will be used to measure the small-scale anisotropy, the SZE, and ultimately the CMB polarization.

**11.5 The Sunyaev-Zel'dovich Effect**

The Sunyaev-Zel'dovich Effect (SZE), the scattering of CMB photons by the hot gas in galaxy clusters, was one of the earliest CMB distortions to be detected. The techniques and technology used for SZE measurements are similar to those used to detect primordial CMB anisotropies and indeed in some cases SZ programs served as technology development and test environments for later experiments. Figure 11.3 shows measurements of the SZE made by the BIMA and SuZIE experiments. BIMA's HEMT amplifiers were designed by NRAO and are the same as those used in the DASI and CBI interferometers and WMAP, which have mapped both temperature and polarization anisotropies. The spider-web NTD germanium bolometers in SuZIE were development precursors to the bolometers which were used in BOOMERanG, and which will fly in Planck. That synergy remains, as many of the techniques that are required for CMB polarization measurements (large-format bolometer arrays, low-loss optical components and filters) will first be used in upcoming ground-based experiments that survey the SZ effect.

The SZ signal must be distinguished from other astrophysical signals, particularly CMB fluctuations and sub-millimeter emission from galaxies, which generally requires observations at more than one frequency. This requirement is shared by CMB polarization measurements.

Moreover, the science goals of the SZ surveys – to chart the emergence of cosmic structure with the aim of clarifying the cosmological model (especially the dark sector) – are intimately linked to those of CMB polarization science. In this spirit, the goals of the SZ program should be fully integrated into the program to map the CMB polarization.

**11.6 Sunyaev-Zel'dovich Science**

The SZE offers a means to detect and map the hot gas in galaxy clusters. SZ measurements are complementary to both X-ray data, and to optical lensing measurements that probe the dark matter in the cluster (figure 11.4). There are two components to the SZ spectrum, each with a distinct spectral shape. The thermal component is proportional to the temperature of the electrons in the plasma, and produces a decrement in brightness at low frequencies and an increment in brightness at high frequencies, with a null near 217 GHz, as shown in figure 11.3.

Relativistic corrections cause the shape of the thermal SZ spectrum to depend weakly on electron temperature, but to first order the spectral shape is independent of other cluster parameters. The kinematic component of the SZ effect, also shown in figure 11.3, is the result of the bulk motion of the plasma with respect to the Hubble expansion and is the only method that can be used to directly measure cluster peculiar velocities without using distance indicators.

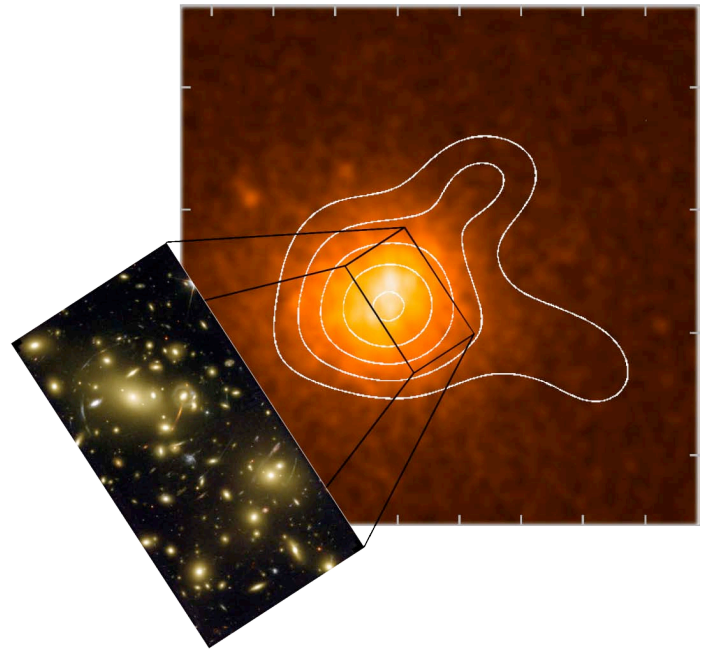

*Figure 11.4: A montage showing the SZ effect (white contours) from the rich cluster of galaxies A2218, measured by BIMA. The SZ effect is well correlated with the hot gas in the cluster (orange) as measured using X-ray emission by the XMM satellite. The inset shows the location of the galaxies in the cluster. Figure courtesy of John Peterson.*

**11.7 SZ Surveys as a Probe of Dark Energy**

Surveys of clusters of galaxies have been proposed as a means to measure key parameters of dark energy, including the equation of state parameter, w, and its evolution, if any, with time. The number density of clusters depends on two functions, both with cosmological dependence: (i) the redshift dependence of the size of a volume element, which depends on the evolution of the scale factor with time and (ii) the cluster mass distribution function which depends on the growth of structure and thus on the relative abundances of matter and dark energy as a function of time. Cluster samples obtained by using the SZ effect as the selection criterion are particularly attractive for two reasons. First, the brightness of the effect is independent of distance, so clusters can be found over a large range of redshifts. Second, it is expected that the correlation between the SZ flux and the cluster mass, which is the quantity of interest, is especially tight. A recent measurement by the SuZIE experiment showed a strong correlation between SZ flux and X-ray temperature (a mass indicator) indicating the reasonableness of this assumption.

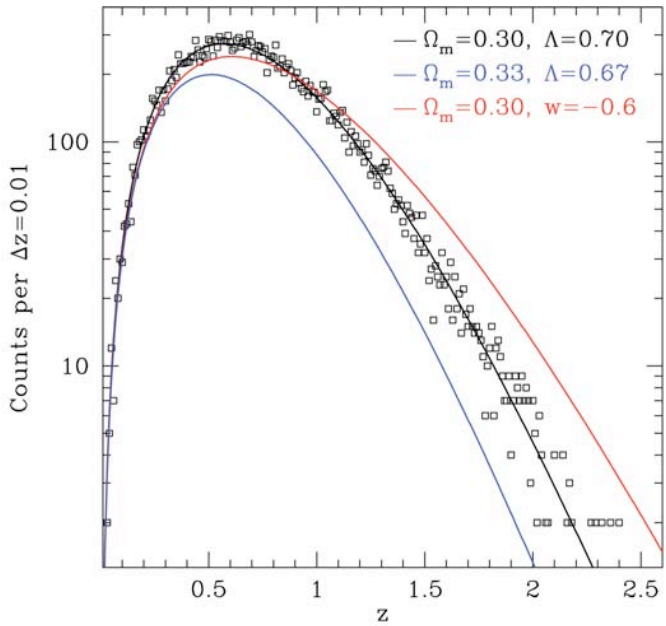

*Figure 11.5: A simulation of an SZ and redshift cluster survey. The number of clusters as a function of distance (parameterized by redshift, z) is sensitive to the cosmic density of dark matter ($\Omega_m$), the effective density of dark energy ($\Lambda$), and the equation of state of dark energy (w), which determines how the density of dark energy changes as the Universe expands. The open squares are simulated data points from a 4000 square degree survey. Figure courtesy of the SPT team.*

Figure 11.5 shows how the number of clusters as a function of distance depends on dark energy parameters. This crucial function will be measured by multiple experiments. Radio interferometers are likely to play a key role in this field. The SZA, a 30 GHz interferometer with 8 elements, will come on line in 2005 and is expected to detect approximately 100 new clusters. Bolocam, a 151-element NTD germanium bolometer arrays has already been used and is expected to yield new results. The APEX-SZ experiment, a bolometric array receiver to be mounted on the 12m APEX telescope in Chile, is anticipated to be the first application of large format TES bolometer arrays to CMB observations and can detect many hundreds of clusters. The ACT and SPT will also measure the SZE, and Planck will measure the SZE of the most massive clusters across the full sky.

**11.8 Cluster Peculiar Velocities**

The peculiar velocity of a cluster – its motion relative to the overall Hubble expansion of the Universe – is a tracer of mass. Consequently, peculiar velocity surveys can be used to probe the density and distribution of dark matter in the Universe with little or no dependence on the behavior of dark energy. The kinematic SZE can be used to directly measure peculiar velocities, in contrast to other techniques that compare the measured velocity from redshift data to the expected velocity from the Hubble flow based on a distance determination. Additionally the brightness of the kinematic SZE is redshift independent, allowing the peculiar velocity field to be measured at any redshift. The technique requires spectral measurements of the SZE at three or more different frequencies to separate the thermal and kinematic components from each other and from astrophysical confusion. In principle, a sample of N peculiar velocities, measured with a precision of 400 km/s, can be used to determine the density parameter, $\Omega_m$, to a precision of $\Omega_m = 0.02 \sqrt{(400/N)}$.

**Appendix A: Acronyms used in this report**

ACBAR: Arcminute Cosmology Bolometer Array Receiver (a current South Pole CMB experiment)

ACT: Atacama Cosmology Telescope (a current ground CMB experiment)

ADP: Astrophysics Data Program (a NASA grant program)

ADR: Adiabatic Demagnetization Refrigerators

ALMA: Atacama Large Millimeter Array (an NSF radio telescope under construction)

APEX: Atacama Pathfinder EXperiment

APEX-SZ: SZE experiment on APEX

ATP: Astrophysics Theory Program (a NASA grant program)

BICEP: Background Imaging of Cosmic Extragalactic Polarization (a past South Pole CMB experiment)

BIMA: Berkeley-Illinois Millimeter Array

BLIP: Background Limited Infrared Photodetector (the best performance possible)

BOOMERanG: Balloon-borne Observations Of Millimetric Extragalactic Radiation and Geophysics (a current and future CMB experiment)

B2K: BOOMERanG 2000

BRAIN: Background RAdiation INterferometer (a future ground CMB experiment)

CAPMAP: Cosmic Anisotropy Polarization MAPper (a current ground CMB experiment)

CBI: Cosmic Background Imager (a current ground CMB experiment)

CCD: Charge Coupled Device (a detector technology)

ClOVER: $C_l$ ObserVER, aka CMB Polarization Observer (a future ground CMB experiment)

CMB: Cosmic Microwave Background

CMBFAST: not an acronym – software to compute predictions of the CMB anisotropy spectrum

CMBPOL: CMB polarization mission (straw man *Beyond Einstein* concept)

COBE: Cosmic Background Explorer (a past space CMB experiment)

COMPASS: COsmic Microwave Polarization At Small Scales (a past ground CMB experiment)

DARPA: Defense Advanced Research Projects Agency

DASI: Degree Angular Scale Interferometer (a past South Pole CMB experiment)

DoE: Department of Energy

EBEX: The E and B EXperiment (a future balloon CMB experiment)

EMI: Electromagnetic Interference

EMC: Electromagnetic Compatibility

ESA: European Space Agency

GB6: Sixth Green Bank survey (at 4.85 GHz)

GSFC: Goddard Space Flight Center

GUT: Grand Unification Theories

HAWC: High-resolution Airborne Wide bandwidth Camera (a far-infrared camera on SOFIA)

HEALPix: Hierarchical Equal Area isoLatitude Pixelisation of the sphere (a spherical pixelization scheme used by many CMB investigations)

HEMT: High Electron Mobility Transistor (a detector technology)

HST: Hubble Space Telescope

JCMT: James Clerk Maxwell Telescope (a UK millimeter telescope)

JFET: Junction Field Effect Transistor (an electronics technology)

JPL: Jet Propulsion Laboratory

KSZ: Kinetic Sunyaev-Zel'dovich effect

KUPID: Ku-band Polarization IDentifier

LBNL: Lawrence Berkeley National Laboratory

LLNL: Lawrence Livermore National Laboratory

LTSA: Long Term Space Astrophysics (a NASA grant program)

MADCAP: Microwave Anisotropy Dataset Computational Analysis Package

MAXIMA: Millimeter wave Anisotropy eXperiment IMaging Array (a past balloon CMB experiment)

MAXIPOL: not an acronym – a polarization version of MAXIMA

MBI: Millimeter-wave Bolometric Interferometer

MKID: Microwave Kinetic Inductance Detectors

MMIC: Monolithic Microwave Integrated Circuit (an electronics technology)

MSAM2: Medium Scale Anisotropy Measurement (a past balloon CMB experiment)

NASA: National Aeronautics and Space Agency

NET: Noise Equivalent Temperature

NIST: National Institute of Standards and Technology

NRAO: National Radio Astronomy Observatory

NSF: National Science Foundation

NTD: Neutron Transmutation Doped (an electronics technology)

NVSS: NRAO VLA Sky Survey

OMT: Ortho-Mode Transducer

OV: Ostriker-Vishniac

PAPPA: Primordial Anisotropy Polarization Pathfinder Array (a future balloon CMB experiment)

PIQUE: Princeton I Q U Experiment (a past ground CMB experiment)

POLAR: Polarization Observations of Large Angular Regions (a past ground CMB experiment)

PSB: Polarization Sensitive Bolometer

QUaD: QUEST at DASI (a current South Pole CMB experiment)

QUEST: Q and U Extragalactic Submillimeter Telescope

SCUBA2: Submillimeter Common User Bolometer Array 2 (an instrument on the JCMT)

SEQUOIA: SEcond QUabbin Optical Imaging Array (a 3mm receiver for the Mexican Large Millimeter Telescope)

SHARC: Submillimeter High Angular Resolution Camera

SOFIA: Stratospheric Observatory For Infrared Astronomy

SPT: South Pole Telescope

SQUID: Superconducting QUantum Interference Device

SuZIE: Sunyaev-Zel'dovich Infrared Experiment

SZ: Sunyaev-Zel'dovich

SZA: Sunyaev-Zel'dovich Array

SZE: Sunyaev-Zel'dovich Effect

TES: Transition Edge Sensor

TFCR: Task Force on CMB Research (this group)

VLA: Very Large Array (a radio telescope complex in New Mexico)

VSA: Very Small Array (a current ground CMB experiment)

WHAM: Wisconsin H-Alpha Mapper

WMAP: Wilkinson Microwave Anisotropy Probe (a current space CMB experiment)

XMM: X-ray Multi-Mirror mission (now the XMM-Newton observatory)

XQC: X-ray Quantum Calorimeter (a rocket experiment to observe the diffuse X-ray background)

XRS: X-Ray Spectrometer (a future space X-ray experiment)